\documentclass{article}

\usepackage{arxiv}

\usepackage[utf8]{inputenc} 
\usepackage[T1]{fontenc}    
\usepackage{hyperref}       
\usepackage{url}            
\usepackage{booktabs}       
\usepackage{amsfonts}       
\usepackage{nicefrac}       
\usepackage{microtype}      
\usepackage{lipsum}		
\usepackage{graphicx}
\usepackage{natbib}
\usepackage{doi}
\usepackage{caption}
\usepackage{subcaption}
\usepackage{algorithm}
\usepackage{algorithmic}

\title{A Modified de Casteljau Subdivision\\ that Supports Smooth Stitching with \\Hierarchically Organized Bicubic B\'{e}zier Patches}

\date{October 5, 2023}	

\author{Saied Zarrinmehr \\
	Computer Science and Engineering Department,\\ Texas A\&M University, College Station, TX, 77831\\
	\texttt{szarinmehr@gmail.com } \\
 \And
 \href{https://orcid.org/0000-0003-3618-4166}{\includegraphics[scale=0.06]{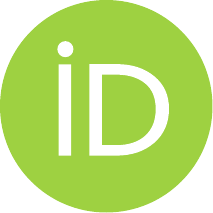}\hspace{1mm}Ergun Akleman}\thanks{Joint with Computer Science and Engineering Department.} \\
	Visual Computing \& Computational Media,\\ Texas A\&M University, College Station, TX, 77831\\
	\texttt{ergun@tamu.edu} \\
	   \And
Jianer Chen\\
 Computer Science and Engineering, \\
 College of Engineering\\
 Texas A\&M University, \\College Station, TX, 77831\\
	\texttt{chen@cs.tamu.edu} \\
}

\renewcommand{\shorttitle}{Smooth Stitching}

\hypersetup{
pdftitle={A template for the arxiv style},
pdfsubject={q-bio.NC, q-bio.QM},
pdfauthor={David S.~Hippocampus, Elias D.~Striatum},
pdfkeywords={First keyword, Second keyword, More},
}

\begin{document}
\maketitle

\begin{abstract}
One of the theoretically intriguing problems in computer-aided geometric modeling comes from the stitching of the tensor product B\'{e}zier patches. When they share an extraordinary vertex, it is not possible to obtain continuity $C^1$ or $G^1$ along the edges emanating from that extraordinary vertex.  Unfortunately, this stitching problem cannot be solved by using higher degree or rational polynomials. In this paper, we present a modified de Casteljau subdivision algorithm that can provide a solution to this problem.
The main advantage of the modified subdivision is that the continuity $C^1$ on a given boundary edge does not depend on the positions of the control points on other boundary edges. The modified subdivision allows us to obtain the desired $C^1$ continuity along the edges emanating from the extraordinary vertices along with the desired $G^1$ continuity in the extraordinary vertices (see Figures~\ref{fig_3-valent} and~\ref{fig_10-valent}). 
Our modified de Casteljau subdivision, when combined with topological modeling, provides a framework for interactive real-time modeling of piecewise smooth manifold meshes with arbitrary topology. Based on this approach, we have extended an interactive modeling system to topologically update a smooth surface in real-time by opening or closing holes, creating handles, and combining and disconnecting surfaces (see Figure~\ref{fig_OpeningHole} for an example). 
The smooth surfaces resulting from any given orientable two-manifold polynomial mesh have the following properties: (1) These surfaces consist of hierarchically organized $C^2$ continuous bicubic B\'{e}zier patches. (2) They provide visual smoothness by providing geometric $G^1$ continuity in the extraordinary vertices and $C^1$ continuity along the edges emanating from the extraordinary vertices. (3) They can be updated interactively in real time by directly evaluating the underlying B\'{e}zier patches.

\end{abstract}

\section{Discussion}
\label{sec_discussion}

\begin{figure}[htpb]
    \centering  
    \begin{subfigure}[t]{0.490\textwidth}
        \includegraphics[width=1.0\textwidth]{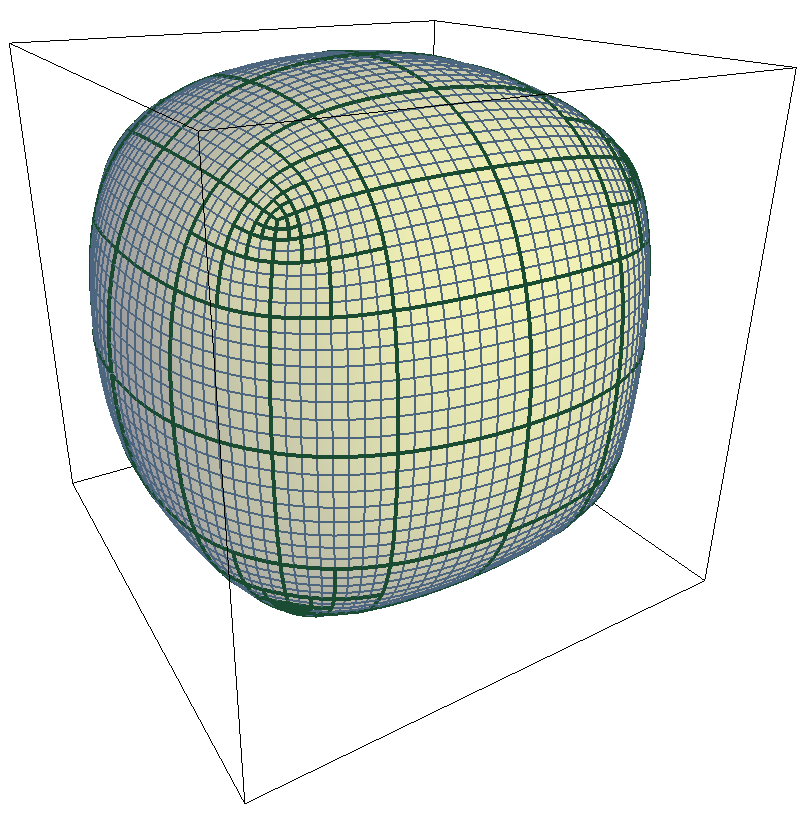}
        \caption{ Smoothed cube with our method.}
        \label{fig_cube_6}
    \end{subfigure}
    \hfill  
    \begin{subfigure}[t]{0.490\textwidth}
        \includegraphics[width=1.0\textwidth]{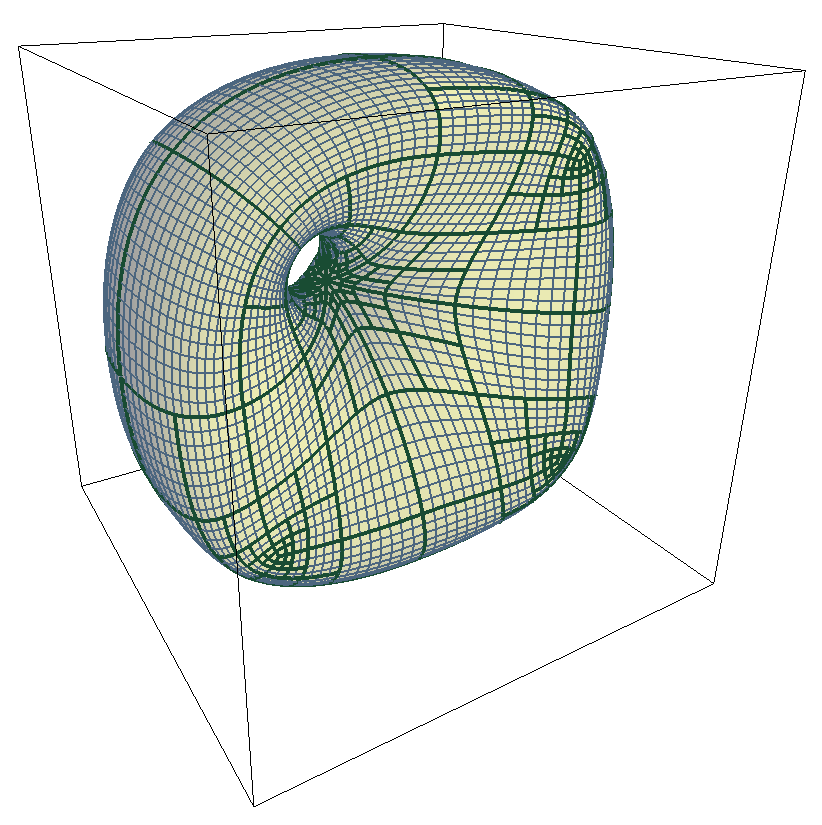}
        \caption{ Opening a hole by inserting an edge.}
        \label{fig_10SidedFace_6}
    \end{subfigure}
    \hfill 
\caption{Two images from our interactive modeling session with modified de Casteljau subdivision. Users can visually experience that inserting an edge between two opposite faces of the cube opens a $G^1$ smooth hole by creating a 10-sided face. }
\label{fig_OpeningHole}
\end{figure}

Tensor product parametric polynomials and rational parametric polynomials are some of the simplest representations to define surfaces. Using basis functions such as B\'{e}zier or B-spline, parametric polynomial representations of the tensor product become intuitive to describe the shapes of the final surfaces \cite{bartels1987}. The coefficients of the B\'{e}zier or B-spline functions are simply considered to be the control points of the final shape of the surface. B\'{e}zier patches are popular since they are particularly intuitive \cite{decasteljau1963,bezier1974}. The resulting surface interpolates four of the control points, which are the corners of the patch. The derivatives and second derivatives in the corners of the patch are also intuitively defined by a set of neighboring control points. 

Despite their simplicity and intuitiveness, there are significant problems with the forms of tensor products in terms of their representational power. Since their domain is simply a 2D square, they can only represent a very limited set of topologically distinct two-manifold surfaces. They can represent only those that can be given as maps from the 2D square domain, namely genus-0 or genus-1 surfaces. Moreover, with tensor product forms, it is not possible to include non-quadrilateral faces or non-four-valent vertices, which are usually called extraordinary vertices. This limitation makes it harder to model even surfaces of genus-0 or genus-1 \cite{akleman2006}.

\begin{figure}[htpb!]
    \begin{subfigure}[t]{0.490\textwidth}
        \includegraphics[width=1.0\textwidth]{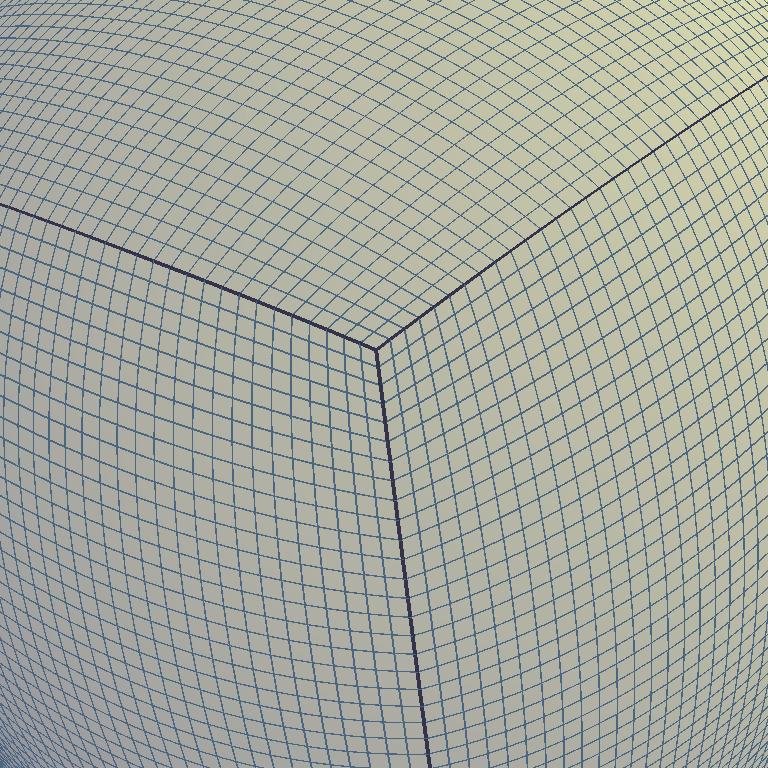}
        \caption{ Detail using method in \cite{akleman2017}.}
        \label{fig_tetra0}
    \end{subfigure}
    \hfill     
    \begin{subfigure}[t]{0.490\textwidth}
        \includegraphics[width=1.0\textwidth]{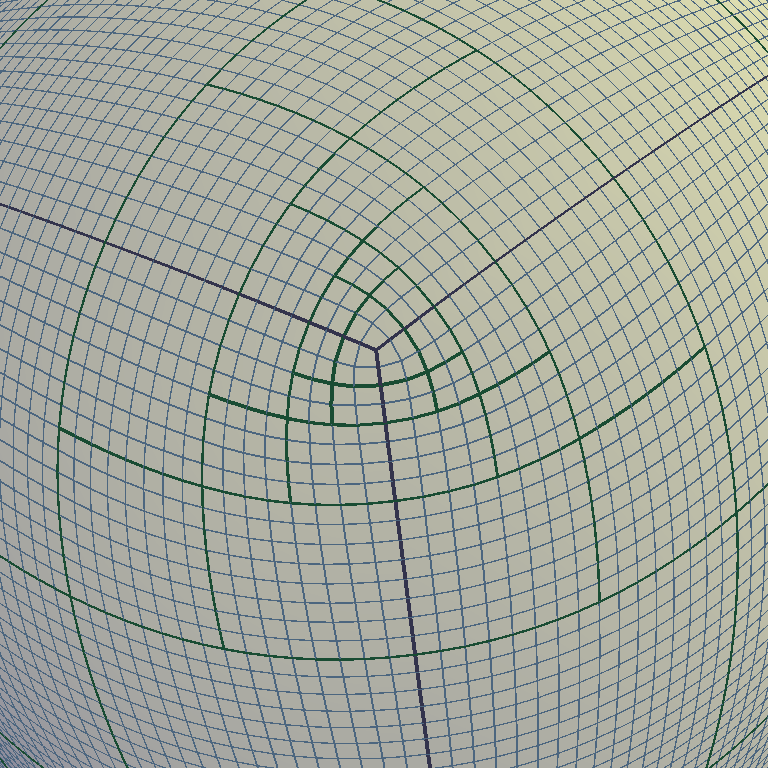}
        \caption{ Detail using our method.}
        \label{fig_tetra1}
    \end{subfigure}
    \hfill 
\caption{Comparison our method with AS\&C-method \cite{akleman2017} around a three-valent extraordinary vertex. As it can be seen in this example, AS\&C-method creates broken lines that cause $C^1$ discontinuity along the edges emanating from that extraordinary vertex while our method removes $C^1$ discontinuities by using hierarchically organized B\'{e}zier patches that are constructed by our modified de Casteljau subdivision. }
\label{fig_3-valent}
\end{figure}

To overcome this problem, one approach has always been to smoothly stitch quadrilateral patches. Martin Newell's famous teapot consists of smoothly stitched bicubic B\'{e}zier patches \cite{newell1975}. Smoothly stitched bicubic B\'{e}zier patches were also common during the late 1980s and early 1990s. Geometric modeling practitioners in the special effects and animation industry used smooth stitched quadrilateral patches to create complicated surfaces\footnote{Private communications and observations of professional modelers who worked in special effects and computer animation companies during the 1990s.}. Visual smoothness along the patch boundaries was achieved primarily by manually moving the control vertices. 

It is fascinating that professional modelers still manually obtained nice-looking shapes in that time with this approach despite this being a laborious and time-consuming process that is prone to errors. Because it is not theoretically possible to obtain derivative continuity around the extraordinary vertices, the success of early modelers poses an intriguing and captivating question. A recent paper by Akleman, Srinivasan, and Chen (AS\&C) \cite{akleman2017} has contextualized the ground for this discussion and provided insight into this question \cite{peters2017,akleman2017r}. 

AS\&C demonstrated that it is easy to construct bicubic B\'{e}zier patch control polyhedra that appear to be stitched with $G^1$ continuity regardless of the underlying mesh topology \cite{akleman2017}. AS\&C-method provided better visual smoothness than the Catmull-Clark subdivision \cite{catmull1978} for extraordinary vertices of very high valence by providing $G^1$ continuity at these points. However, subsequent discussion about this issue revealed that regardless of how we position control vertices,  it is still not possible to obtain $C^1$ or $G^1$ continuity along the edges emanating from that extraordinary vertex \cite{peters2017,akleman2017r}. On the other hand, it turned out that their process causes only subtle discontinuities. Moreover, these subtle discontinuities of $G^1$ occur on continuous curves $G^2$, although it is not often even possible to see the problem from the silhouette of the shape.

\begin{figure}[htpb!]
    \begin{subfigure}[t]{0.490\textwidth}
        \includegraphics[width=1.0\textwidth]{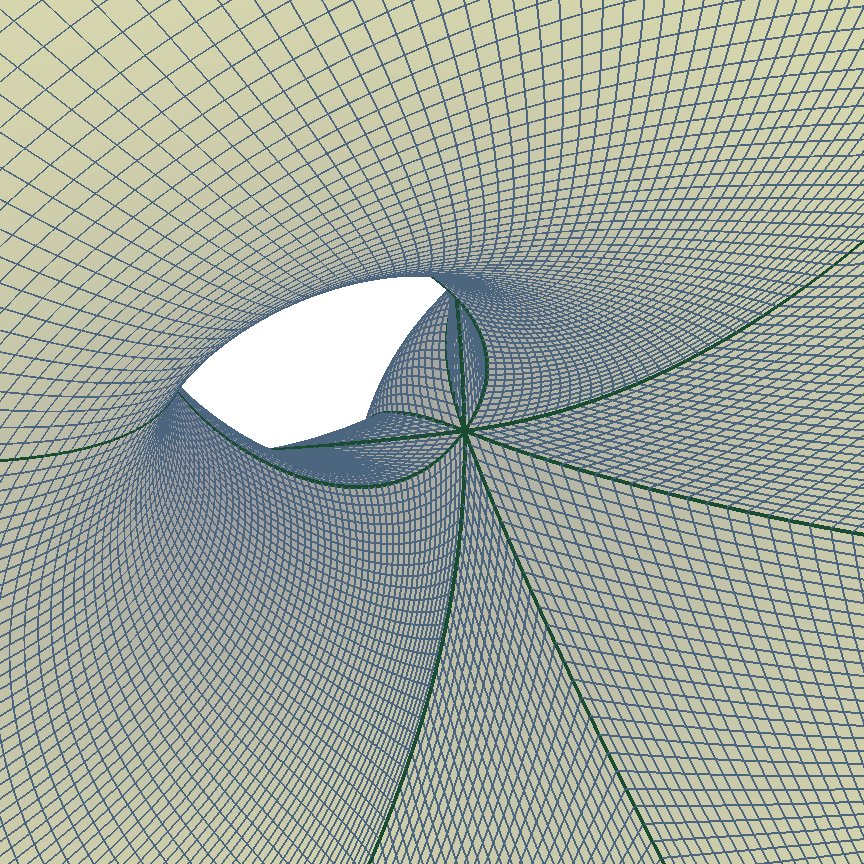}
        \caption{ A detailed view around a 10-valent extraordinary vertex obtained with the method in \cite{akleman2017}.}
        \label{fig_cubehole0}
    \end{subfigure}
    \hfill     
    \begin{subfigure}[t]{0.490\textwidth}
        \includegraphics[width=1.0\textwidth]{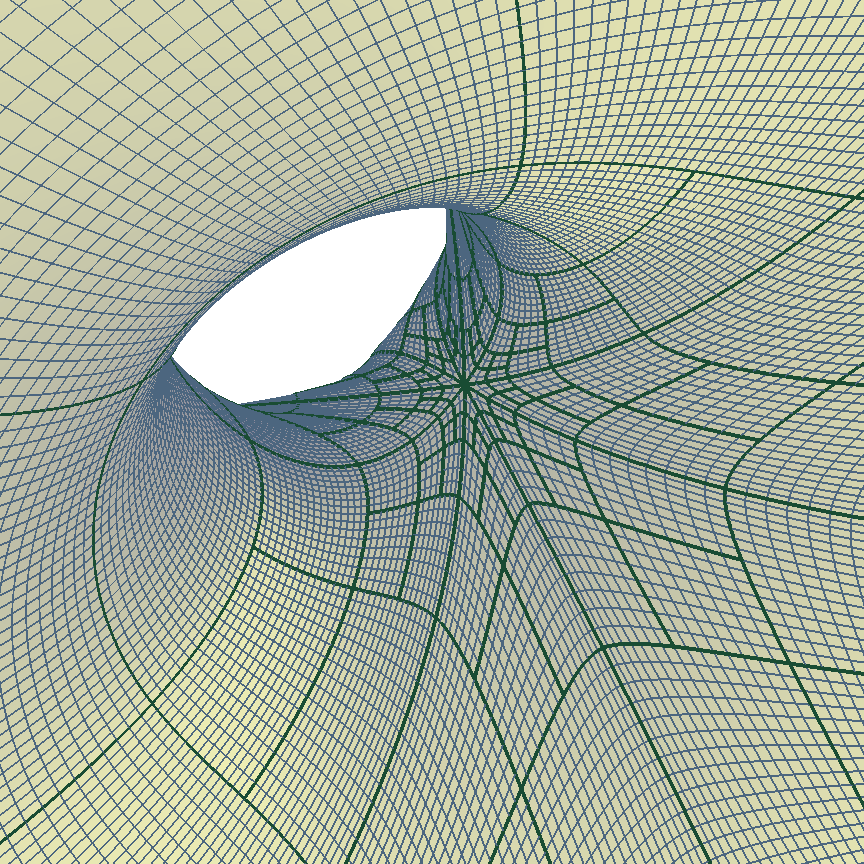}
        \caption{ The same view of the same neighborhood was obtained with our method.}
        \label{fig_cubehole2}
    \end{subfigure}
    \hfill
        \begin{subfigure}[t]{0.490\textwidth}
        \includegraphics[width=1.0\textwidth]{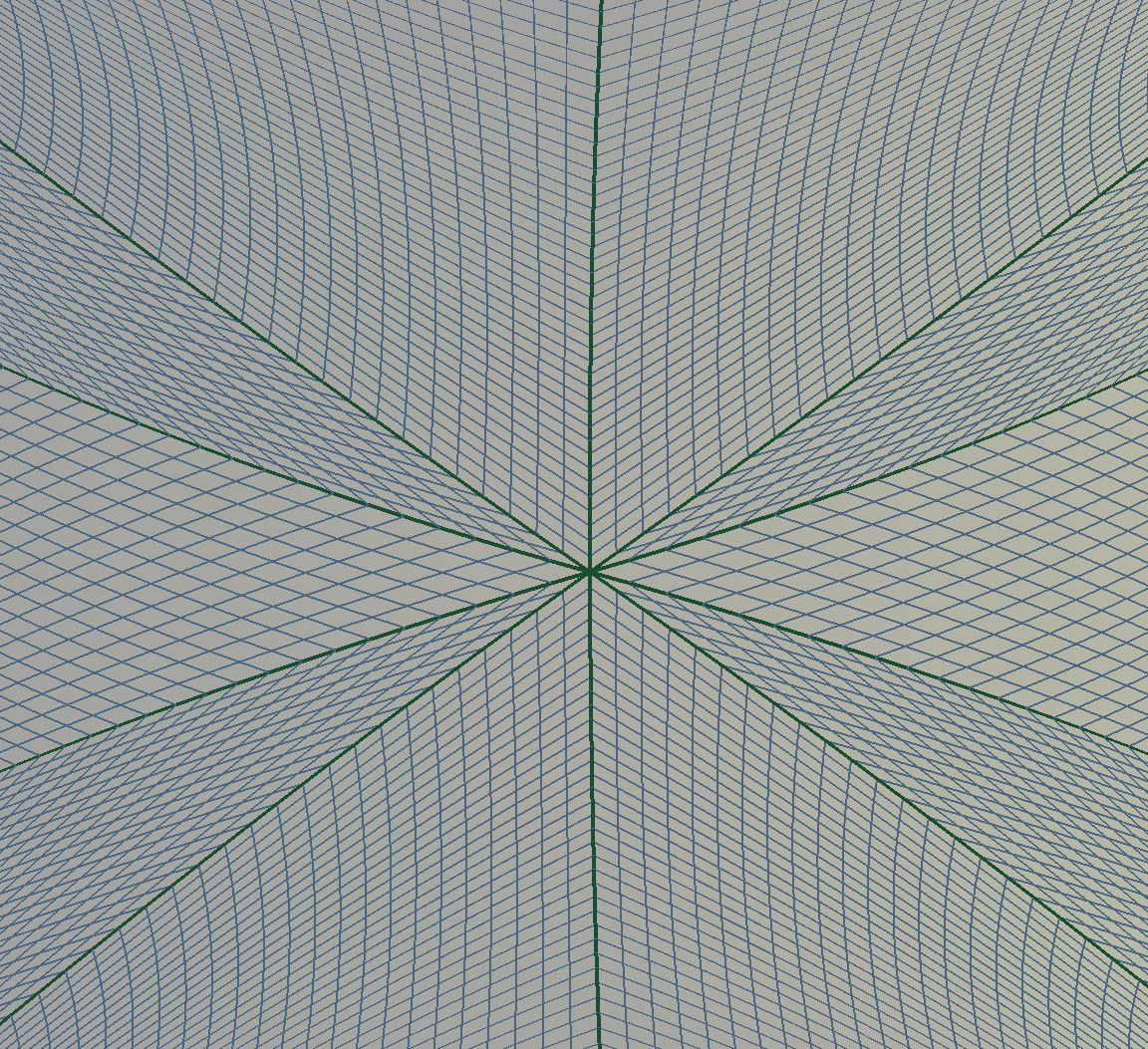}
        \caption{ Another detailed view around the same extraordinary vertex obtained with AS\&C method \cite{akleman2017}.}
        \label{fig_cubehole3}
    \end{subfigure}
    \hfill     
    \begin{subfigure}[t]{0.490\textwidth}
        \includegraphics[width=1.0\textwidth]{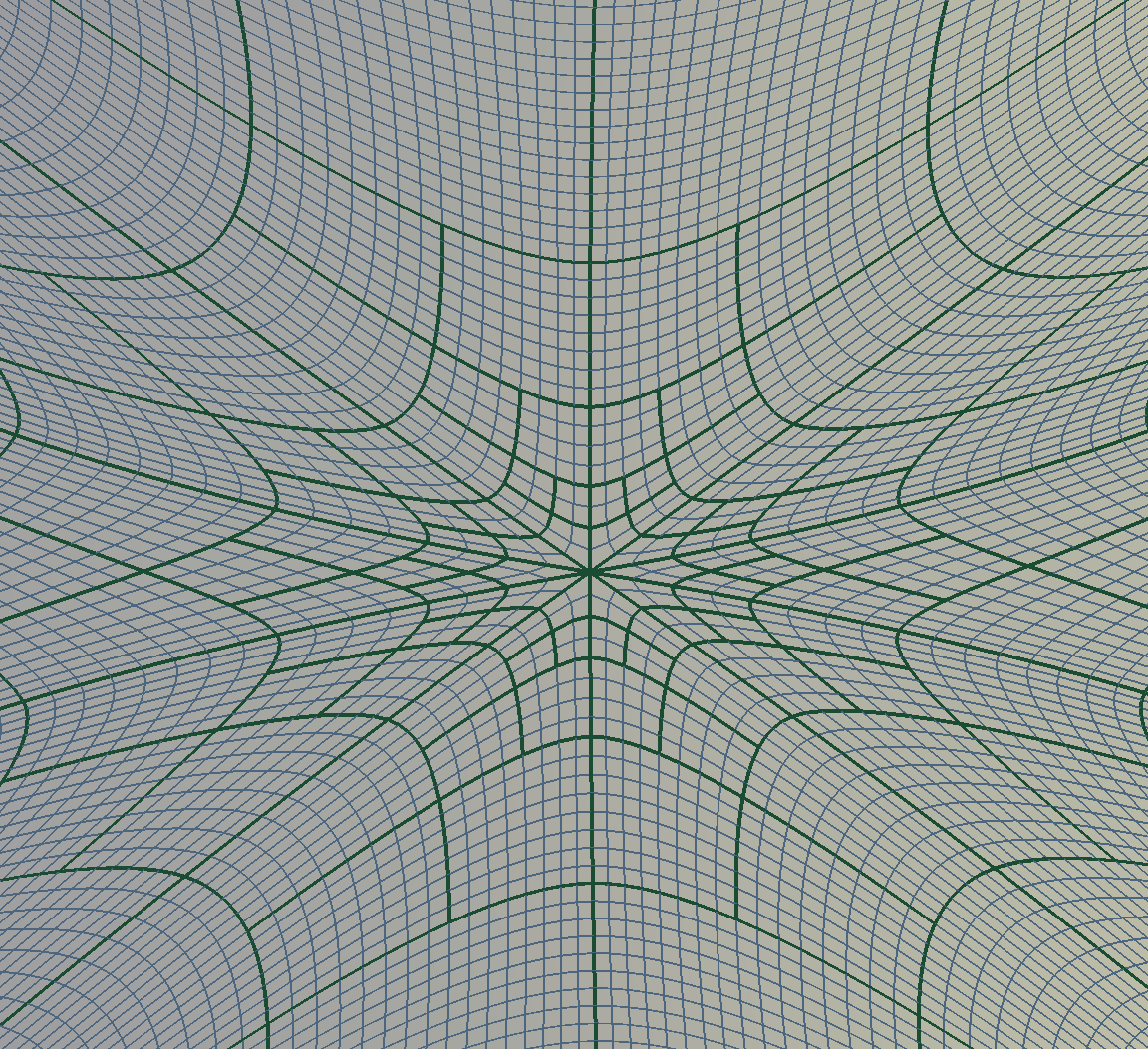}
        \caption{ The same view of the same neighborhood was obtained with our method.}
        \label{fig_cubehole5}
    \end{subfigure}
    \hfill 
\caption{Comparison our method with AS\&C-method \cite{akleman2017} around the complicated neighborhood of an extraordinary vertex with 10-valent. This extraordinary vertex comes from a ten-sided face that is obtained by combining two quadrilateral faces by inserting a single edge shown in Figure~\ref{fig_10SidedFace_6}. The insert edge operation creates the hole shown here. As can be seen in this example, our method removes the discontinuity $C^1$ along the boundary edges emanating from the extraordinary vertex. }
\label{fig_10-valent}
\end{figure}

These observations provide a possible explanation for the early popularity of stitched bicubic B\'{e}zier patches among practitioners. Most likely, the process was not as laborious and time consuming for professional modelers. They might have obtained visually smooth-looking stitched surfaces with relative ease. But once modeling tools for Catmull-Clark subdivision \cite{catmull1978} were introduced \cite{derose1998} and an exact evaluation method for Catmull-Clark subdivision was developed for any arbitrary point \cite{stam1998}, most practitioners could have converted to subdivision surfaces. Moreover, in modeling practice, people do not deal with very high valence vertices, which is where smoothness problems with the Catmull-Clark subdivision become more apparent \cite{akleman2017}.

\section{Introduction and Motivation}
\label{sec_motivation}

This discussion establishes the need for a solution to obtain continuity $C^1$ or $G^1$ along the edges emanating from that extraordinary vertex for smooth stitched patches. Although AS\&C method \cite{akleman2017} provides a simple solution for interactive applications, the resulting surfaces cannot be used in applications in which at least $C^1$ continuity should be guaranteed everywhere. Since tensor-product polynomials or rational polynomials are proven not to have a solution, there is a need to replace polynomials with analytic functions around extraordinary vertices. Since analytic functions can be represented as an infinite series of polynomials, the need for analytic functions suggests that we need a subdivision approach to solve this problem.  

In this paper, we present a new subdivision approach based on a minor modification of the original de Casteljau subdivision \cite{decasteljau1963}. A notable property of this new subdivision is that each patch can be subdivided independently from other patches once the initial control polyhedra are correctly established. The only problem with this method is that the number of patches exponentially increases with the number of iterative applications of the de Casteljau subdivision.

An important property of our modified subdivision is that it allows the number of patches to be reduced. Assume that after one application of the modified de Casteljau subdivision, we can evaluate each of the four resulting patches with the standard bicubic B\'{e}zier formulation. Although the four resulting patches do not form the same shape as the original bicubic B\'{e}zier, the whole structure is still $C^2$ continuous everywhere. This suggests that if none of the corners of a patch is extraordinary, we can simply evaluate it with bicubic B\'{e}zier formulation.

Using this property, we have developed a hierarchical evaluation process that significantly reduces the number of patches by providing high resolution around extraordinary vertices, as shown in Figures~\ref{fig_3-valent} and~\ref{fig_10-valent}). This hierarchical evaluation process is similar to Jos Stam's exact evaluation of the Catmull-Clark subdivision at any arbitrary point \cite{stam1998}.

To explain our hierarchical evaluation process, we need to provide a few definitions. Let a B\'{e}zier patch that does not share any extraordinary vertex with any other B\'{e}zier patch be called a regular patch and let the rest be called extraordinary patches. If a B\'{e}zier patch is regular, we can simply render it using standard B\'{e}zier evaluation or original de Casteljau subdivision. 

On the other hand, if the patch is not regular, we apply a modified de Casteljau subdivision that preserves $C^1$ continuity in the edge boundaries. The modified de Casteljau algorithm is essentially similar to the original de Casteljau algorithm. It subdivides a $4\times 4$ control polyhedron of the extraordinary patch into four new $4\times 4$ control polyhedra by producing four new bicubic B\'{e}zier patches that stitch together with $C^2$ continuity. Moreover, like the original subdivision, the modified subdivision preserves the desired continuity $G^1$ in the boundary vertices, which is given by a position and a tangent plane for each boundary vertex.

The main advantage of the modified subdivision is that the continuity $C^1$ in a given boundary edge does not depend on the positions of the control points on other boundary edges. This allows us to obtain the desired continuity $C^1$ to properly stitch two patches that share an extraordinary point.

In each iteration of this process, some of the new patches become regular patches, and we can simply evaluate them using the standard B'{e}zier evaluation. For the rest of the extraordinary patches, we continue to apply the modified de Casteljau subdivision. This process produces hierarchically organized bicubic B\'{e}zier patches that are stitched with $G^1$ continuity in extraordinary vertices and $C^1$ continuity along the edges emanating from extraordinary vertices.  

\subsection{Practical Contributions}

There are several practical contributions that can originate from this theoretical framework. We have identified three main practical contributions, such as fast computation, greater control of the shapes, and two-manifold mesh modeling.  

\paragraph{Fast Computation:} The modified subdivision operates only on a single B\'{e}zier control polyhedron. Therefore, each patch can be evaluated independently of the other patches. Once one of the subdivided patches is turned into a regular patch, it can be evaluated directly using the B\'{e}zier formulation, allowing real-time rendering. 

\paragraph{Greater Control of the Shape:} We can provide greater control of the final shape, since with our algorithm it is possible to provide control on the local region around any extraordinary vertex using a planar polygon. The number of sides of the planar polygon must be the same as the valence of the extraordinary vertex. The center and normal of the planar polygon define the positions and tangents of extraordinary vertices, similar to controlling the B'{e}zier surfaces. 

\paragraph{Smooth two-manifold Modeling:} Practically the key attribute of this method is that it can be used to convert any quad mesh into a piecewise smooth surface by converting every quad into a B\'{e}zier patch. Since any manifold mesh can be converted to a quadrilateral mesh using a single application of vertex insertion, which is the remeshing algorithm of Catmull-Clark subdivision \cite{catmull1978}, or $\sqrt{2}$ subdivision \cite{li2004}, which is the dual of the simplest subdivision \cite{peters1997}, this process allows interactive smooth shape modeling along with topology changes. In this paper, for interactive smooth surface modeling, we used vertex insertion in our implementation. 

\subsection{Organization of the Rest of the Paper}

The rest of the paper is organized as follows. We first give a subdivision-based analysis of the root cause of the $C^1$ or $G^1$ discontinuity problem along the edges emanating from the extraordinary vertices in Section~\ref{sec_DC}. We later provide a modified de Casteljau subdivision that solves the discontinuity problem in $C^1$ or $G^1$ along the edges emanating from the corresponding extraordinary vertices in Section~\ref{sec_MDC}. In Section~\ref{sec_implementation}, we provide a high-level summary of implementation. Finally, in Section~\ref{sec_conclusion}, we discuss implementation details and possible extensions. 

\section{Preliminaries \label{sec_DC}}

To provide a simple explanation for why B\'{e}zier patches cannot provide the continuity $C^1$ or $G^1$ along the edges, we revisit both the topology and geometry of the de Casteljau algorithm. Although most of the results covered in this section are known, the analysis of the problem as a preservation of properties through subdivision, as far as we know, is new. This new method is critical for the formal development of the modified de Casteljau algorithm. For other related analysis methods, please refer to \cite{peters2017}.

\begin{figure}[htpb]
    \centering  
    \begin{subfigure}[t]{0.490\textwidth}
        \includegraphics[width=1.0\textwidth]{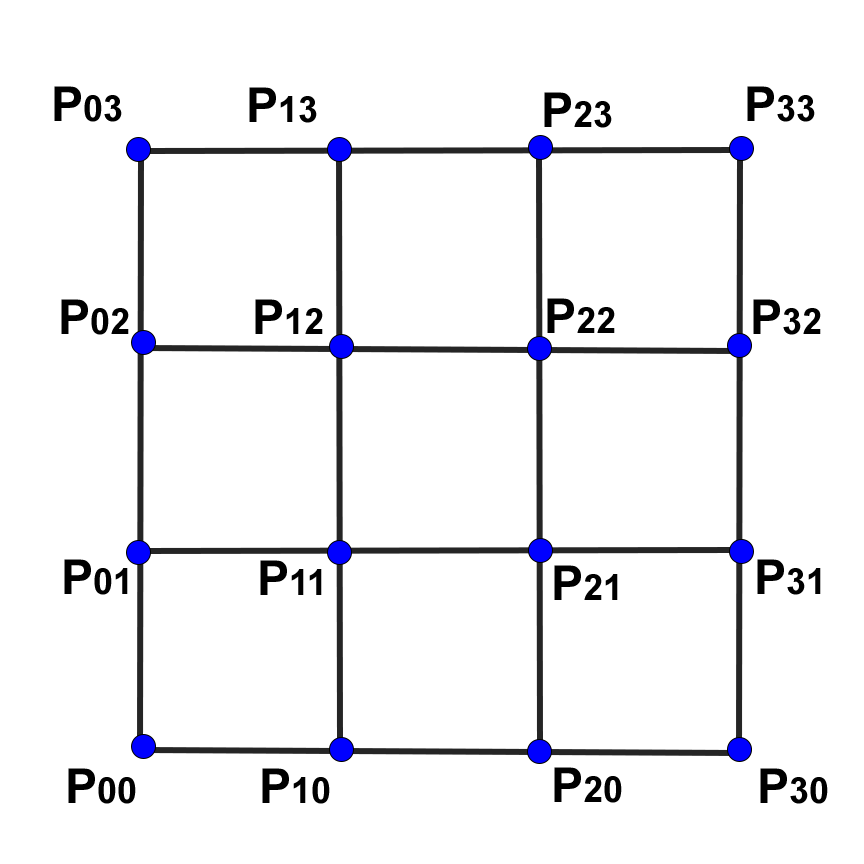}
        \caption{ Elements of $4 \times 4$ control polyhedron.}
        \label{fig_Patch0}
    \end{subfigure}
    \hfill  
        \begin{subfigure}[t]{0.490\textwidth}
        \includegraphics[width=1.0\textwidth]{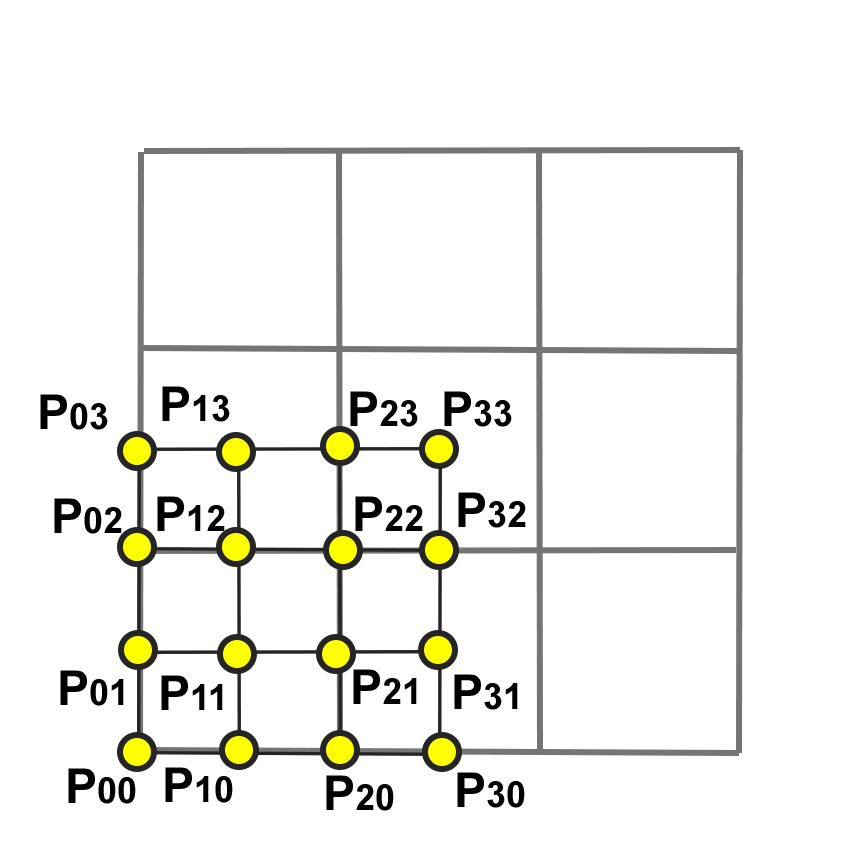}
        \caption{ Application of the De-Casteljau subdivision.}
        \label{fig_Patch1}
    \end{subfigure}
    \hfill  
\caption{The de Casteljau subdivision splits the original control polyhedron into four topologically identical copies. Here, the control points of the original $4 \times 4$ control polyhedron are shown in blue and the control points of one of the copies are shown with yellow control points. The positions of the control points of each of these copies produce the original B\'{e}zier formula. }
\label{fig_Patch}
\end{figure}

The control polyhedron of any tensor product B\'{e}zier surface is always a $N \times M $ polyhedral grid, where $N$ and $M$ can be any integer.  Figure~\ref{fig_Patch} shows what happens topologically with an application of the de Casteljau subdivision - without loss of generality - on the $4 \times 4$ polyhedral grid. The classical de Casteljau algorithm always creates four copies of the original $N \times M$ polyhedral grid. Each of the new polyhedral grids defines the same B\'{e}zier equation. Therefore, we can use the de Casteljau subdivision to analyze the behavior of the B\'{e}zier surfaces. 

\begin{figure*}[htpb]
    \centering  
    \begin{subfigure}[t]{0.24\textwidth}
        \includegraphics[width=1.0\textwidth]{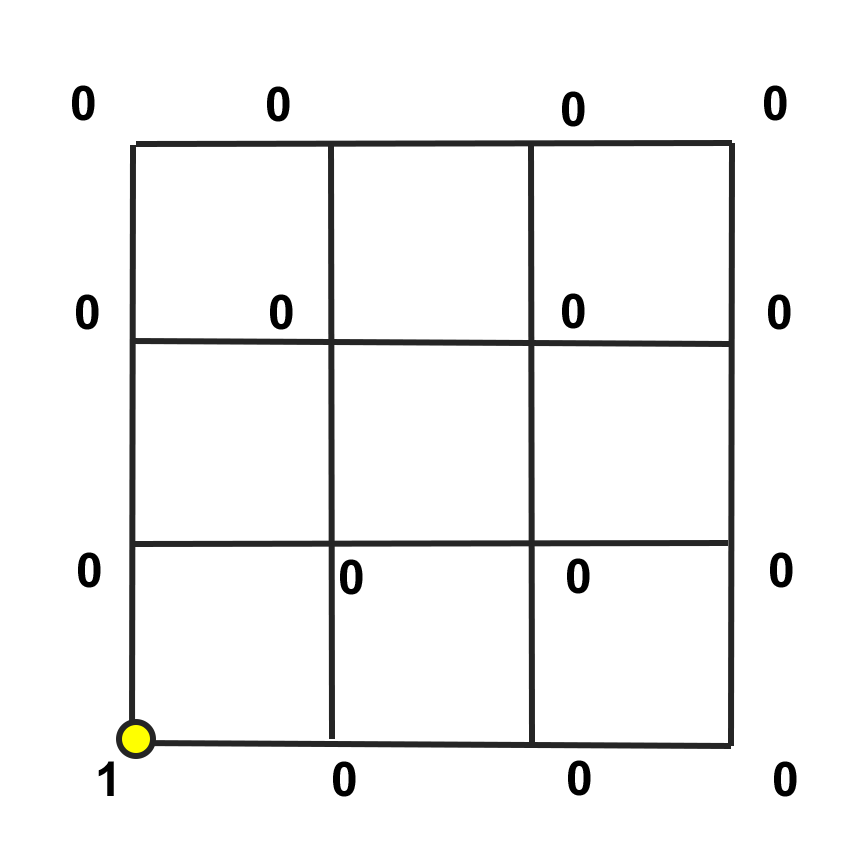}
        \caption{ New position of $p_{00}$.}
        \label{fig_DC00}
    \end{subfigure}
    \hfill  
    \begin{subfigure}[t]{0.24\textwidth}
        \includegraphics[width=1.0\textwidth]{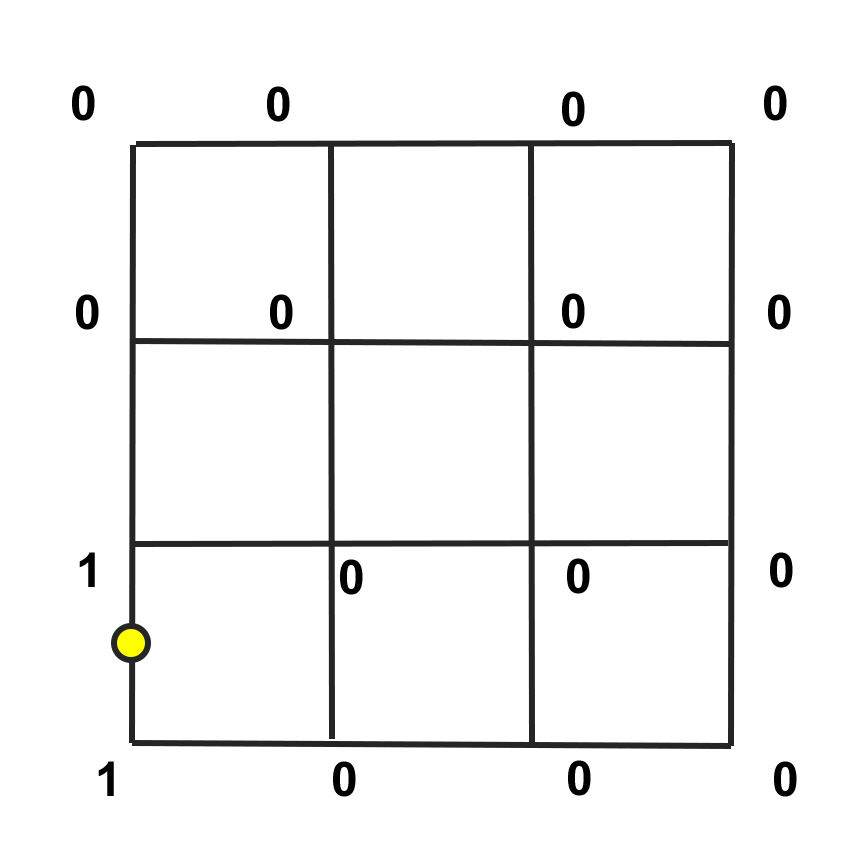}
        \caption{ New position of $p_{01}$.}
        \label{fig_DC01}
    \end{subfigure}
    \hfill 
    \begin{subfigure}[t]{0.24\textwidth}
        \includegraphics[width=1.0\textwidth]{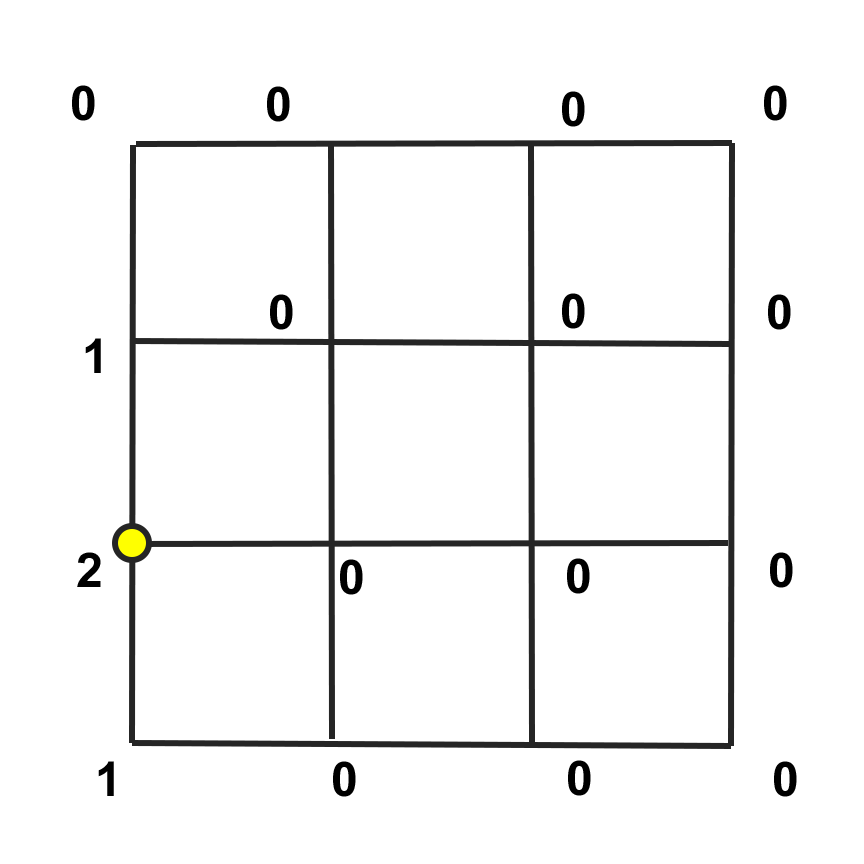}
        \caption{ New position of $p_{02}$.}
        \label{fig_DC02}
    \end{subfigure}
    \hfill
    \begin{subfigure}[t]{0.24\textwidth}
        \includegraphics[width=1.0\textwidth]{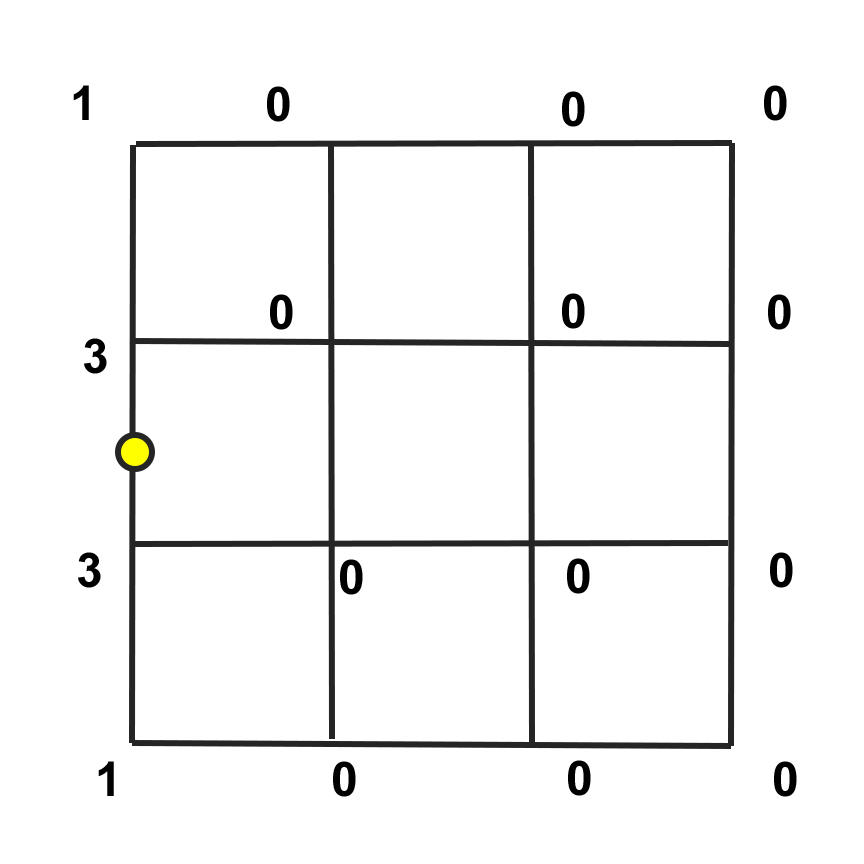}
        \caption{ New position of $p_{03}$.}
        \label{fig_DC03}
    \end{subfigure}
    \hfill
        \begin{subfigure}[t]{0.24\textwidth}
        \includegraphics[width=1.0\textwidth]{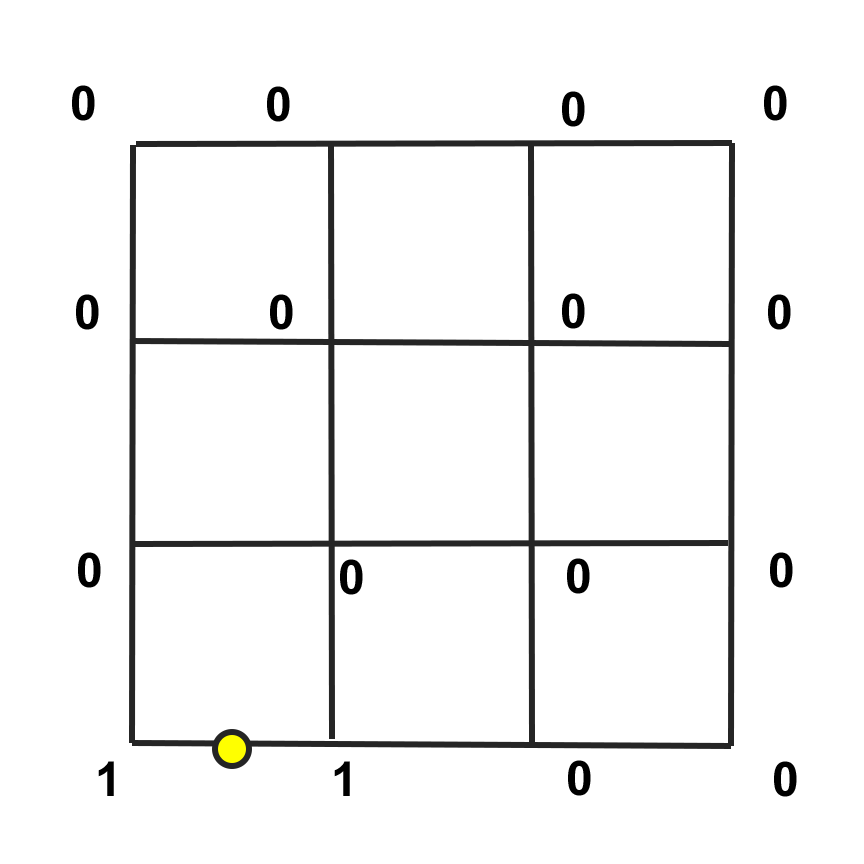}
        \caption{ New position of $p_{10}$.}
        \label{fig_DC10}
    \end{subfigure}
    \hfill  
    \begin{subfigure}[t]{0.24\textwidth}
        \includegraphics[width=1.0\textwidth]{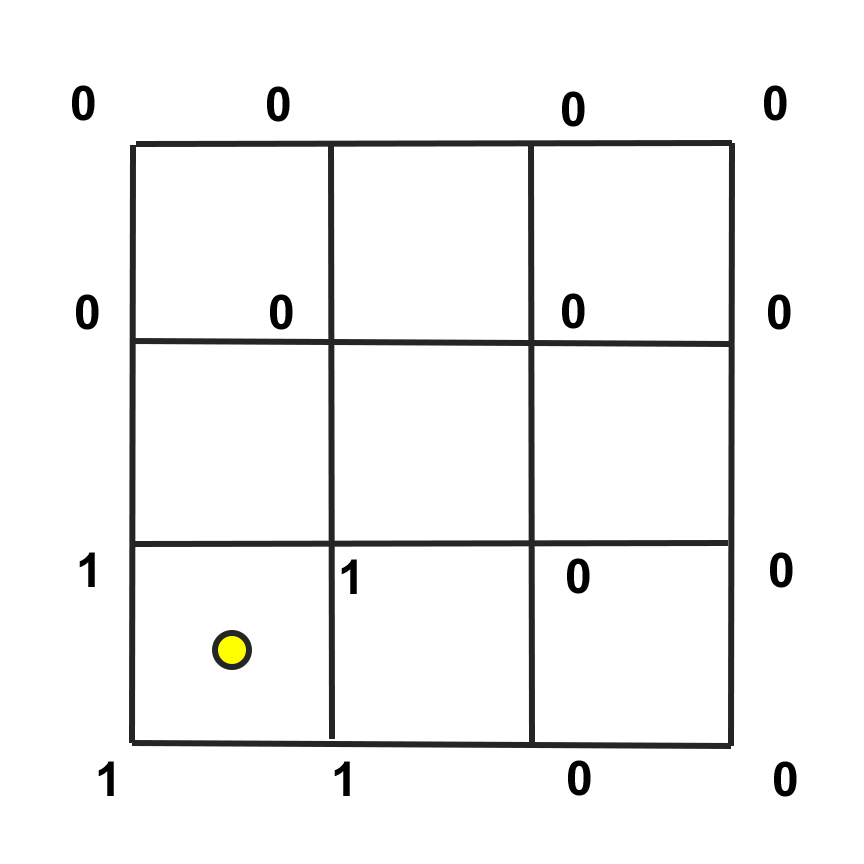}
        \caption{ New position of $p_{11}$.}
        \label{fig_DC11}
    \end{subfigure}
    \hfill 
    \begin{subfigure}[t]{0.24\textwidth}
        \includegraphics[width=1.0\textwidth]{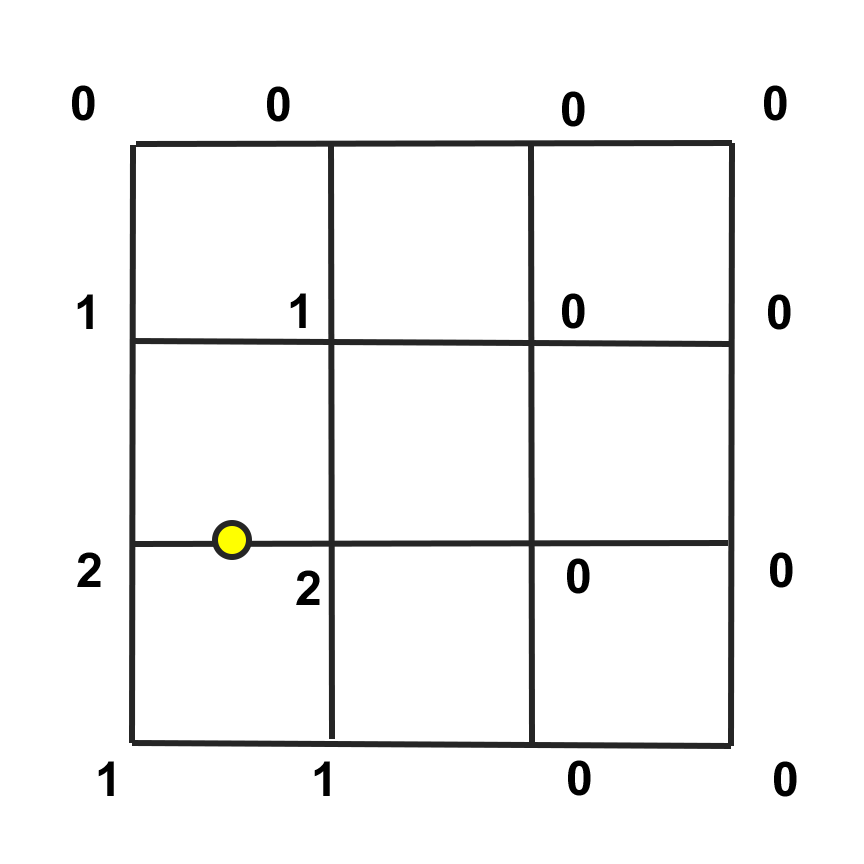}
        \caption{ New position of $p_{12}$.}
        \label{fig_DC12}
    \end{subfigure}
    \hfill
    \begin{subfigure}[t]{0.24\textwidth}
        \includegraphics[width=1.0\textwidth]{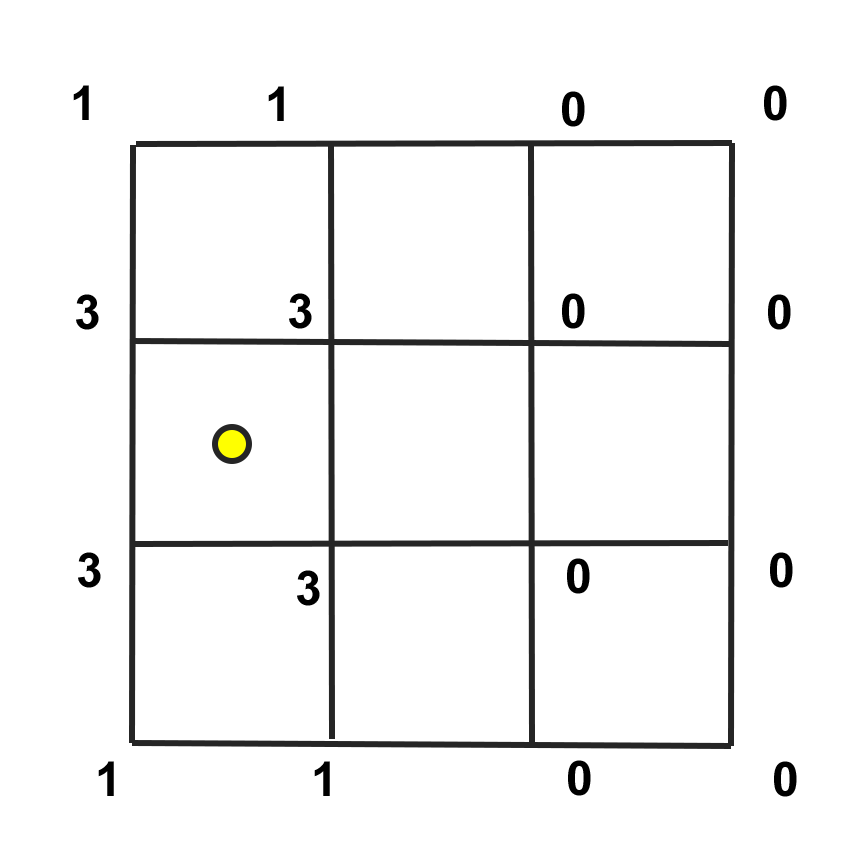}
        \caption{ New position of $p_{13}$.}
        \label{fig_DC13}
    \end{subfigure}
    \hfill

        \centering  
    \begin{subfigure}[t]{0.24\textwidth}
        \includegraphics[width=1.0\textwidth]{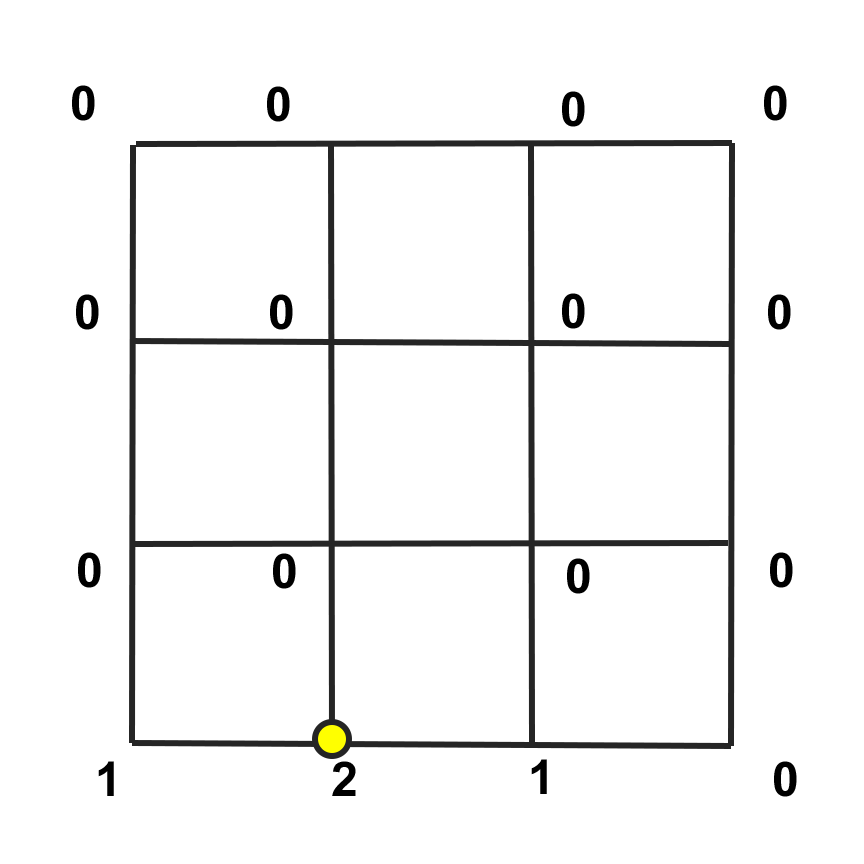}
        \caption{ New position of $p_{20}$.}
        \label{fig_DC20}
    \end{subfigure}
    \hfill  
    \begin{subfigure}[t]{0.24\textwidth}
        \includegraphics[width=1.0\textwidth]{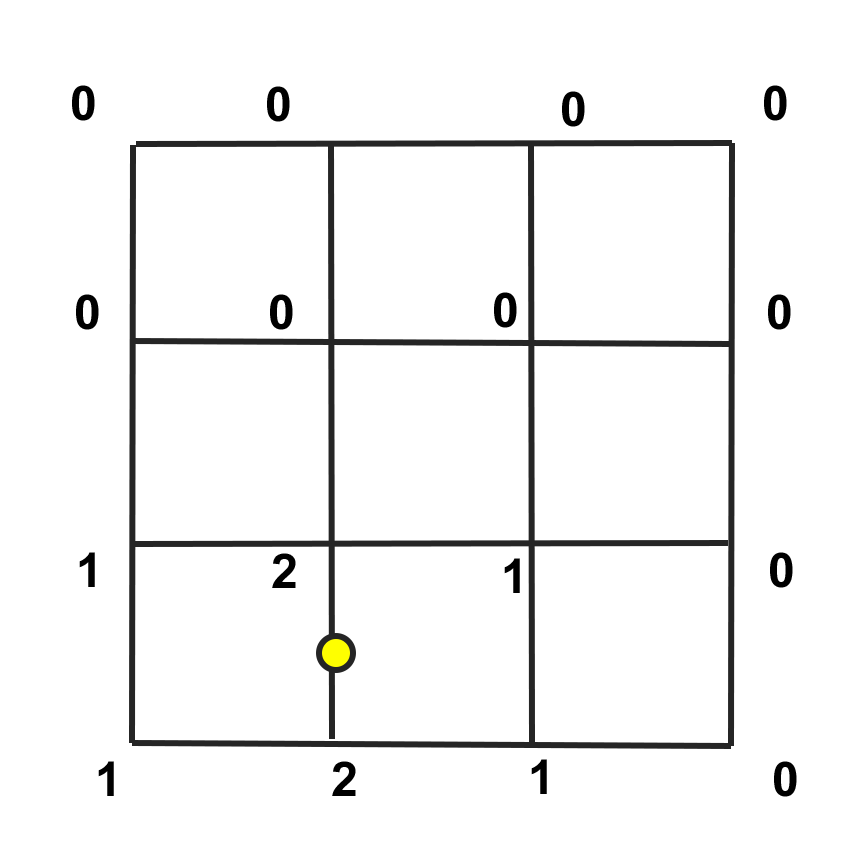}
        \caption{ New position of $p_{21}$.}
        \label{fig_DC21}
    \end{subfigure}
    \hfill 
    \begin{subfigure}[t]{0.24\textwidth}
        \includegraphics[width=1.0\textwidth]{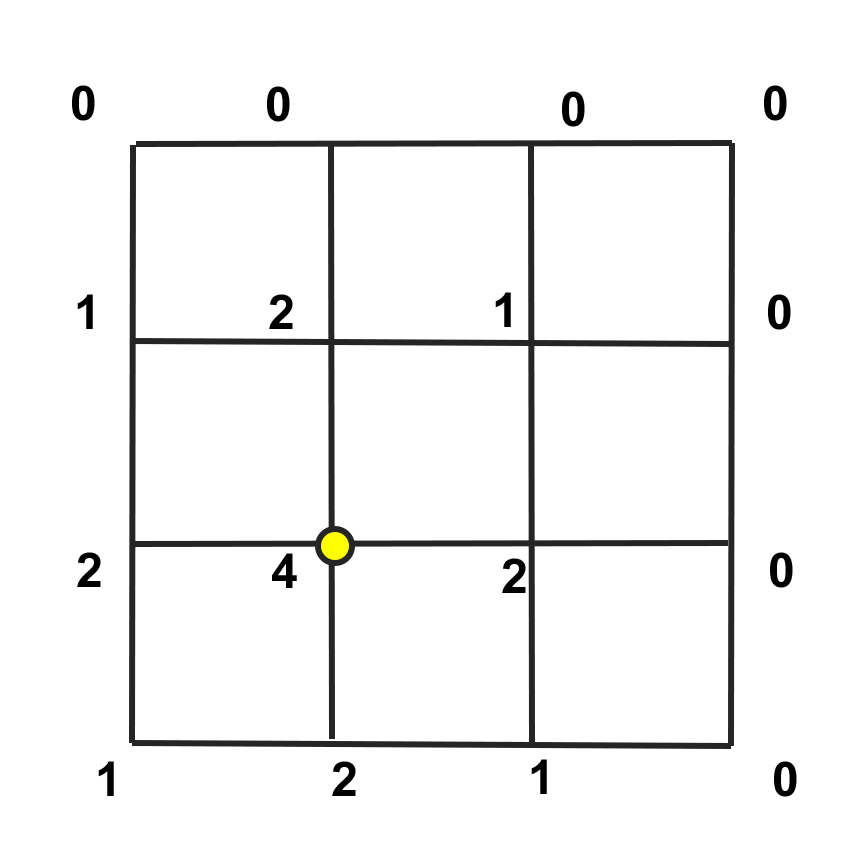}
        \caption{ New position of $p_{22}$.}
        \label{fig_DC22}
    \end{subfigure}
    \hfill
    \begin{subfigure}[t]{0.24\textwidth}
        \includegraphics[width=1.0\textwidth]{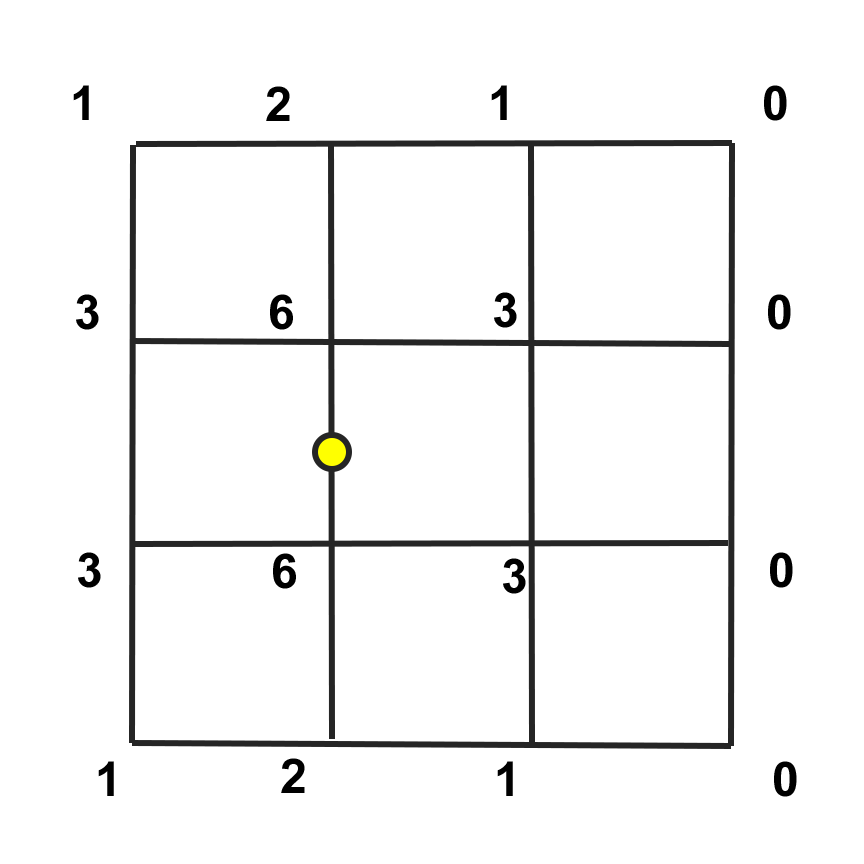}
        \caption{ New position of $p_{23}$.}
        \label{fig_DC23}
    \end{subfigure}
    \hfill
        \begin{subfigure}[t]{0.24\textwidth}
        \includegraphics[width=1.0\textwidth]{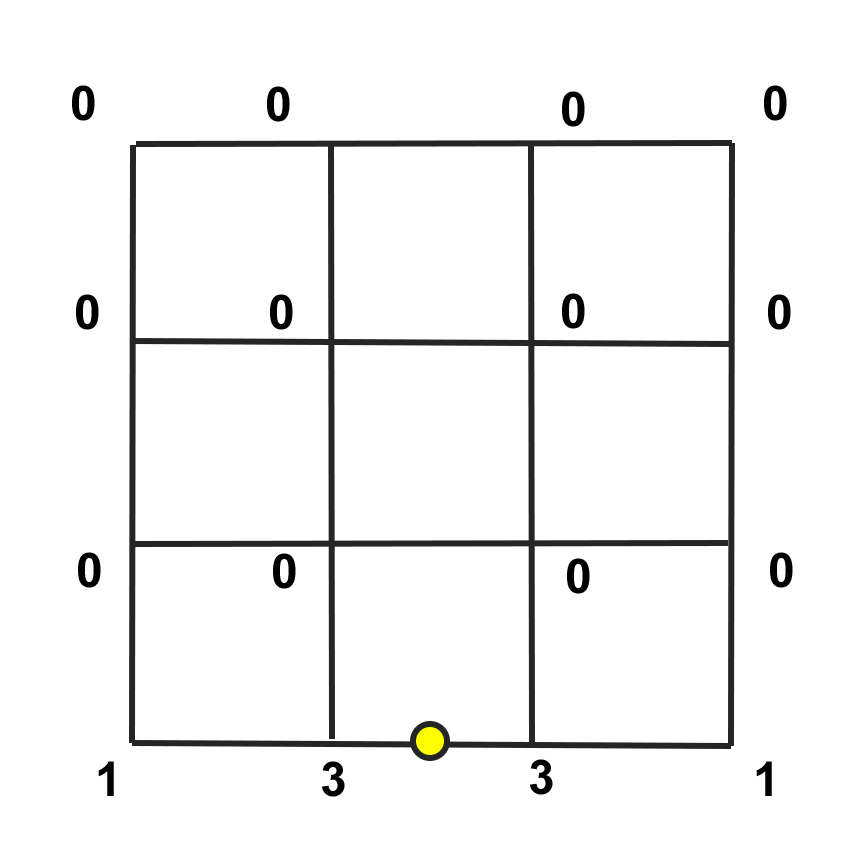}
        \caption{ New position of $p_{30}$.}
        \label{fig_DC30}
    \end{subfigure}
    \hfill  
    \begin{subfigure}[t]{0.24\textwidth}
        \includegraphics[width=1.0\textwidth]{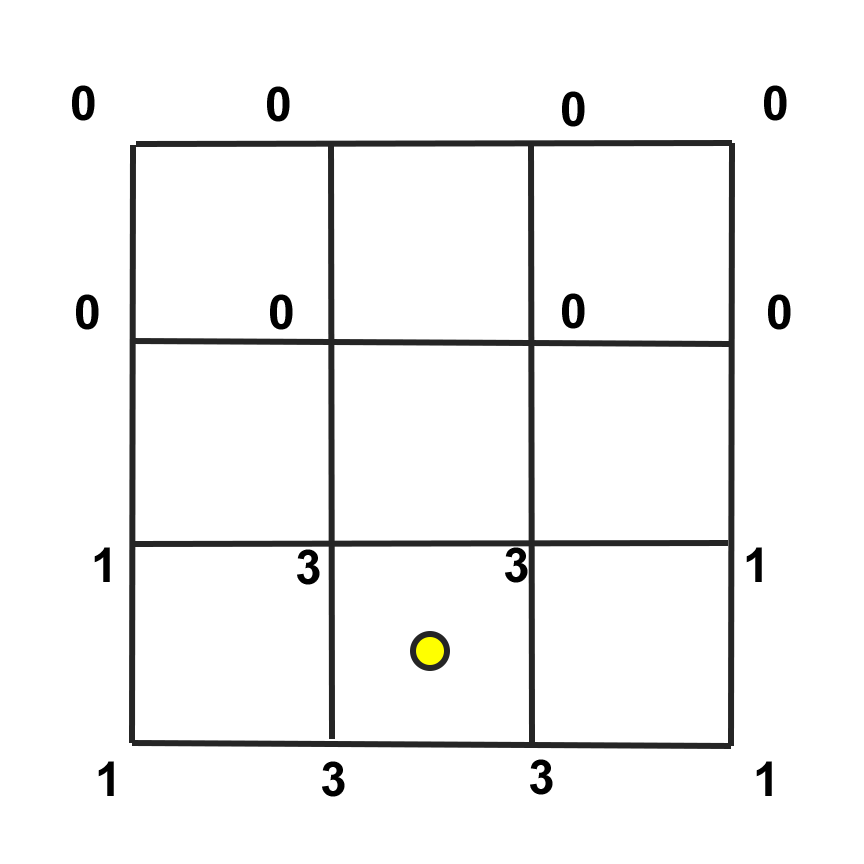}
        \caption{ New position of $p_{31}$.}
        \label{fig_DC31}
    \end{subfigure}
    \hfill 
    \begin{subfigure}[t]{0.24\textwidth}
        \includegraphics[width=1.0\textwidth]{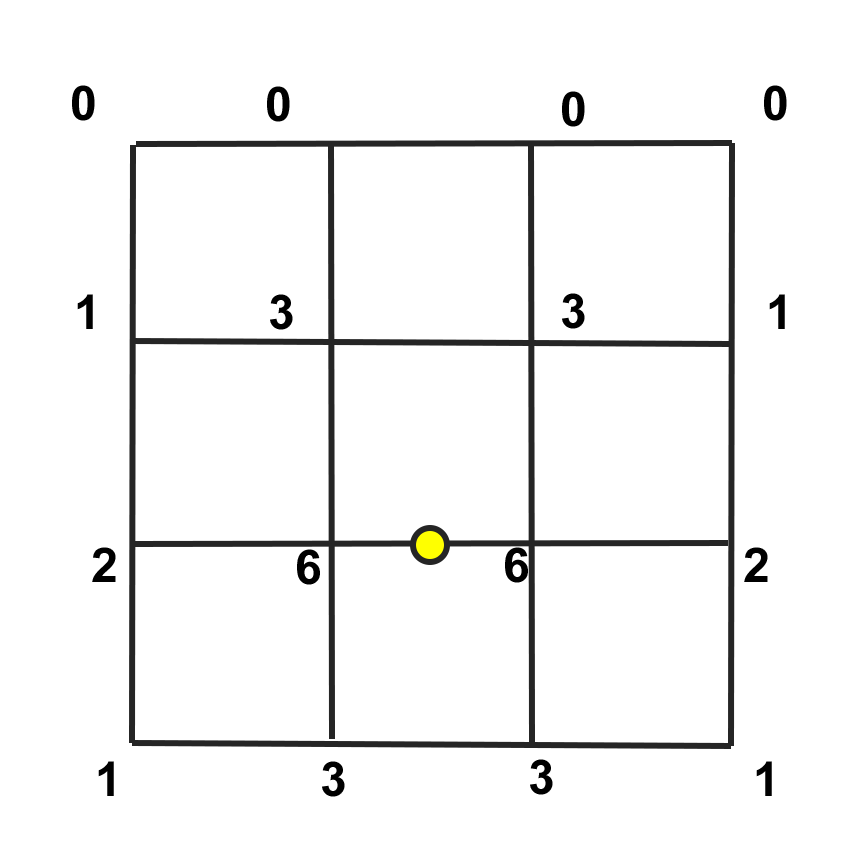}
        \caption{ New position of $p_{32}$.}
        \label{fig_DC32}
    \end{subfigure}
    \hfill
    \begin{subfigure}[t]{0.24\textwidth}
        \includegraphics[width=1.0\textwidth]{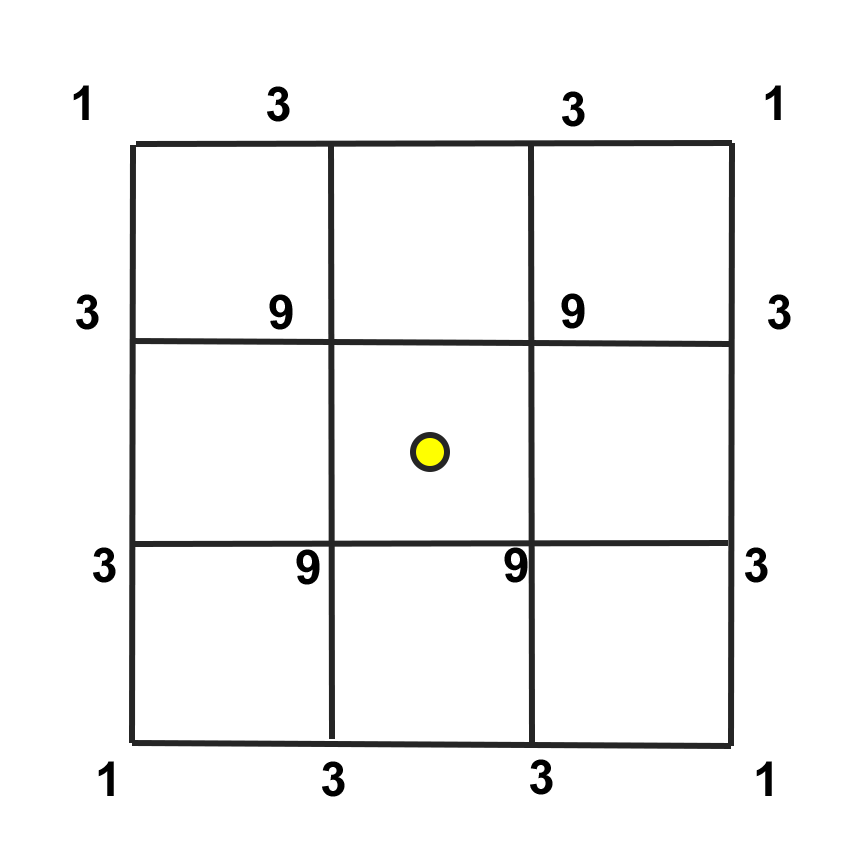}
        \caption{ New position of $p_{33}$.}
        \label{fig_DC33}
    \end{subfigure}
    \hfill
\caption{This figure shows the De-Casteljau kernels to produce each control. }
\label{fig_DC}
\end{figure*}

To perform the analysis, we will provide geometric effects of the de Casteljau subdivision in an unusual form, as shown in Figure~\ref{fig_DC}. In this case, instead of using the classical recursive definition of de Casteljau, we provided kernels to produce each new control point for one of the patches. Since in practice we are interested in bicubic B\'{e}zier surfaces, we provide kernels for the $4 \times 4$ grid.

Since the algorithm is symmetric, the kernels for other points can be obtained directly by simply renaming control points. The following equation shows how the kernel is used to compute the new $P'_{mn}$'s (that is, the positions of yellow points shown in Figure~\ref{fig_Patch1}) from the old $P'_{ij}$ (that is, the positions of blue points shown in Figure~\ref{fig_Patch0}).   
$$P'_{mn} =  \displaystyle \sum_{i=0}^{3} \sum_{j=0}^{3} \left( \frac{ w_{mnij} }{\sum \sum w_{mnij}} \right) P_{ij}, $$
where $w_{mnij}$'s are the weights given in Figure~\ref{fig_DC}. 

Now, let two B\'{e}zier patches be given by a set of points $P^0_{ij}$ and $P^1_{ij}$. 
Using classical B\'{e}zier formulation, we can show that these two B\'{e}zier patches that share a boundary edge and have the same $C^1$ continuity in the boundary if the following two conditions hold for all $i$'s: 

\begin{eqnarray}
P^0_{i0} &=& P^1_{i0}  \label{eq:1} \\
P^0_{i1} - P^0_{i0} &=& P^1_{i0} - P^1_{i1}  \label{eq:2} 
\end{eqnarray}
These two conditions are easy to obtain if there are only two patches. We can randomly select all $P^0_{i1}$ and $P^1_{i1}$ and compute $P^0_{i0} = P^1_{i0}$ for all $i$'s as follows: 
$$P^0_{i0} = P^1_{i0} = \frac{P^0_{i1} + P^1_{i1}}{2} $$

\subsection{C1 Condition in Extraordinary Vertex}

\begin{figure}[htpb]
    \centering  
    \begin{subfigure}[t]{0.490\textwidth}
        \includegraphics[width=1.0\textwidth]{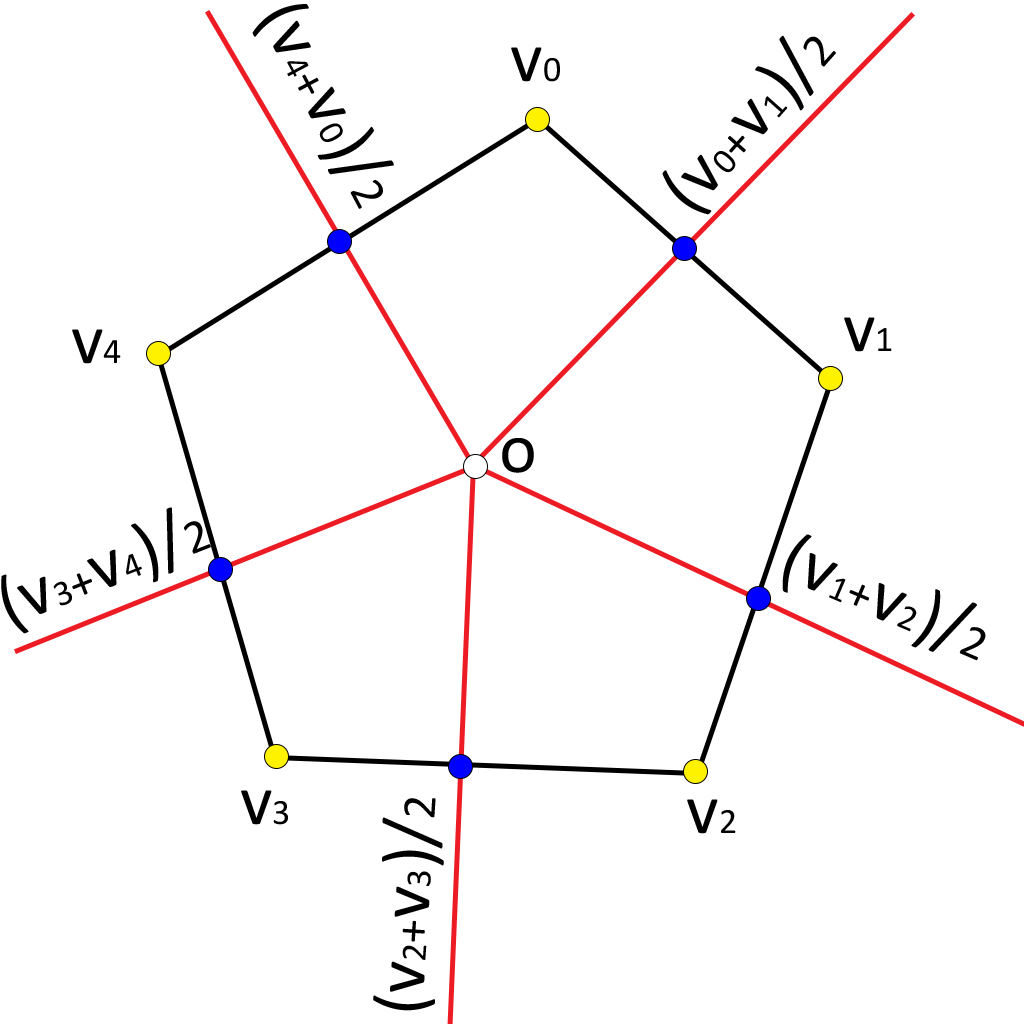}
        \caption{ Planar convex polygon around an extraordinary vertex.}
        \label{fig_example1}
    \end{subfigure}
    \hfill  
        \begin{subfigure}[t]{0.490\textwidth}
        \includegraphics[width=1.0\textwidth]{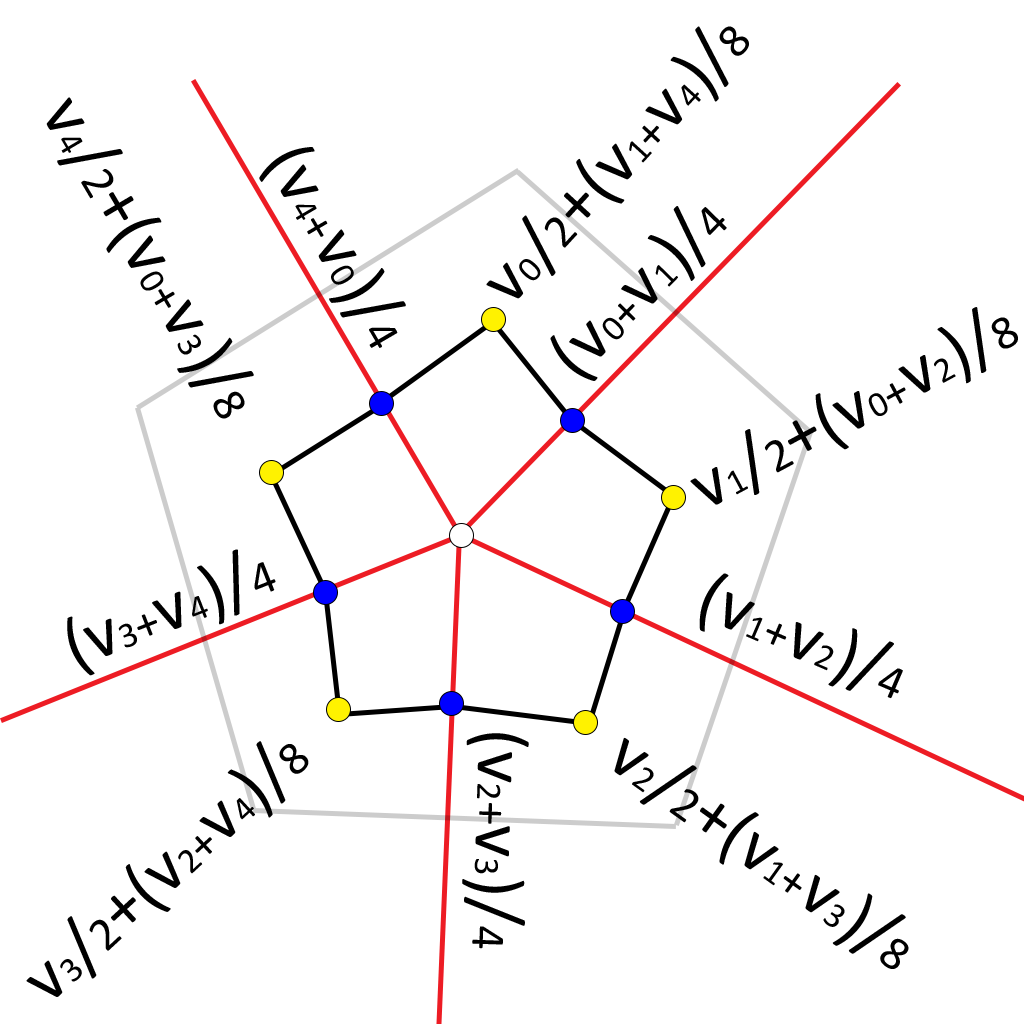}
        \caption{ Application of the first iteration of  De-Casteljau subdivision.}
        \label{fig_example2}
    \end{subfigure}
    \hfill  
\caption{An example that shows five patches that share a 5-valent extraordinary point. Although five initial quadrilaterals that are defined by 20 control points form a convex planar pentagon, the application of the de Casteljau subdivision turns this convex planar pentagon into a star-shaped planar decagon (10-sided polygon) by breaking line segments.}
\label{fig_example}
\end{figure}

Now, let $K$ number of B\'{e}zier patches that are defined by a set of control points, $P^k_{ij}$ where $0 \leq k < K$, share a vertex with continuity $C^1$. In this case, $P^k_{0i}$'s also become boundary vertices, and we cannot freely choose them. The notation we use also becomes cumbersome for this extraordinary case. Therefore, in order to do the analysis, we first need to simplify our notation. Let $P^k_{00}$ be the same for all $k$'s and the origin, and let $\vec{V}_{k} =P^{k}_{11}-P^k_{00}$. These vectors form a polygon with $K$ sides that is not necessarily planar (see Figure~\ref{fig_example1}), which is defined by $K$ vectors such as $\vec{V}_0,\vec{V}_1,\ldots \vec{V}_{K-1}$ that are denoted by yellow circles in Figure~\ref{fig_example1}. Based on condition~\ref{eq:2}, for all consecutive vectors $\vec{V}_{k}$ and $\vec{V}_{k+1}$, a new average vector is calculated as $\frac{(\vec{V}_{k}+\vec{V}_{k+1})}{2}$, which are denoted by blue circles in Figure~\ref{fig_example1}. These vectors give us the blue points that are control points on the boundaries between patches (see the red lines in Figure~\ref{fig_example1}). Since each blue point must be in the middle of the line that connects two yellow points, the control points still visually appear like the original polygon with $K$ sides as shown in Figure~\ref{fig_example1}. 

According to the condition~\ref{eq:2}, the average of $\frac{(\vec{V}_{k}+\vec{V}_{k+1})}{2}$ and $\frac{(\vec{V}_{k+2}+\vec{V}_{k+3})}{2}$ must be the origin. 
This gives us the following condition for all $k$'s:
$$\vec{V}_{k}+\vec{V}_{k+1} + \vec{V}_{k+2}+\vec{V}_{k+3} =0 $$
This set of equations can have a non-zero solution only when $K=4$. This observation suggests that $C^1$ continuity cannot be achieved at any extraordinary point. Although it is not possible to satisfy conditions~\ref{eq:1} and~\ref{eq:2} for $K \neq 4$, if four patches share a vertex satisfying these conditions, it becomes trivial. We can freely choose positions of $P^{0}_{11}$, $P^{1}_{11}$, $P^{2}_{11}$, and $P^{3}_{11}$. Then $P^k_{00}$ is simply chosen as their average\footnote{Choosing these control points freely can cause self-intersections, but, we can ignore that issue.}. This well-known result is practically useless since this arrangement of control points again produces a $N \times M$ polyhedral grid as control polyhedra for four B\'{e}zier patches.  

\subsection{G1 Condition in Extraordinary Vertex}

As a second option, we can identify the conditions on the positions of $P^k_{0i}$'s for obtaining $G^1$ continuity at extraordinary points when $K \neq 4$. Using de Casteljau for analysis consists of two steps. We first need to identify an initial configuration of $P^k_{0i}$'s to satisfy the desired property. Second, we need to demonstrate that this property is preserved under the de Casteljau subdivision. For the conditions $G^1$, the control points of the neighboring quadrilaterals around an extraordinary point must belong to the same implicit surface. After the application of the de Casteljau subdivision, the transformed control points of the neighboring quadrilaterals around an extraordinary point must belong to the same - but not necessarily to the original - implicit surface.

In the general case, we need to deal with implicit versions of bilinear surfaces formed by neighboring quadrilaterals, which is not that easy. A simple approach is to ensure that they share the same plane. In this case, it is easy to satisfy the same affine equation by simply moving all points in the neighboring quadrilaterals to the same plane. In the simple vector notation, we used around extraordinary vertices, all $K$ vectors as $\vec{V}_0,\vec{V}_1,\ldots \vec{V}_{K-1}$ must be perpendicular to the same vector. To avoid self-intersections, the origin and the vectors must form a star shape, that is, any line emanating from the origin must intersect with the polygon that is formed by $\vec{V}_0,\vec{V}_1,\ldots \vec{V}_{K-1}$ only once. 

To further simplify the analysis, let $\vec{V}_0,\vec{V}_1,\ldots \vec{V}_{K-1}$ form a convex polygon and let $P^k_{00}$ be the origin (i.e. $\vec{V}_0+\vec{V}_1+\ldots+ \vec{V}_{K-1}=0$). This can simply be achieved by providing a planar convex polygon and using its corners as $P^k_{11}$'s.We can still enforce the condition~\ref{eq:2} and for all consecutive vectors $\vec{V}_{k}$ and $\vec{V}_{k+1}$, we can compute a new average vector as $\frac{(\vec{V}_{k}+\vec{V}_{k+1})}{2},$ which are again denoted by blue circles in Figure~\ref{fig_example1}. This is basically similar to the previous case. The only difference is that the $K$-sided polygon is now planar.

\begin{figure}[htb!]
    \centering  
    \begin{subfigure}[t]{0.320\textwidth}
        \includegraphics[width=1.0\textwidth]{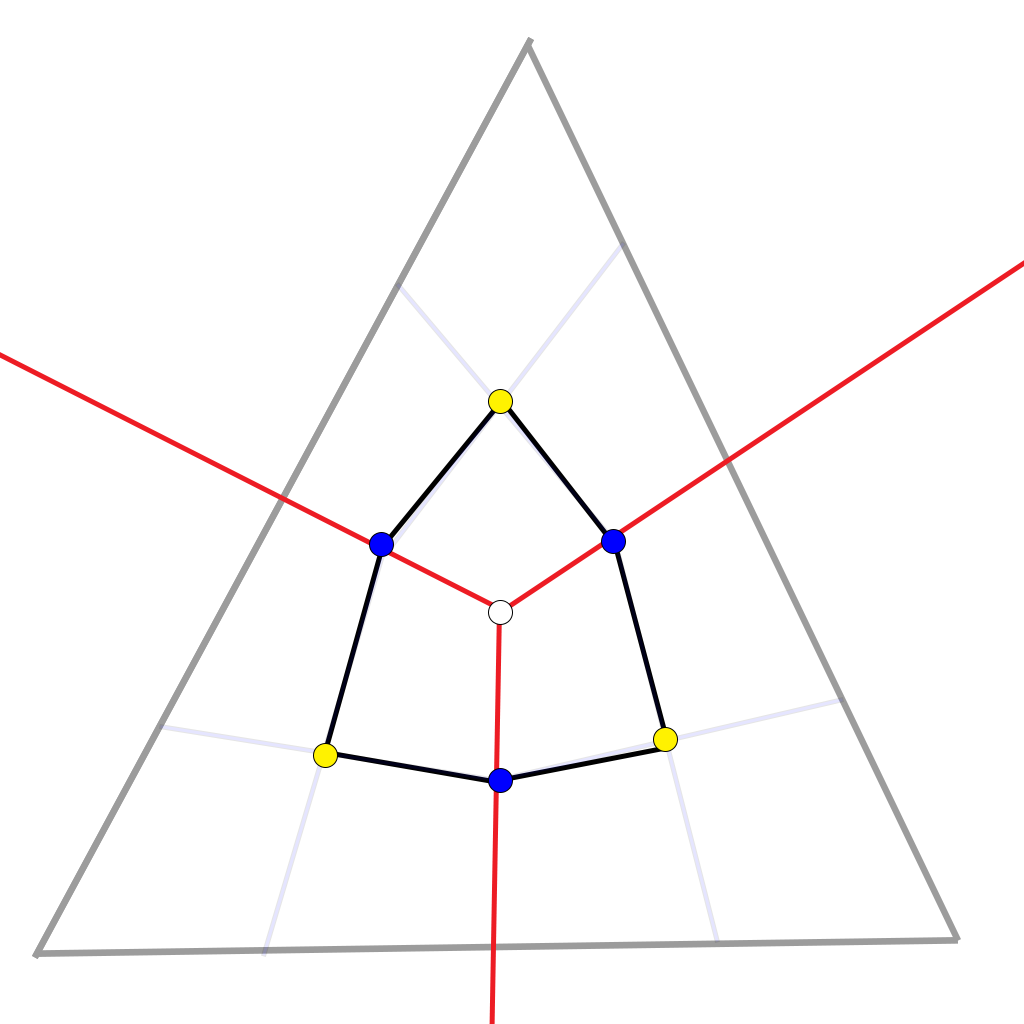}
        \caption{ $n=3$.}
        \label{fig_t}
    \end{subfigure}
    \hfill  
    \begin{subfigure}[t]{0.320\textwidth}
        \includegraphics[width=1.0\textwidth]{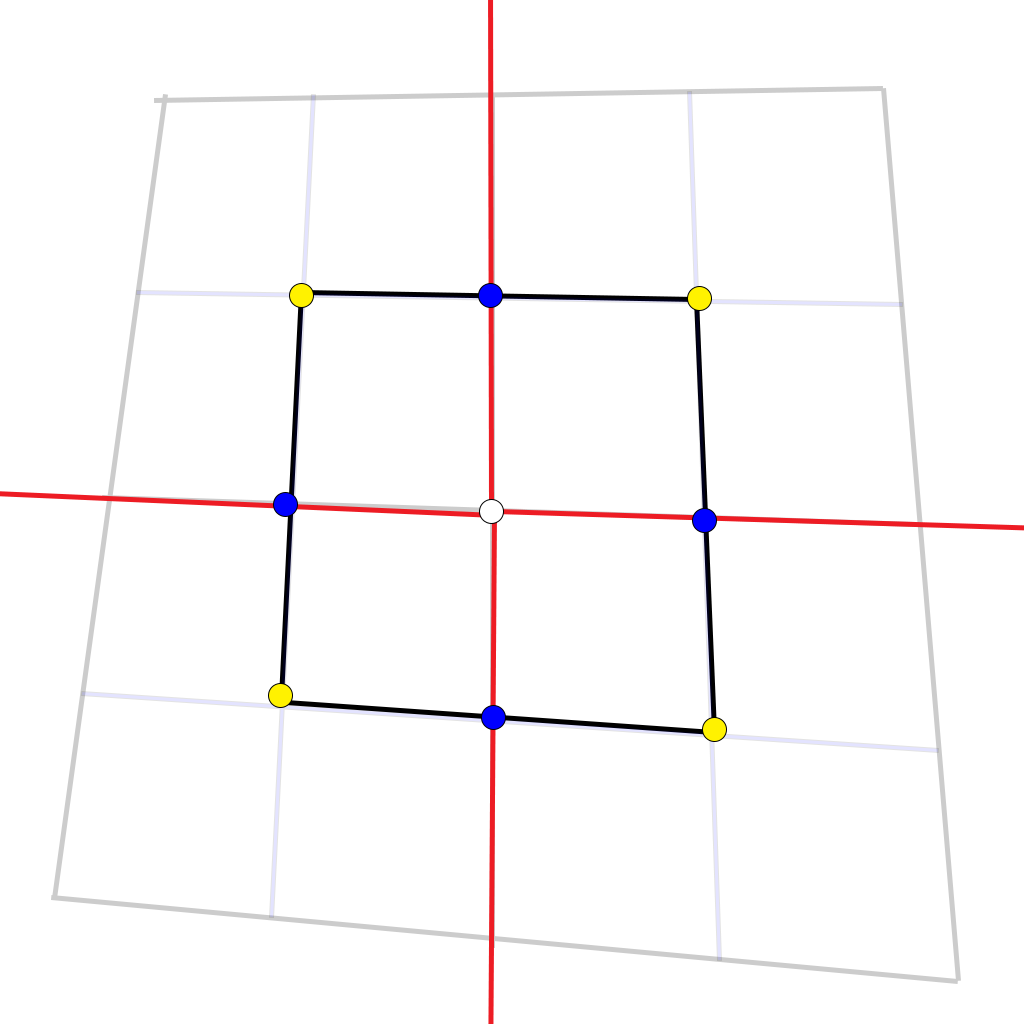}
        \caption{  $n=4$.}
        \label{fig_q}
    \end{subfigure}
    \hfill  
    \begin{subfigure}[t]{0.320\textwidth}
        \includegraphics[width=1.0\textwidth]{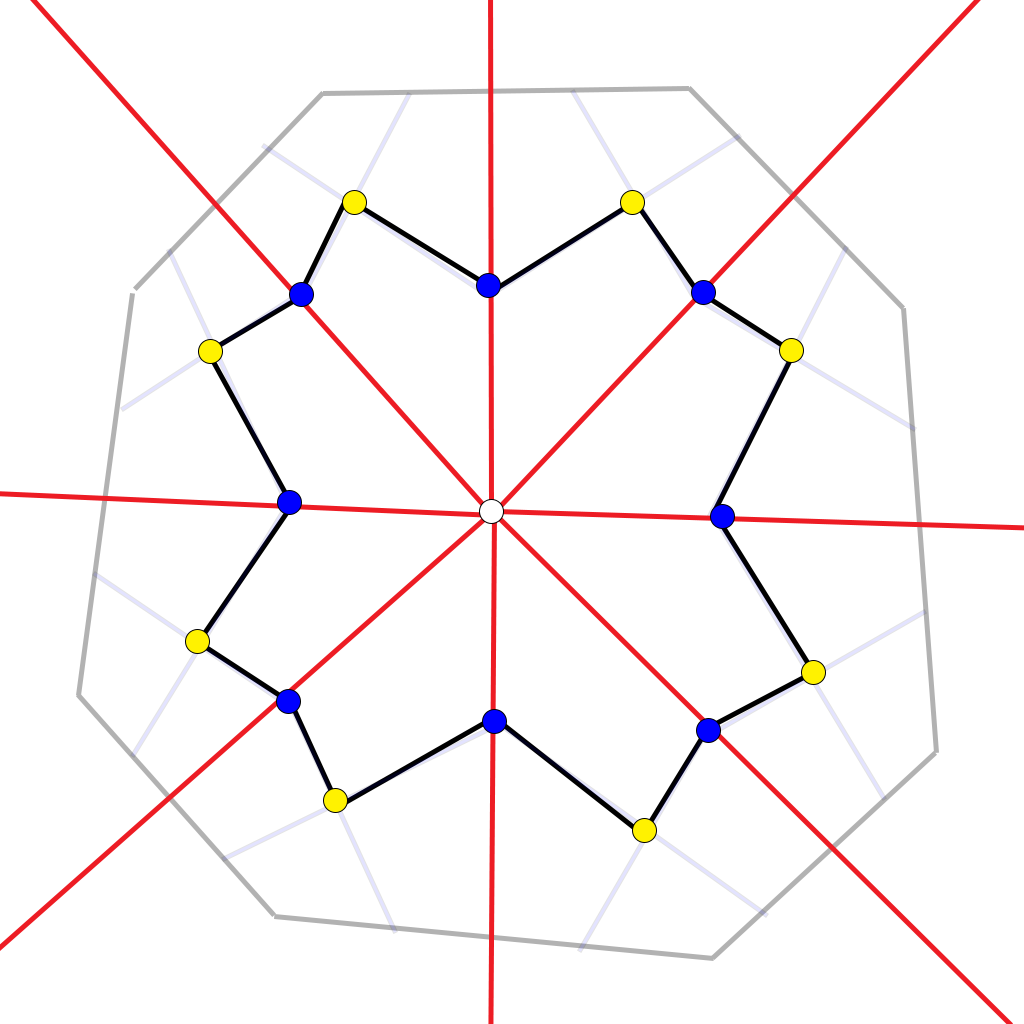}
        \caption{  $n=8$.}
        \label{fig_o}
    \end{subfigure}
    \hfill 
\caption{Examples demonstrate unbroken line property can only be preserved for $n=4$. This is because the initial structure provides the same bilinear configuration for four quadrilaterals. Since the de Casteljau algorithm consists of bilinear transformations, the lines do not break for $n=4$. The structures of the original control points that form convex polygons $n$ are shown in lighter color in (b).}
\label{fig_unbroken}
\end{figure}

This configuration along with the de Casteljau subdivision makes it easy to evaluate $G^1$ continuity around an extraordinary vertex.  It also provides an intuitive understanding of what is happening locally around the extraordinary vertex. All we need to check is whether the de Casteljau subdivision preserves the planarity condition of this polygon to confirm the preservation of the $G^1$ condition in an extraordinary vertex. Note that any linear combination of the vectors $\vec{V}_0,\vec{V}_1,\ldots \vec{V}_{K-1}$ still provides another vector in the same vector space. de Casteljau subdivision just provides such a linear combination of these vectors, therefore, the transformed vectors still define the same vector space. In other words, the planar property is preserved under the de Casteljau subdivision. The fact that the de Casteljau subdivision is simply a linear combination of control vertices suffices to demonstrate that the $G^1$ condition in extraordinary vertices is always preserved. 

$G^1$ continuity is one of the reasons why AS\&C method \cite{akleman2017} provides visually smoother results than the Catmull-Clark subdivision \cite{catmull1978}. The second reason is that the AS\&C method uses polygons formed by the Doo-Sabin subdivision \cite{doo1978}, which is known to provide planar regular and convex polygons in limit. As we discuss later, having more regular and convex polygon is useful to obtain visually desirable results. However, the Doo-Sabin stage may not necessarily be needed, since regular and convex polygons can still be obtained by methods other than Doo-Sabin.  

\subsection{C1 or G1 Condition in Boundaries}

The next question is to check if $C^1$ or $G^1$ conditions are preserved in the boundaries. To preserve the condition $C^1$ in the boundaries, the boundary points must always be the average of two inner points after subdivision. To provide the $G^1$ condition in non-planar boundary regions,  the necessary and sufficient condition is that the line that connects one boundary point and two inner points must be straight, i.e. the boundary point must be a Barycentric average of the two inner points \cite{bartels1987}. 

In this section, we demonstrate that conditions $C^1$ and $G^1$ in the boundaries can never be preserved when we use planar convex polygons to produce control points. Note that the $C^1$ condition requires the blue points to be in the middle of the line connecting two yellow points. Let us call this property {\it the unbroken line property}. If the property of the unbroken line is not preserved by the de Casteljau subdivision, we cannot obtain continuity of $C^1$ or $G^1$. Unfortunately, the lines start to break in the planar region, as shown in Figure~\ref{fig_example2}. These broken lines propagate along the edges when we apply the de Casteljau subdivision shown in Figure~\ref{fig_DC} to other boundary regions.

\begin{figure}[htpb!]
    \centering  
    \begin{subfigure}[t]{0.45\textwidth}
        \includegraphics[width=1.0\textwidth]{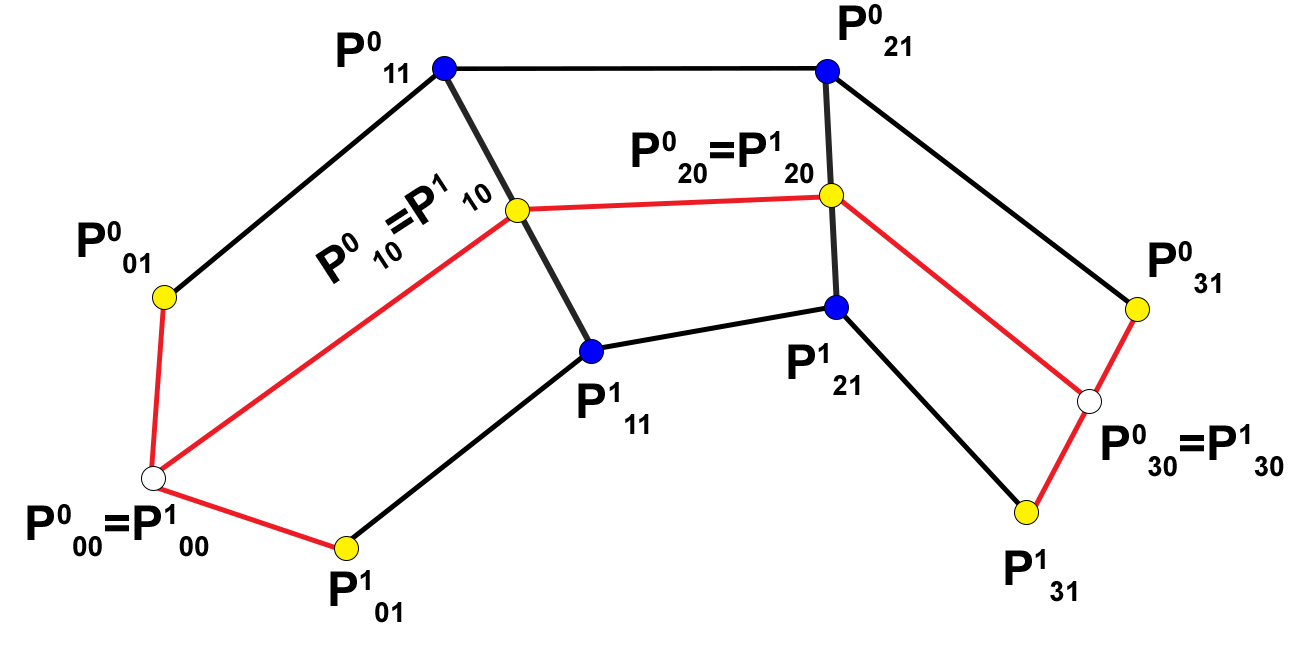}
        \caption{ Two patches that share an extraordinary point. This image shows B\'{e}zier control points on both sides of the boundary. Due to the extraordinary point, $P^0_{01}$, $P^0_{00}$, $P^1_{00}$, and $P^1_{01}$ are not on the same line.}
        \label{fig_C10}
    \end{subfigure}
    \hfill  
    \begin{subfigure}[t]{0.45\textwidth}
        \includegraphics[width=1.0\textwidth]{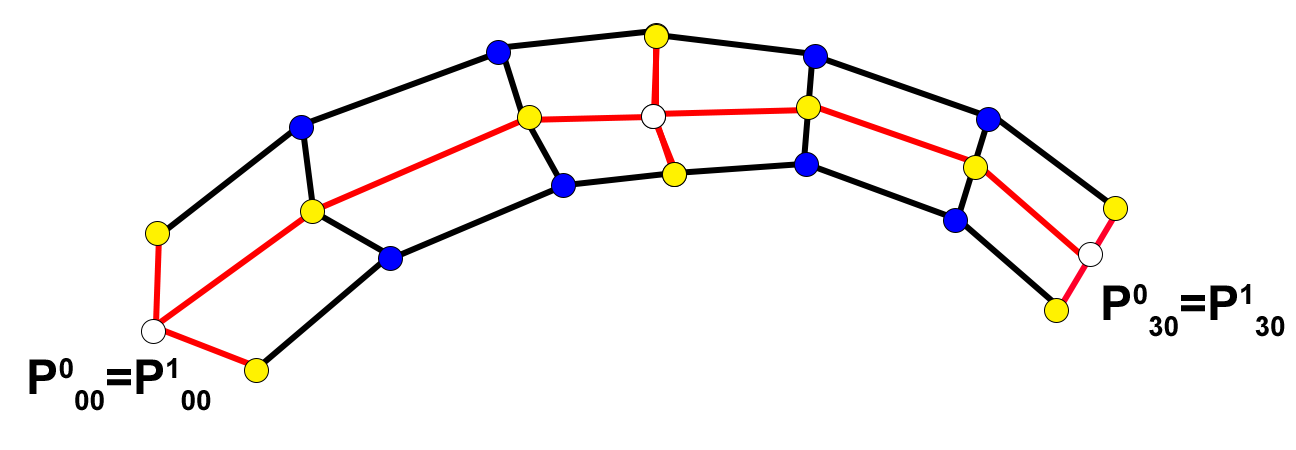}
        \caption{  Control points after the application of de Casteljau subdivision. Note that de Casteljau splits the original patches into two in the boundary.  }
        \label{fig_C11}
    \end{subfigure}
    \hfill  
    \begin{subfigure}[t]{0.65\textwidth}
        \includegraphics[width=1.0\textwidth]{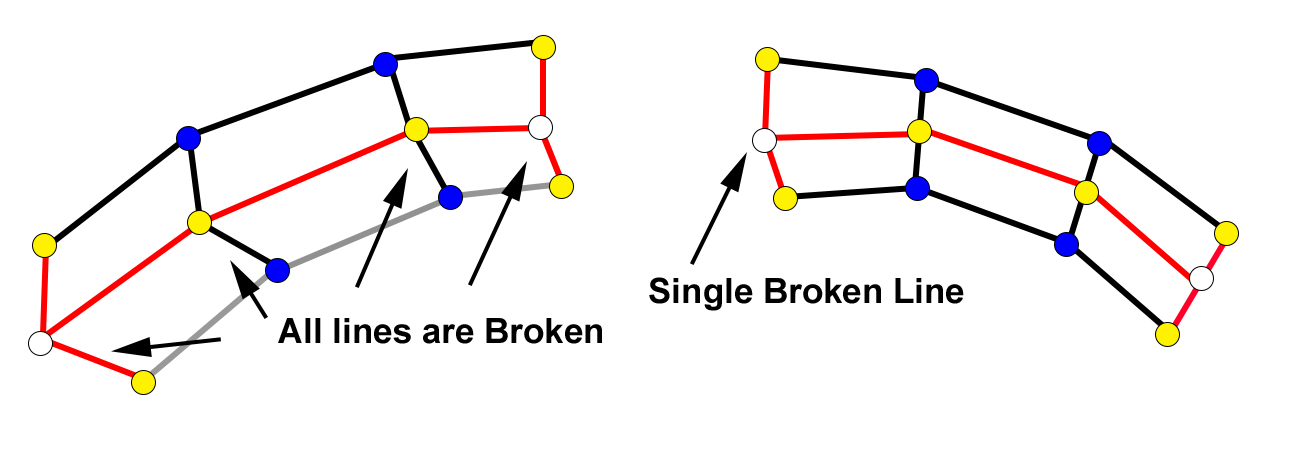}
        \caption{ Separated patches to demonstrate the structure of broken lines.}
        \label{fig_C12}
    \end{subfigure}
\caption{An example demonstrates the change in boundary control points after one application of the original de Casteljau subdivision around a boundary shared by two patches.}
\label{fig_boundaryproblem}
\end{figure}

\begin{figure}[htpb]
    \centering  
    \begin{subfigure}[t]{0.490\textwidth}
        \includegraphics[width=1.0\textwidth]{images/2}
        \caption{ The original de Casteljau around a planar extraordinary region.}
        \label{fig_exampleDC1}
    \end{subfigure}
    \hfill  
        \begin{subfigure}[t]{0.490\textwidth}
        \includegraphics[width=1.0\textwidth]{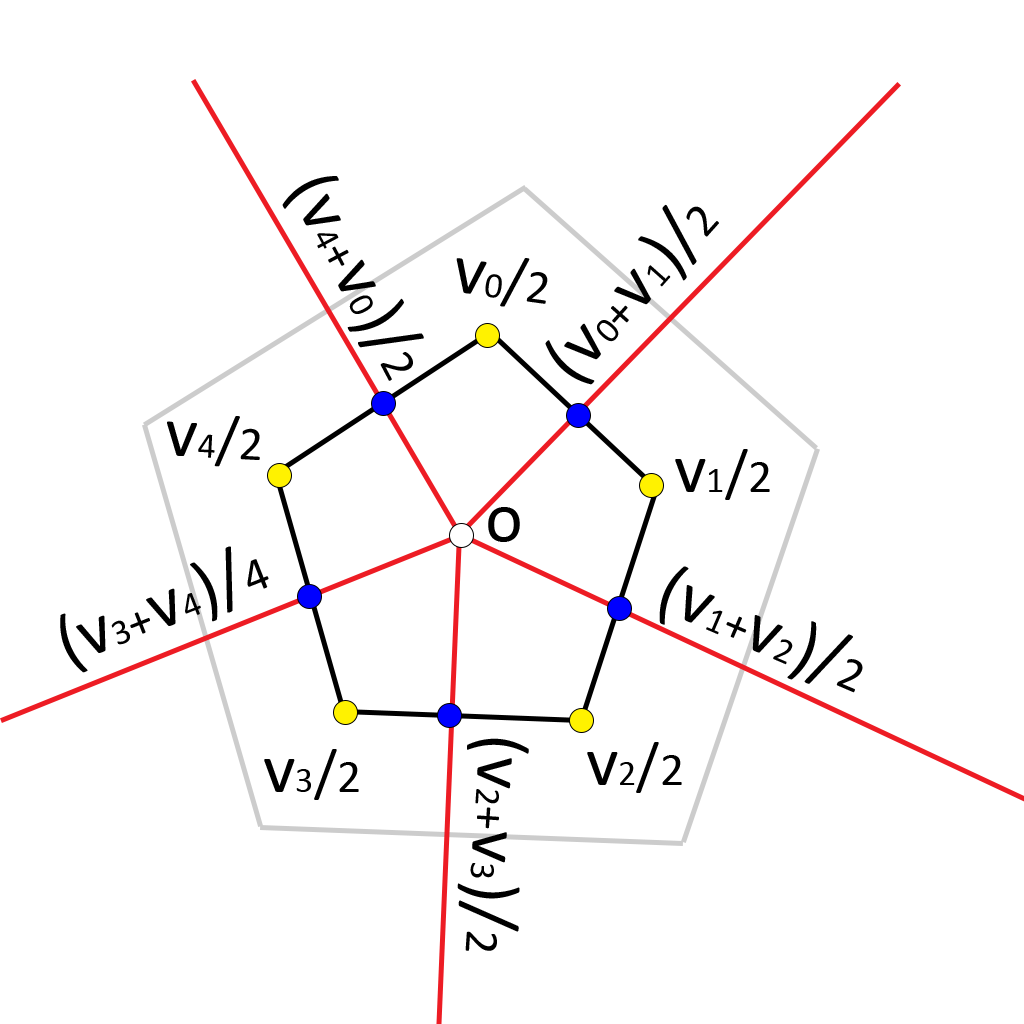}
        \caption{  The modified de Casteljau around the planar extraordinary region.}
        \label{fig_exampleDC2}
    \end{subfigure}
    \hfill  
\caption{An example that demonstrates the difference between the first steps of original and modified de Casteljau algorithms visually.}
\label{fig_exampleDC}
\end{figure}

To demonstrate this problem, we can now apply actual de Casteljau subdivision kernels to these vectors. The vertex at the origin, of course, never moves. In the first step of de Casteljau, the original blue points at $\frac{(\vec{V}_{k}+\vec{V}_{k+1})}{2}$ move to $\frac{(\vec{V}_{k}+\vec{V}_{k+1})}{4}$. On the other hand, the original yellow points at $\vec{V}_{k}$ move to $\frac{\vec{V}_{k}}{2}+
\frac{(\vec{V}_{k-1}+v_{k+1})}{8}$ since the new point is the calculated average of four control points. It is easy to show that the property of the unbroken line is preserved if and only if $n=4$ (see Figure~\ref{fig_unbroken}). For all other cases, the process breaks the lines and the initial planar polygon with $K$ sides becomes a planar polygon with $2K$ sides (see Figures~\ref{fig_exampleDC}, \ref{fig_t} and \ref{fig_o} ). In other words, the straight line segments will be broken originally if $K \neq 4$.

The problem is that these broken lines are not contained in the planar region (see Figure~\ref{fig_boundaryproblem}). They propagate along the boundary-edge. As can be verified by using two de Casteljau kernels in Figures~\ref{fig_DC21} and~\ref{fig_DC31} that are used to calculate $P'_{21}$ and $P'_{31}$, all boundary lines in newly created patches that share the original extraordinary vertex are broken (see Figure~\ref{fig_C12}). An additional problem is that the opposite end of the newly created patches that share the 4-valent vertex now also has a broken line caused by the computation of $P'_{31}$'s. As a result, the broken lines travel further towards the 4-valent vertex in every iteration.

The immediate neighborhood of an extraordinary point is still planar, since the first set of broken line segments is still on the original plane. However, the rest of the broken lines define two inconsistent bilinear surfaces with neighboring control points along the boundary that separates two consecutive patches. These broken lines cause discontinuity $C^1$ and $G^1$ along the boundary between two patches. The problem of broken lines cannot be solved by using higher-degree or rational polynomials, since the kernels for boundary points will still be in the same basic form for both higher-degree and rational cases.

In summary, the broken line segments that are by-products of extraordinary vertices cannot be contained in the planar region around extraordinary points. They create nonplanar bilinear regions by traveling along the boundary in every subsequent step of the de Casteljau algorithm while causing $C^1$ and $G^1$ discontinuity along boundary edges. Although the problem is expected to be worse when $K$, the valence of extraordinary vertex increases, discontinuity along boundary edges $C^1$ and $G^1$ is not visible to a casual viewer during interactive modeling. This is also not unexpected, as these discontinuities occur along $C^2$ continuous curves. This effect is subtle, since the broken lines gradually become more and more straightened by a series of interpolations.

For interactive applications, therefore, AS\&C method \cite{akleman2017} can still be used successfully. This algorithm still provides stitched surfaces that appear to be visually $G^1$ continuous. On the other hand, if the resulting smooth shapes are to
be used for simulation, manufacturing, or 3D printing, there is a need to remove $C^1$ discontinuities that are important for the shape quality. 

\section{Modified de Casteljau Subdivision \label{sec_MDC}}

Although it is not possible to obtain $C^1$ continuity along the edges emanating from extraordinary vertices, we observe that this problem can be resolved by a variety of minor modifications to the de Casteljau subdivision that can guarantee the avoidance of broken lines. These are essentially subdivision algorithms, and we cannot have a closed polynomial form anymore.  

\begin{figure}[htpb]
    \centering  
    \begin{subfigure}[t]{0.320\textwidth}
        \includegraphics[width=1.0\textwidth]{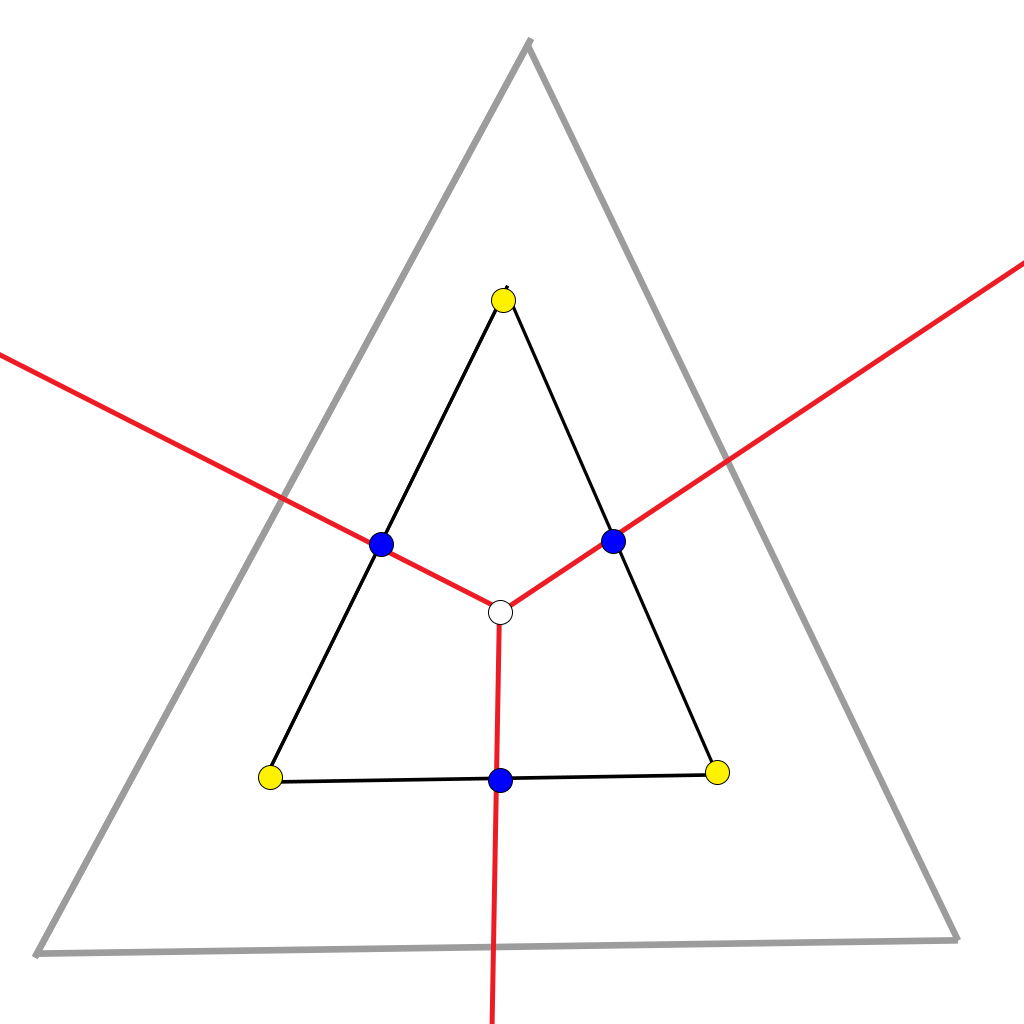}
        \caption{ $n=3$.}
        \label{fig_t1}
    \end{subfigure}
    \hfill  
    \begin{subfigure}[t]{0.320\textwidth}
        \includegraphics[width=1.0\textwidth]{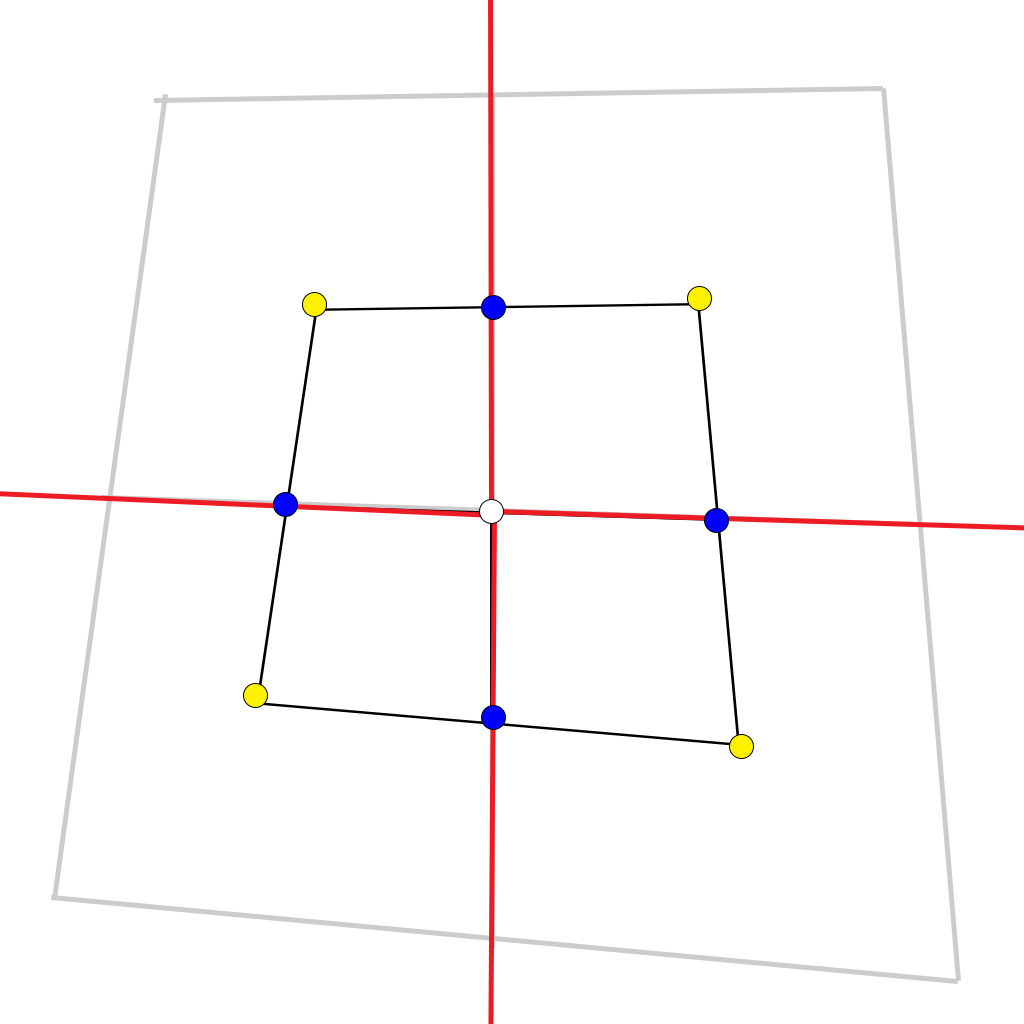}
        \caption{  $n=4$.}
        \label{fig_q1}
    \end{subfigure}
    \hfill  
    \begin{subfigure}[t]{0.320\textwidth}
        \includegraphics[width=1.0\textwidth]{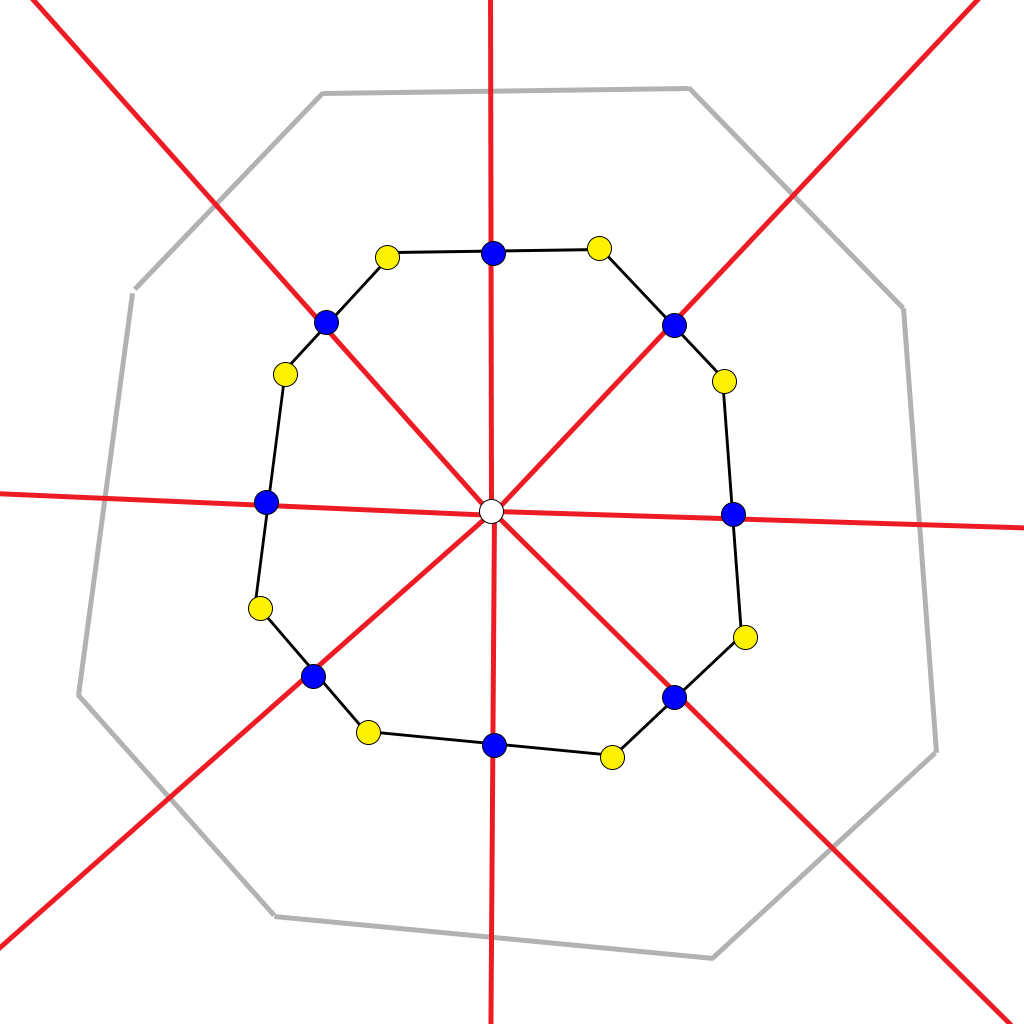}
        \caption{  $n=8$.}
        \label{fig_o1}
    \end{subfigure}
    \hfill 
\caption{Examples demonstrating the first step of modified de Casteljau subdivision. }
\label{fig_modified}
\end{figure}

\begin{figure*}[htpb]
    \centering  
    \begin{subfigure}[t]{0.24\textwidth}
        \includegraphics[width=1.0\textwidth]{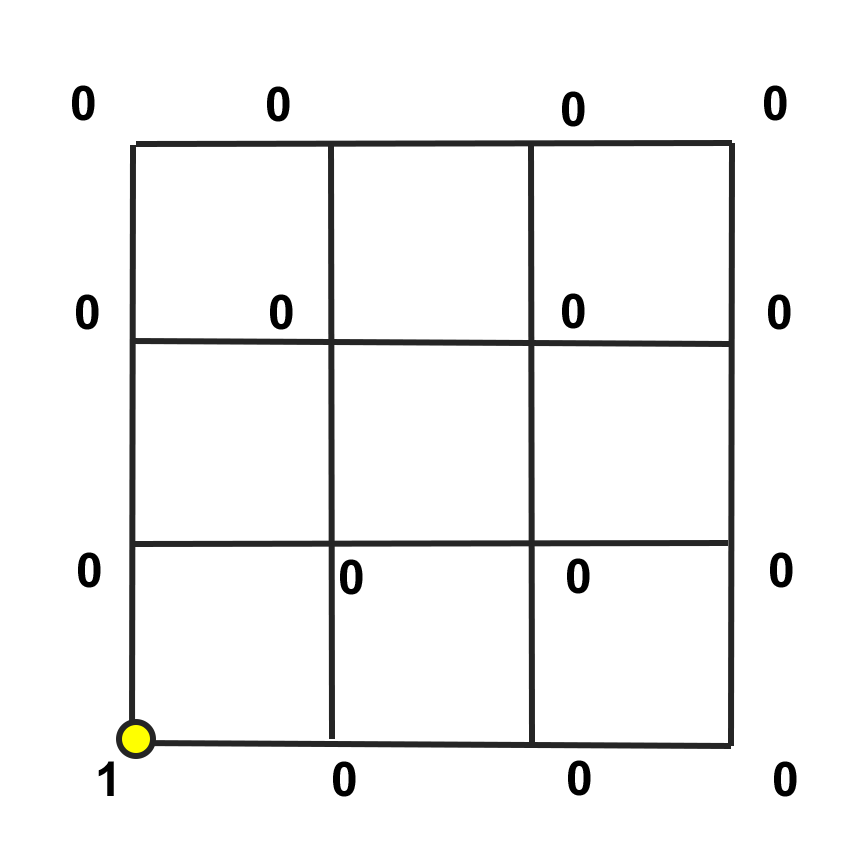}
        \caption{ New position of $p_{00}$.}
        \label{fig_MDC00}
    \end{subfigure}
    \hfill  
    \begin{subfigure}[t]{0.24\textwidth}
        \includegraphics[width=1.0\textwidth]{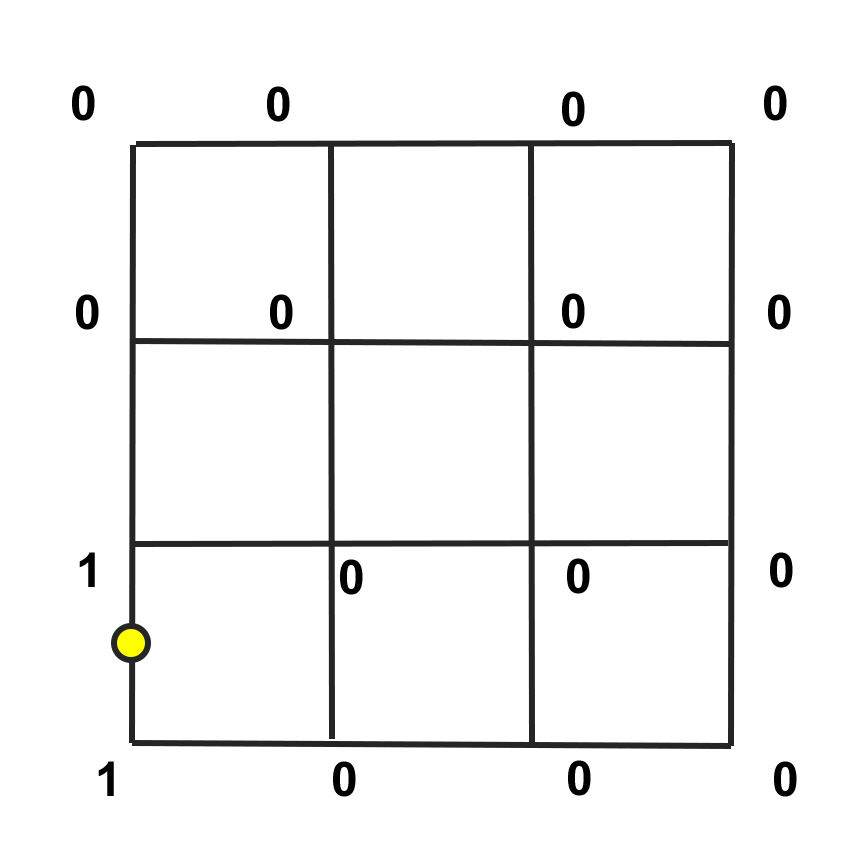}
        \caption{ New position of $p_{01}$.}
        \label{fig_MDC01}
    \end{subfigure}
    \hfill 
    \begin{subfigure}[t]{0.24\textwidth}
        \includegraphics[width=1.0\textwidth]{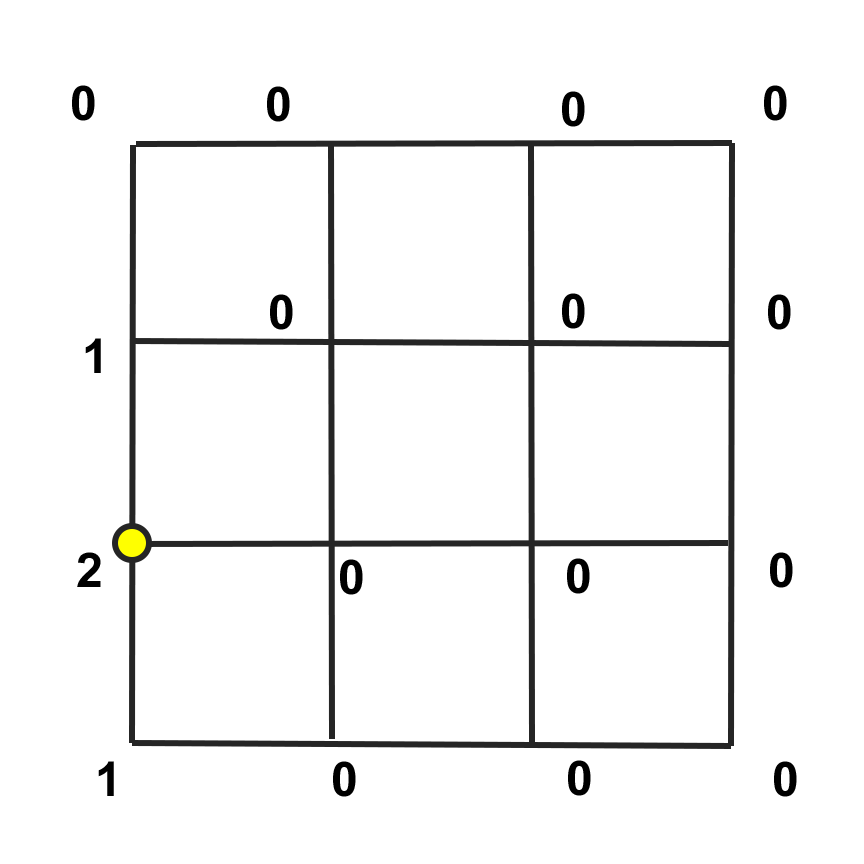}
        \caption{ New position of $p_{02}$.}
        \label{fig_MDC02}
    \end{subfigure}
    \hfill
    \begin{subfigure}[t]{0.24\textwidth}
        \includegraphics[width=1.0\textwidth]{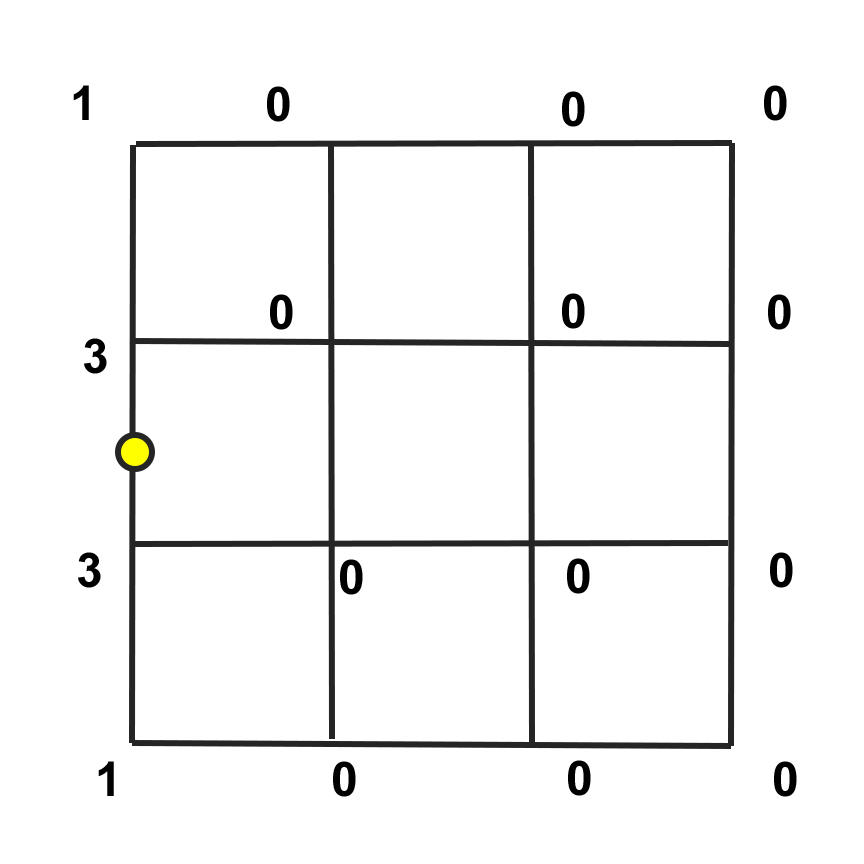}
        \caption{ New position of $p_{03}$.}
        \label{fig_MDC03}
    \end{subfigure}
    \hfill
        \begin{subfigure}[t]{0.24\textwidth}
        \includegraphics[width=1.0\textwidth]{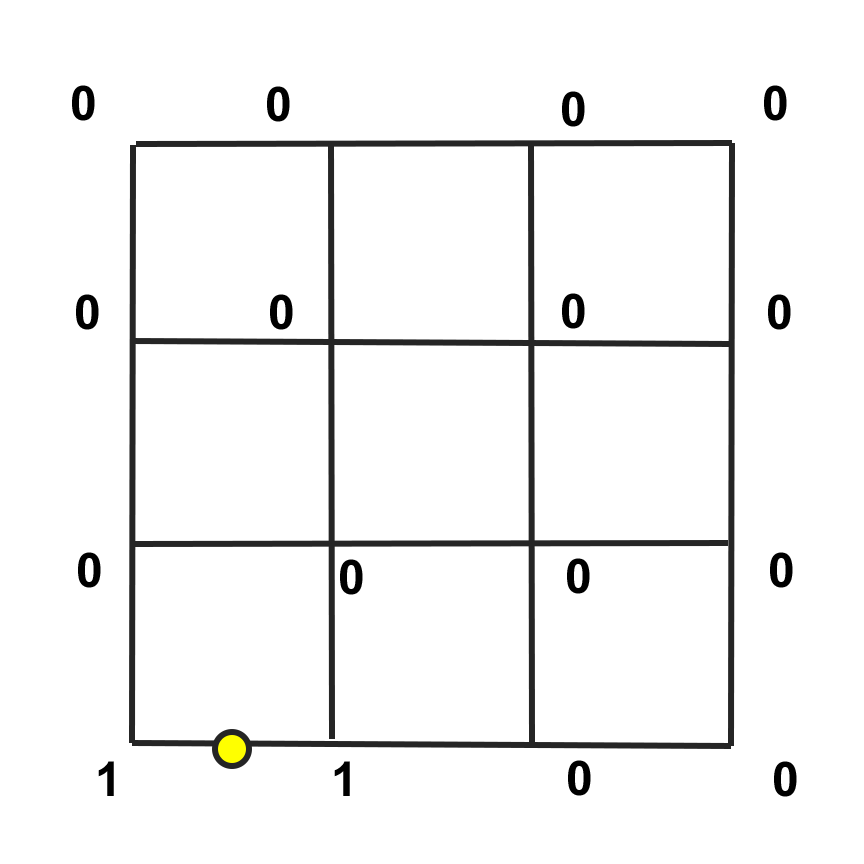}
        \caption{ New position of $p_{10}$.}
        \label{fig_MDC10}
    \end{subfigure}
    \hfill  
    \begin{subfigure}[t]{0.24\textwidth}
        \includegraphics[width=1.0\textwidth]{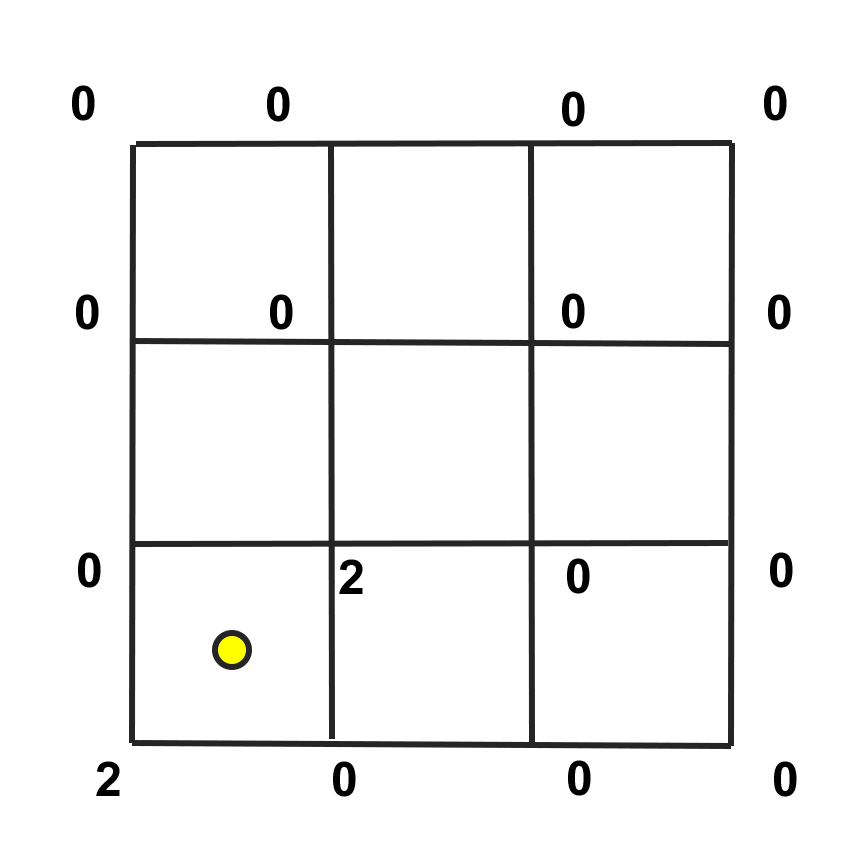}
        \caption{ New position of $p_{11}$.}
        \label{fig_MDC11}
    \end{subfigure}
    \hfill 
    \begin{subfigure}[t]{0.24\textwidth}
        \includegraphics[width=1.0\textwidth]{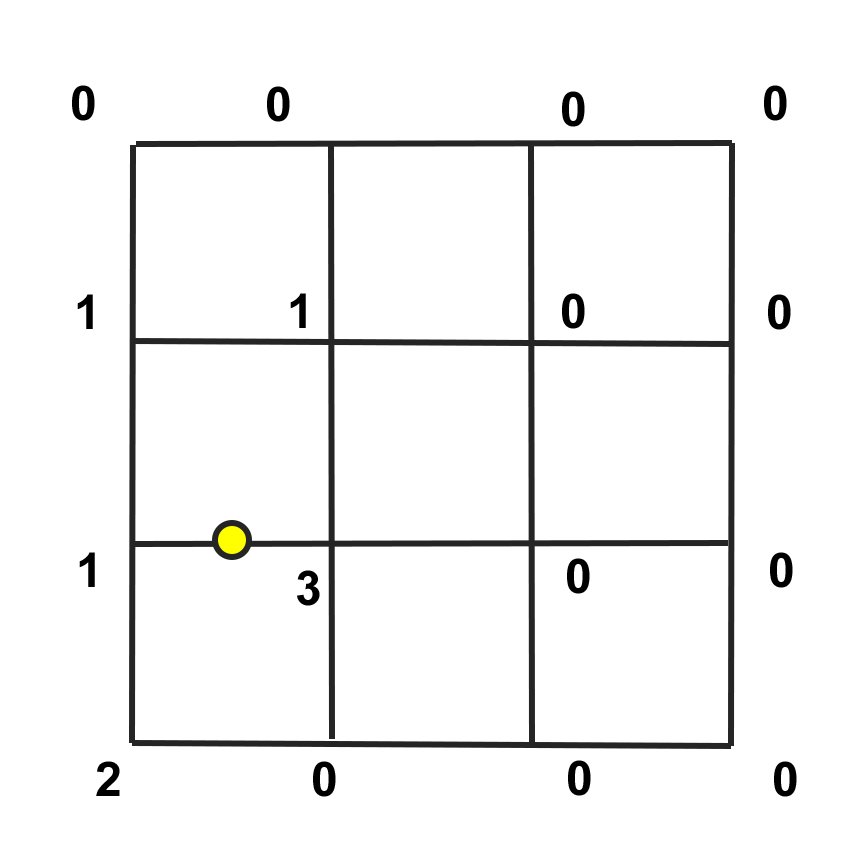}
        \caption{ New position of $p_{12}$.}
        \label{fig_MDC12}
    \end{subfigure}
    \hfill
    \begin{subfigure}[t]{0.24\textwidth}
        \includegraphics[width=1.0\textwidth]{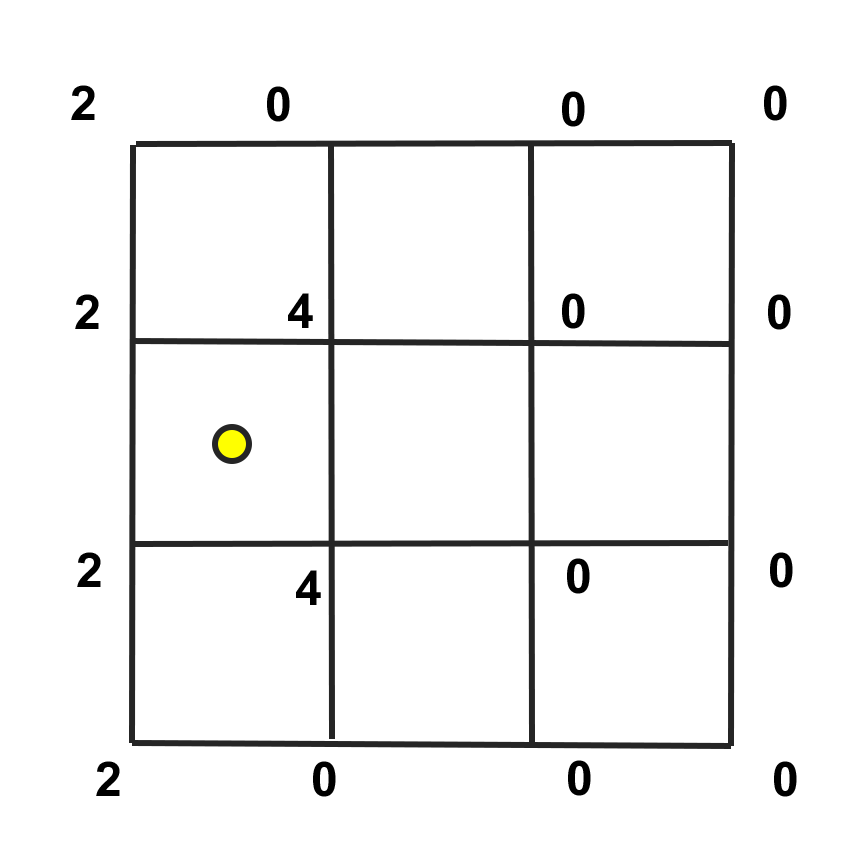}
        \caption{ New position of $p_{13}$.}
        \label{fig_MDC13}
    \end{subfigure}
    \hfill

        \centering  
    \begin{subfigure}[t]{0.24\textwidth}
        \includegraphics[width=1.0\textwidth]{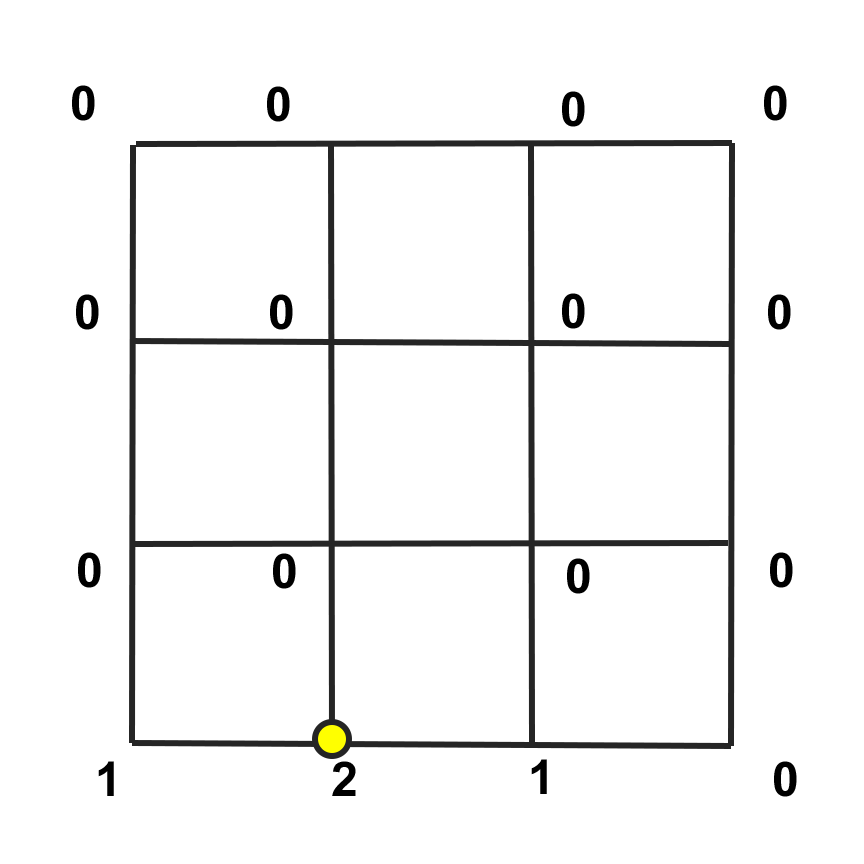}
        \caption{ New position of $p_{20}$.}
        \label{fig_MDC20}
    \end{subfigure}
    \hfill  
    \begin{subfigure}[t]{0.24\textwidth}
        \includegraphics[width=1.0\textwidth]{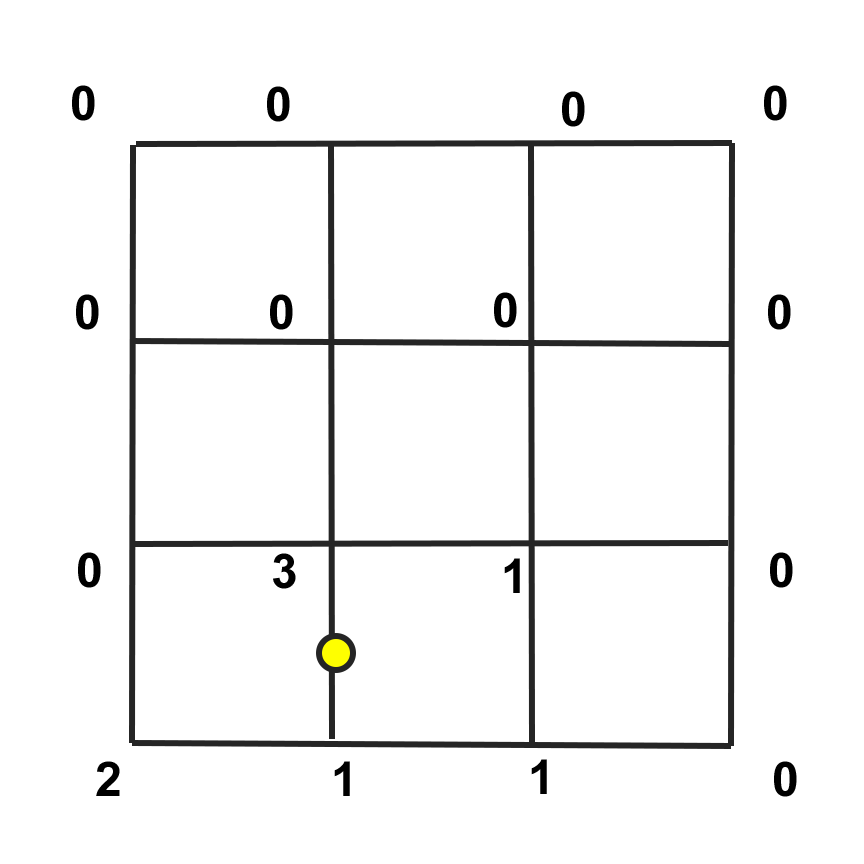}
        \caption{ New position of $p_{21}$.}
        \label{fig_MDC21}
    \end{subfigure}
    \hfill 
    \begin{subfigure}[t]{0.24\textwidth}
        \includegraphics[width=1.0\textwidth]{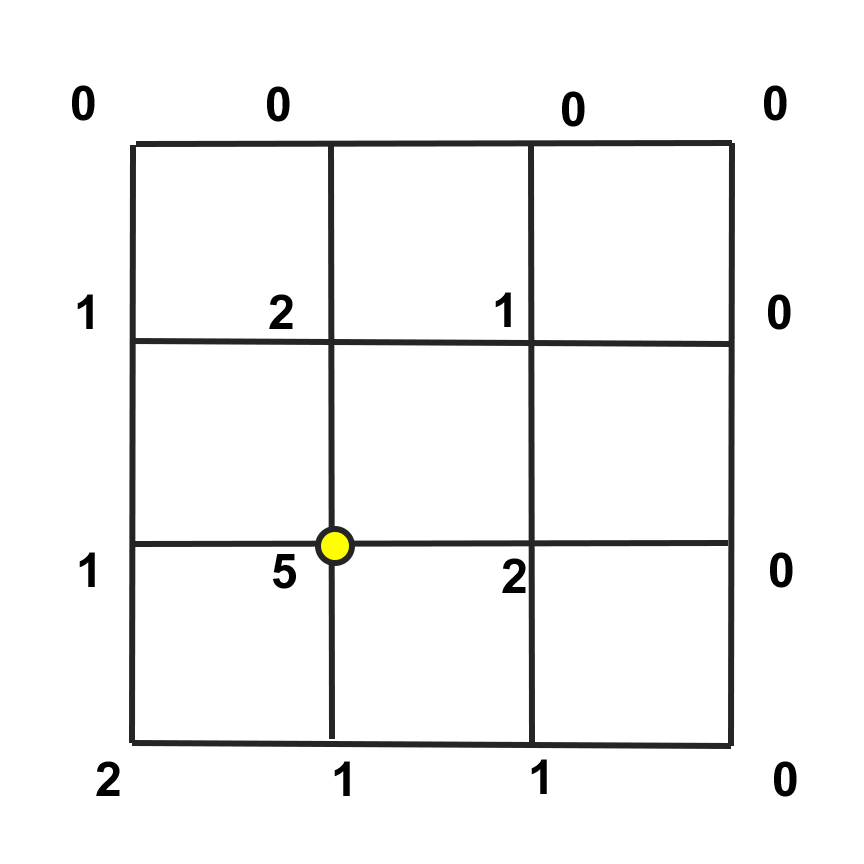}
        \caption{ New position of $p_{22}$.}
        \label{fig_MDC22}
    \end{subfigure}
    \hfill
    \begin{subfigure}[t]{0.24\textwidth}
        \includegraphics[width=1.0\textwidth]{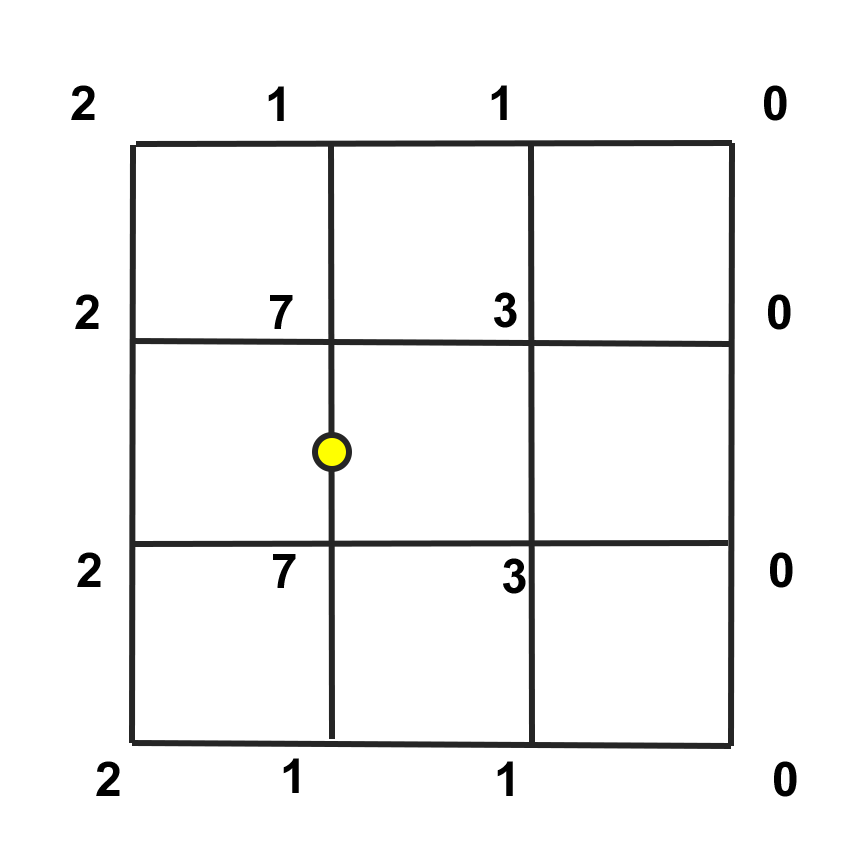}
        \caption{ New position of $p_{23}$.}
        \label{fig_MDC23}
    \end{subfigure}
    \hfill
        \begin{subfigure}[t]{0.24\textwidth}
        \includegraphics[width=1.0\textwidth]{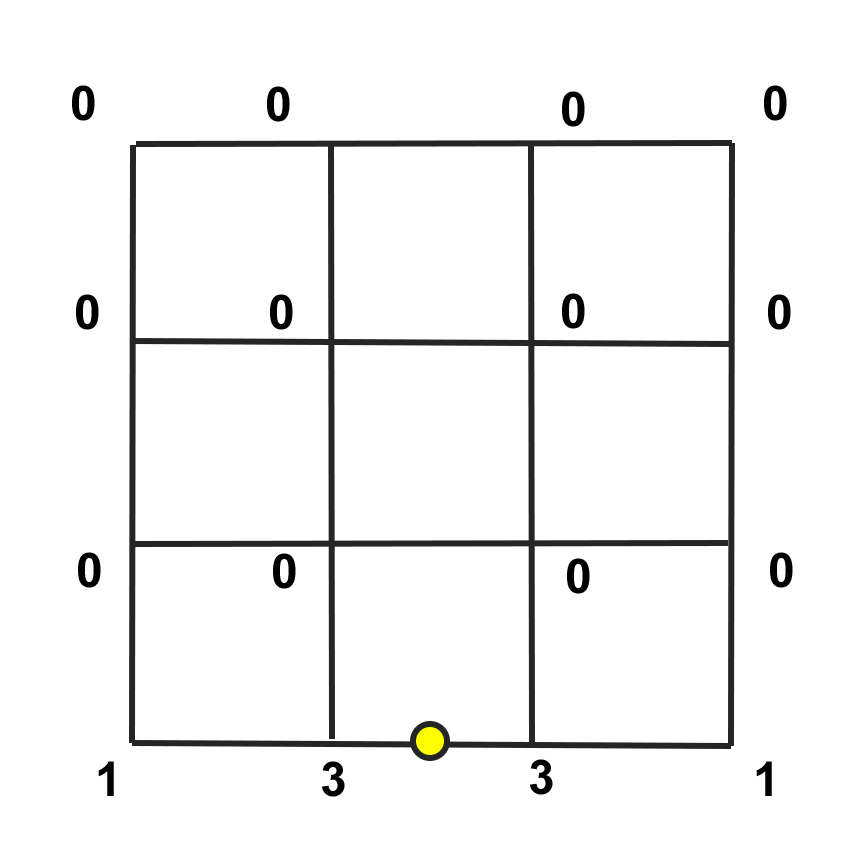}
        \caption{ New position of $p_{30}$.}
        \label{fig_MDC30}
    \end{subfigure}
    \hfill  
    \begin{subfigure}[t]{0.24\textwidth}
        \includegraphics[width=1.0\textwidth]{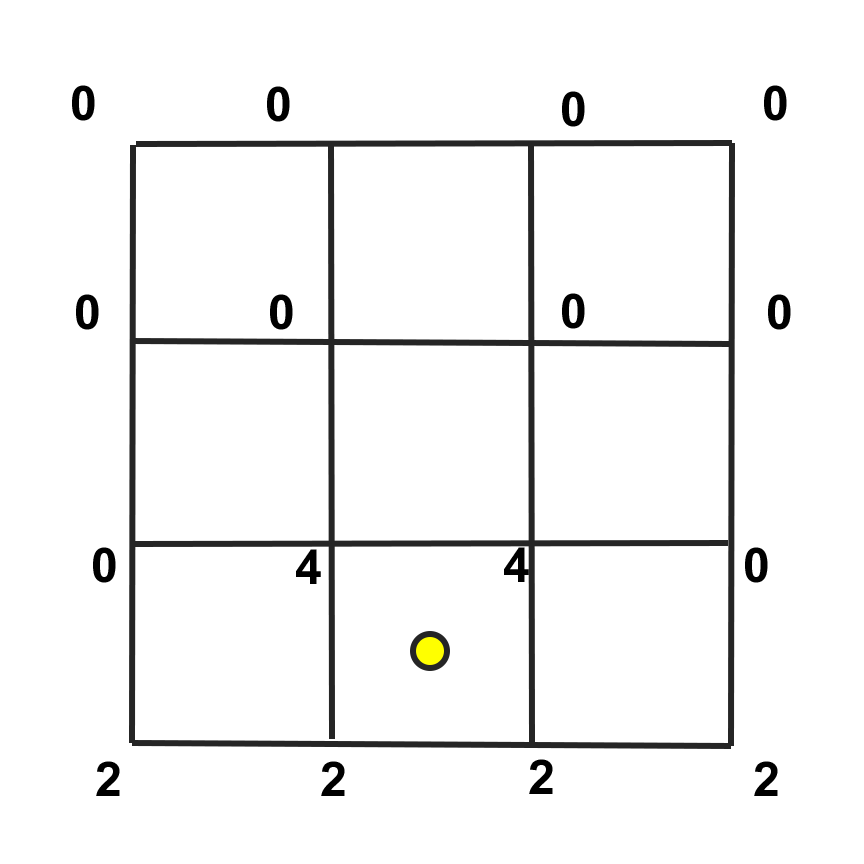}
        \caption{ New position of $p_{31}$.}
        \label{fig_MDC31}
    \end{subfigure}
    \hfill 
    \begin{subfigure}[t]{0.24\textwidth}
        \includegraphics[width=1.0\textwidth]{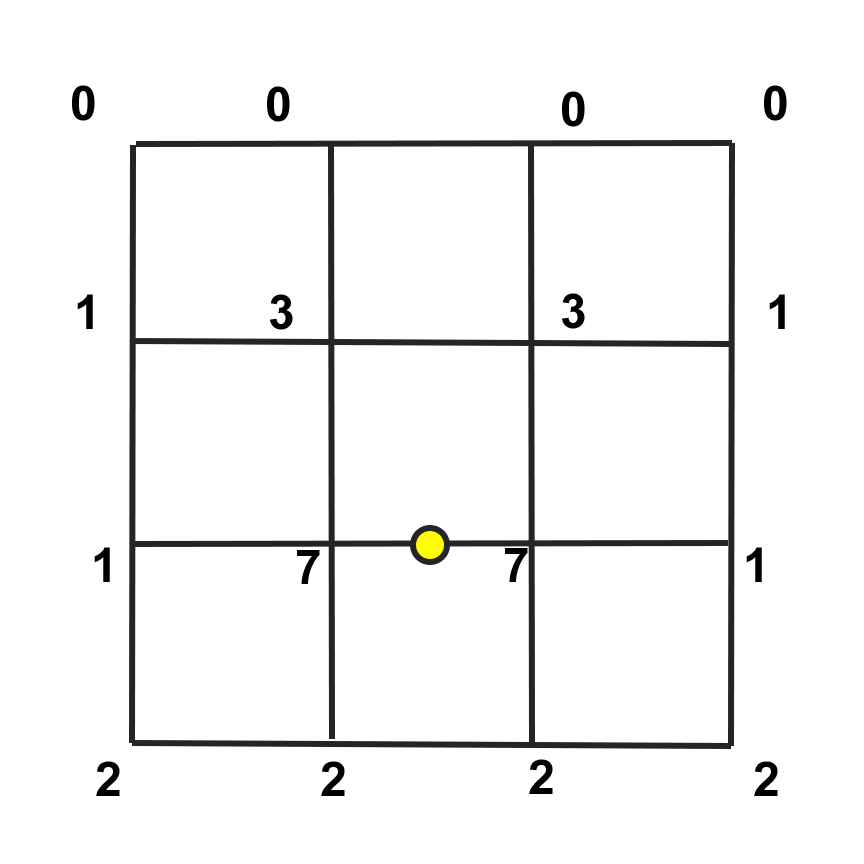}
        \caption{ New position of $p_{32}$.}
        \label{fig_MDC32}
    \end{subfigure}
    \hfill
    \begin{subfigure}[t]{0.24\textwidth}
        \includegraphics[width=1.0\textwidth]{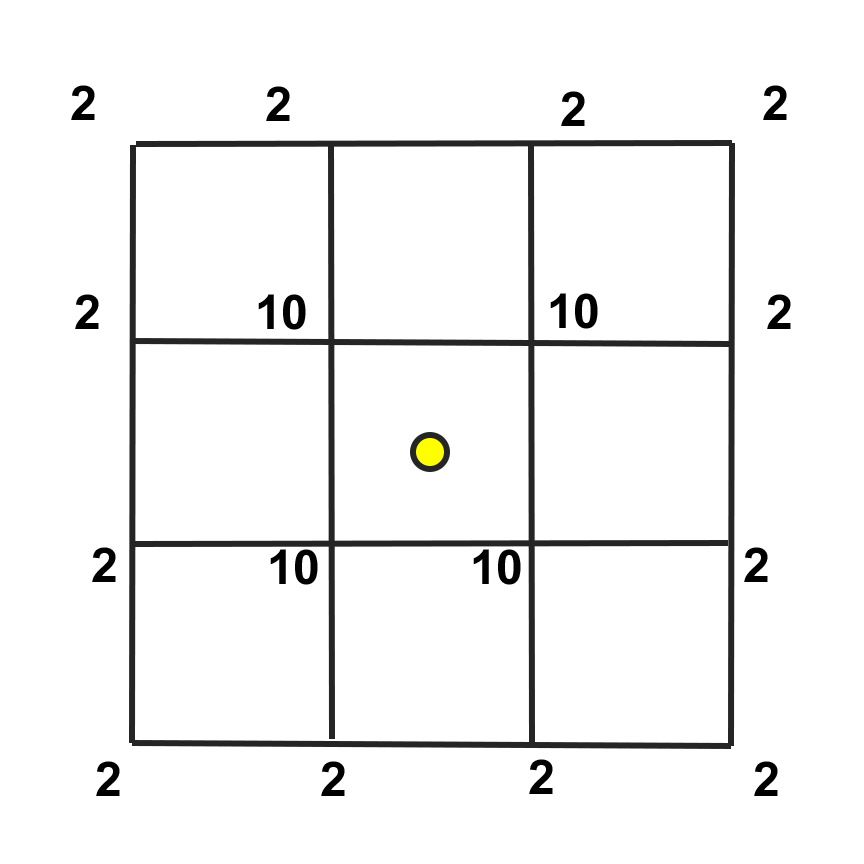}
        \caption{ New position of $p_{33}$.}
        \label{fig_MDC33}
    \end{subfigure}
    \hfill
\caption{This figure shows the Modified De-Casteljau kernels to produce each control point. }
\label{fig_MDC}
\end{figure*}

Among these solutions, one of them, in particular, appears to be much simpler and more powerful than others. Notice that the problem of broken lines stems from the additional term $\frac{(\vec{V}_{k-1}+\vec{V}_{k+1})}{8}$, we simply remove it and move the original yellow points at $\vec{V}_{k}$, to $\frac{\vec{V}_{k}}{2}$ (see~\ref{fig_exampleDC2}). This simple change can replace the original bilinear transformation in Figure~\ref{fig_DC11} with a uniform scale around the origin in Figure~\ref{fig_MDC11} to calculate $P'_{11}$. 

Some other de Casteljau kernels must also change to avoid line breaks. As discussed above, the kernels in Figures~\ref{fig_DC21} and~\ref{fig_DC31}, which are used to calculate $P'_{21}$ and $P'_{31}$, also produce broken lines. We want to point out that the problems in these two kernels come from the interpolation of the bilinear transformation that is used to calculate $P'_{11}$. If we replace the original bilinear term with scaling, we obtain new kernels for $P'_{21}$ and $P'_{31}$ as shown in Figures~\ref{fig_MDC21} and~\ref{fig_MDC31}. Due to the symmetry, we have to use the same kernels for $P'_{12}$ and $P'_{13}$ as shown in Figures~\ref{fig_MDC12} and~\ref{fig_MDC13}. 

We want the boundary curves to stay the same. Therefore, the kernels for computing $P'_{00}$, $P'_{01}$, $P'_{02}$, $P'_{03}$, $P'_{10}$, $P'_{20}$ and $P'_{30}$ will remain the same. The kernels for $P'_{22}$, $P'_{23}$, $P'_{32}$, and $P'_{33}$ can also remain the same. However, since we do not use the original de Casteljau subdivision, we do not create bicubic B\'{e}zier Patches anymore. In other words, we no longer get $C^2$ continuity. If we keep the original kernels for these four points, the newly created patches will no longer be merged with continuity $C^2$, since the positions of $P'_{12}$, $P'_{21}$, $P'_{31}$, and $P'_{13}$ are not the same as before. Therefore, to obtain $C^2$ again between newly created patches, we need to readjust the kernels corresponding to $P'_{22}$, $P'_{23}$, $P'_{32}$, and $P'_{33}$.

To compute new kernels for these four points, we again used the original de Casteljau process, which provides these kernels through an iterative process. When we replace the boundary kernels in an iterative process, we obtain the kernels given in Figures~\ref{fig_MDC22},~\ref{fig_MDC23},~\ref{fig_MDC32}, and~\ref{fig_MDC33}. All these kernels can be obtained simply using the iterative process of de Casteljau just by replacing the original bilinear kernels of $P'_{11}$, $P"_{13}$, $P'''_{31}$, and $P''''_{33}$ where the prime notation refers to four different subpatches of de Casteljau.

Since we keep the boundary kernels the same, this modification does not change the shapes of the boundary curves. It only changes the partial derivative at the extraordinary point. Therefore, the difference in positions is really negligible except for a significant visual improvement in mesh quality, as shown in Figures~\ref{fig_3-valent} and~\ref{fig_10-valent}. Another useful property of this particular modification is that we do not have to consider neighborhood patches. Each patch can still be evaluated independently. 

\section{Implementation, Discussion and Results \label{sec_implementation}}

In this paper, we introduce an extremely simple modification of the de Casteljau subdivision that can guarantee the production of continuity $C^1$ and $G^1$ for any extraordinary vertex. However, the modified de Casteljau subdivision only guarantees that the limit surface is free of discontinuities $G^1$ or $C^1$. In any resolution, if we compute B\'{e}zier surface with direct evaluation using higher resolution control points obtained from the de Casteljau subdivision, there will still be $G^1$ discontinuity that could be visible in resolutions higher than current resolutions. Therefore, it is better to obtain the desired resolution directly with the modified de Casteljau subdivision if there is an extraordinary vertex.  

The modified de Casteljau can be applied to $n=4$ cases even when the control points are not in the same plane. However, in this case, the resulting surface is not a polynomial. Therefore, it is better to apply original de Casteljau algorithms for regular patches whose corners are shared only by $4$ other patches. With each application of the de Casteljau subdivision, each of the original patches will be subdivided into four patches. This process will populate surfaces with regular bicubic B\'{e}zier patches that do not need to be further subdivided. We have to continue subdivision only around extraordinary vertices. This process creates hierarchically organized bicubic B\'{e}zier patches as shown in Figures~\ref{fig_3-valent} and~\ref{fig_10-valent}. 

\begin{figure*}[htpb]
    \centering  
    \begin{subfigure}[t]{0.49\textwidth}
        \includegraphics[width=1.0\textwidth]{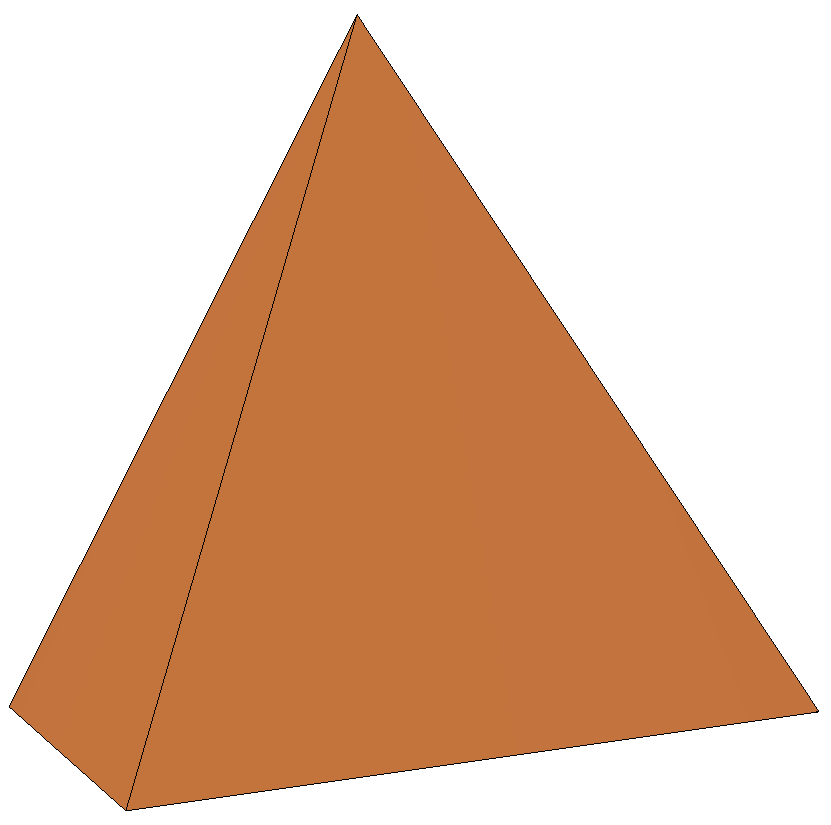}
        \caption{ Start with an orientable two-manifold polygonal mesh.}
        \label{fig_Procedure/tetrahedron_0}
    \end{subfigure}
    \hfill  
    \begin{subfigure}[t]{0.48\textwidth}
        \includegraphics[width=1.0\textwidth]{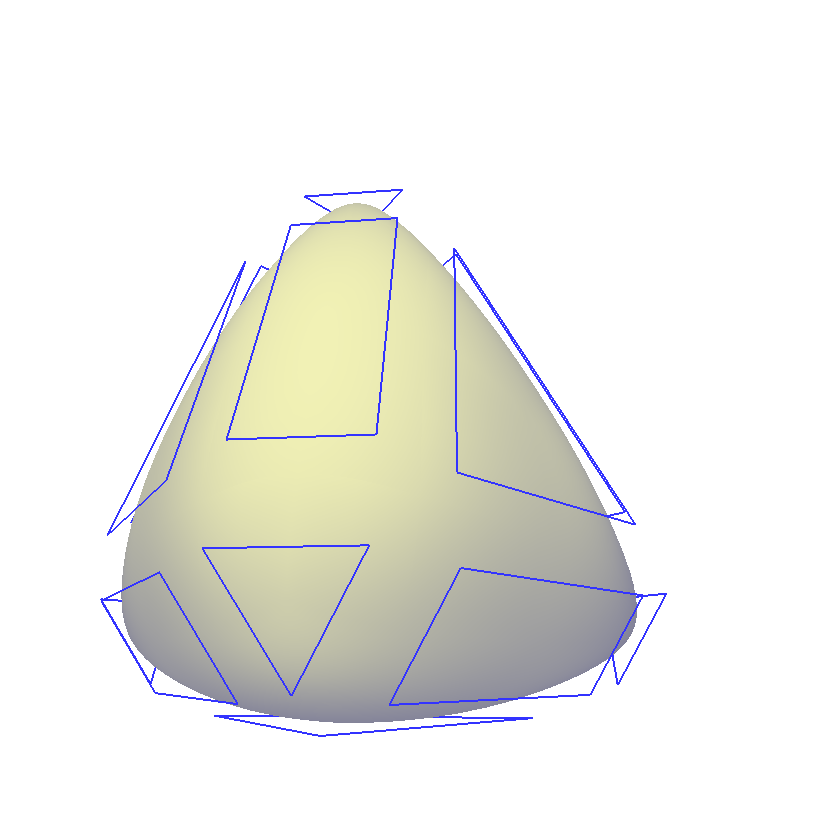}
        \caption{ Assign K-sided polygons to every face, vertex, and edge of the initial mesh.}
        \label{fig_Procedure/tetrahedron_1}
    \end{subfigure}
    \hfill  
    \begin{subfigure}[t]{0.48\textwidth}
        \includegraphics[width=1.0\textwidth]{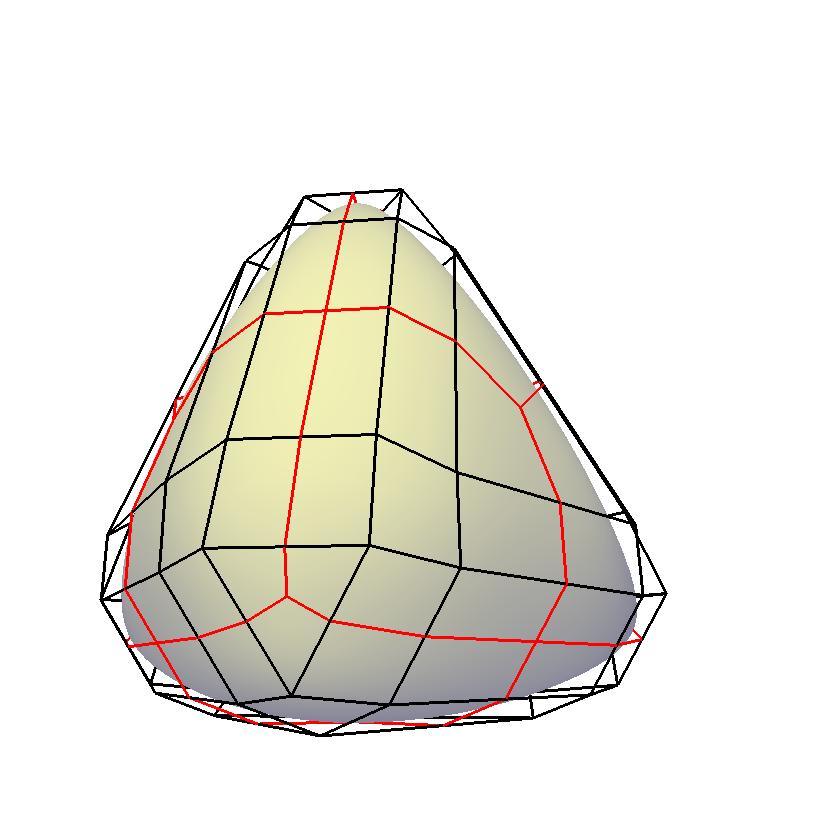}
        \caption{ Create B\'{e}zier control polyhedra using K-sided polygons.}
        \label{fig_Procedure/tetrahedron_3}
    \end{subfigure}
    \hfill  
    \begin{subfigure}[t]{0.48\textwidth}
        \includegraphics[width=1.0\textwidth]{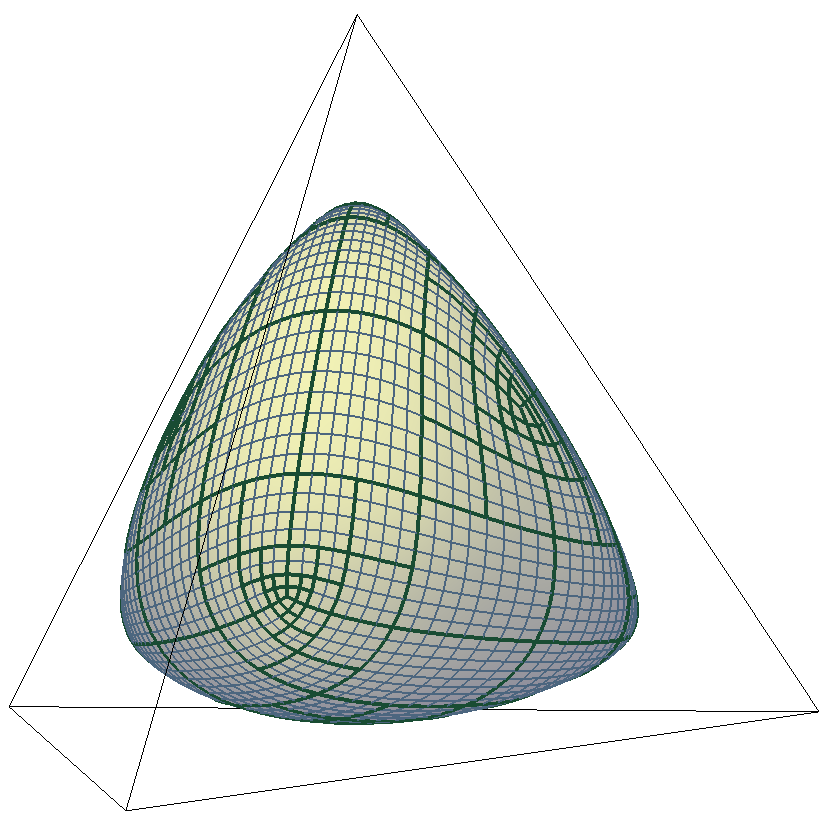}
        \caption{ Obtain a smooth surface that consist of bicubic B\'{e}zier patches.}
        \label{fig_Procedure/tetrahedron_6}
    \end{subfigure}
\caption{The steps of our process. In this case, the initial two-manifold polygonal mesh is a regular tetrahedron. }
\label{fig_Procedure}
\end{figure*}

\begin{figure*}[htbp!]
    \centering  
    \begin{subfigure}[t]{0.46\textwidth}
        \includegraphics[width=1.0\textwidth]{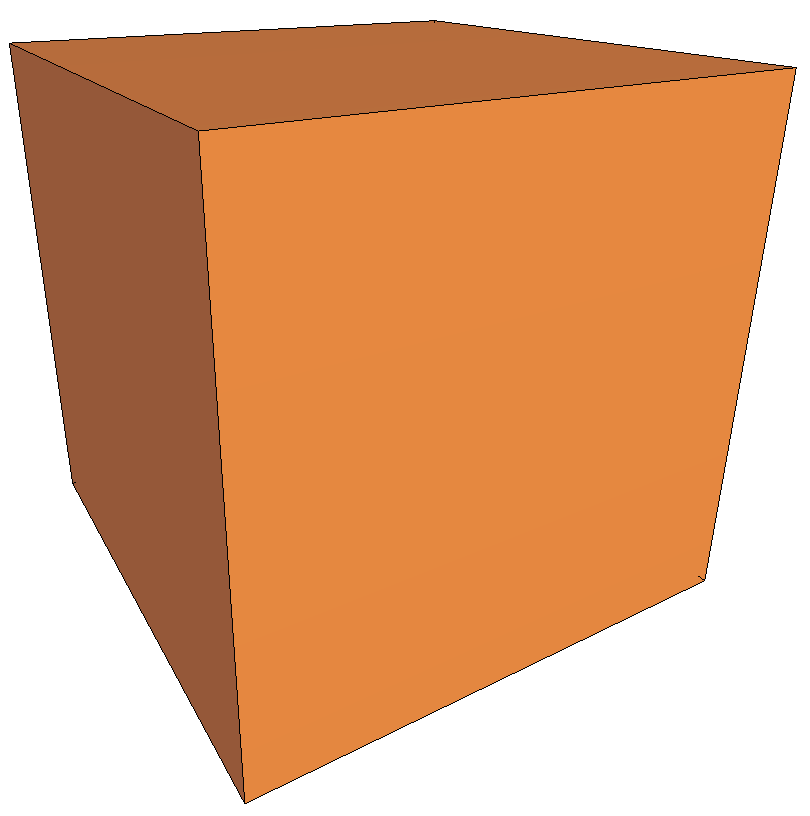} 
        \caption{ Initial two-manifold mesh.}
        \label{fig_Procedure2/0}
    \end{subfigure}
    \hfill  
    \begin{subfigure}[t]{0.48\textwidth}
        \includegraphics[width=1.0\textwidth]{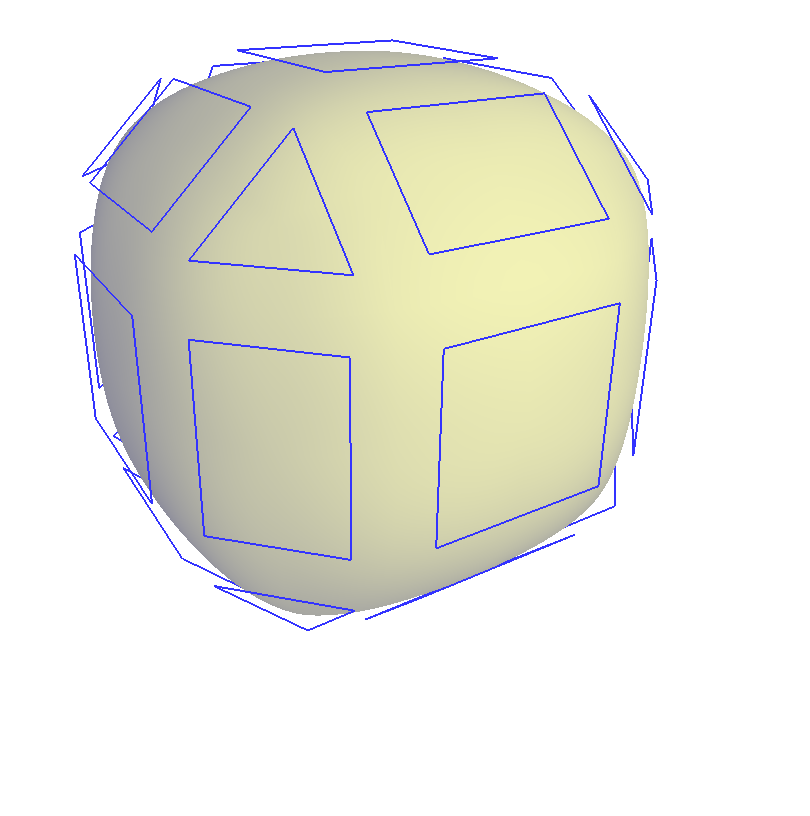}
        \caption{ Assign a polygon to each face, vertex, and edge.}
        \label{fig_Procedure2/1}
    \end{subfigure}
    \hfill  
    \begin{subfigure}[t]{0.48\textwidth}
        \includegraphics[width=1.0\textwidth]{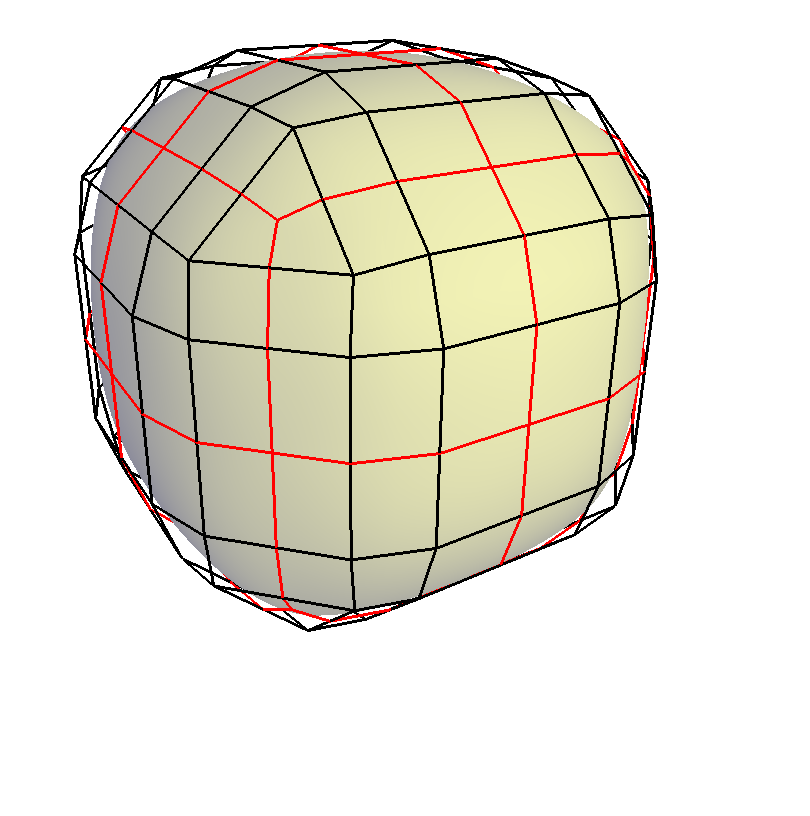}
        \caption{ Create B\'{e}zier control polyhedra.}
        \label{fig_Procedure2/3}
    \end{subfigure}
    \hfill  
    \begin{subfigure}[t]{0.48\textwidth}
        \includegraphics[width=1.0\textwidth]{Procedure/cube_6}
        \caption{ Obtain a smooth surface with modified de Casteljau.}
        \label{fig_Procedure2/6}
    \end{subfigure}
\caption{One more example demonstrating the steps of our process. Note that in the polygonal rendering, these two models appear the same. Once we turn them into smooth models, the difference between them becomes visible, and we can appropriately visualize the edge insertion operation that creates a 10-sided face.  }
\label{fig_Procedure2}
\end{figure*}

\begin{figure*}[htbp!]
    \centering  
    \begin{subfigure}[t]{0.46\textwidth}
        \includegraphics[width=1.0\textwidth]{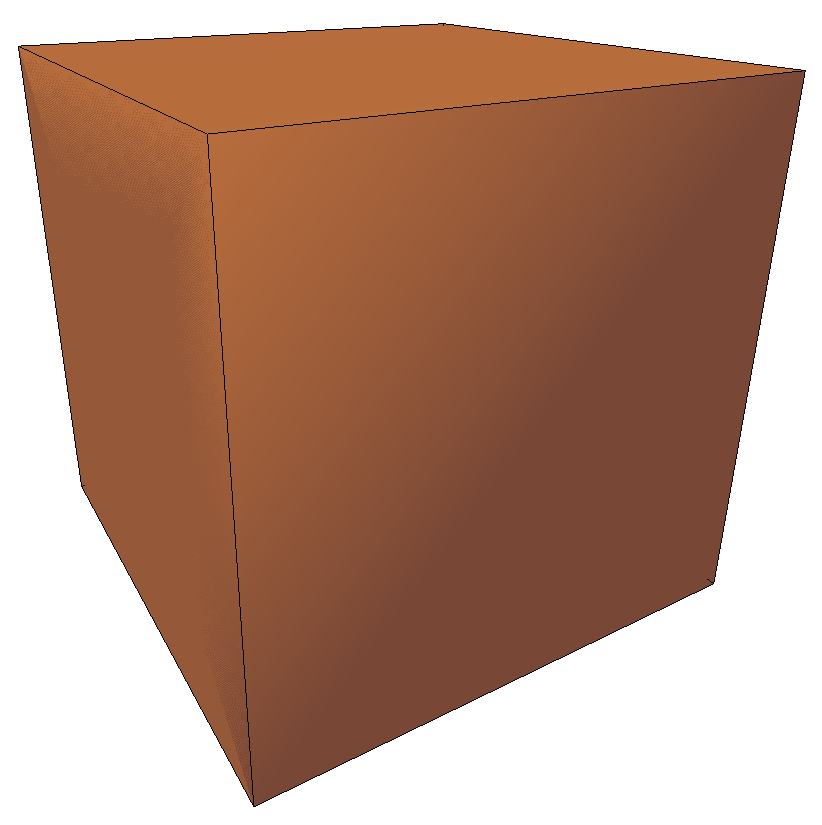}
        \caption{ Initial two-manifold mesh.}
        \label{fig_Procedure3/0}
    \end{subfigure}
    \hfill  
    \begin{subfigure}[t]{0.48\textwidth}
        \includegraphics[width=1.0\textwidth]{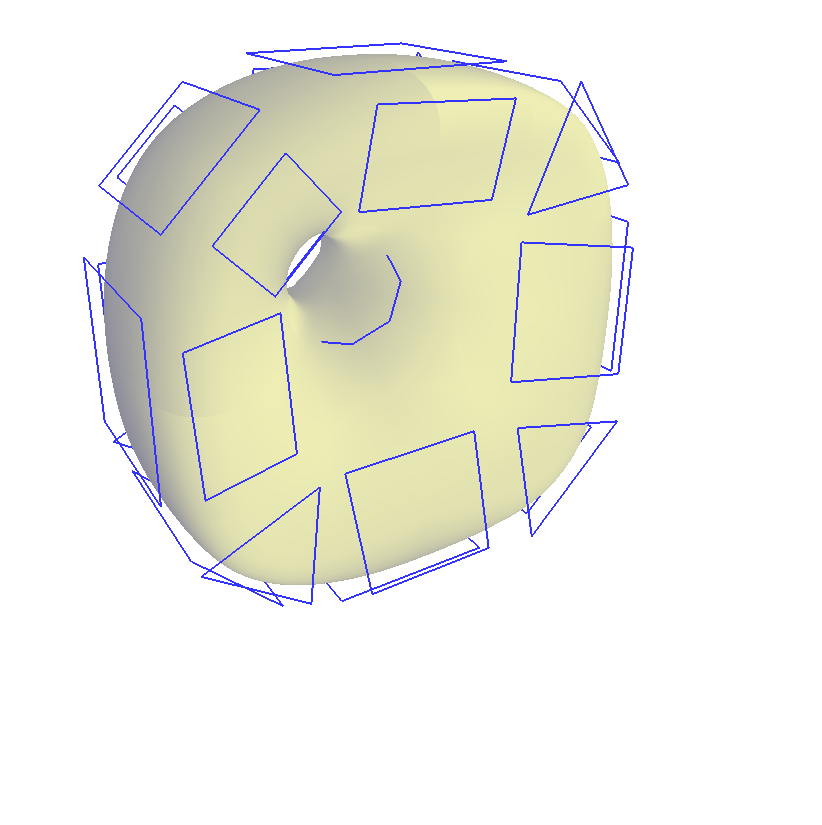}
        \caption{ Assign a polygon to each face, vertex, and edge.}
        \label{fig_Procedure3/1}
    \end{subfigure}
    \hfill  
    \begin{subfigure}[t]{0.48\textwidth}
        \includegraphics[width=1.0\textwidth]{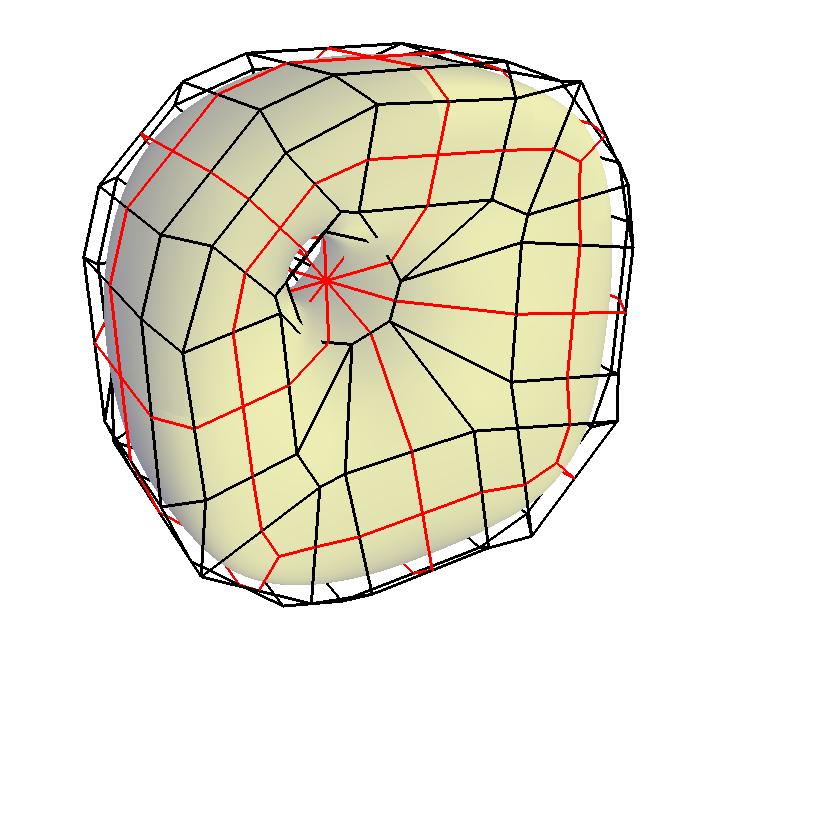}
        \caption{ Create B\'{e}zier control polyhedra.}
        \label{fig_Procedure3/3}
    \end{subfigure}
    \hfill  
    \begin{subfigure}[t]{0.48\textwidth}
        \includegraphics[width=1.0\textwidth]{Procedure/10SidedFace_6}
        \caption{ Obtain a smooth surface with modified de Casteljau.}
        \label{fig_Procedure3/6}
    \end{subfigure} 
\caption{Another example demonstrating the five steps of our process. Note that in the polygonal rendering, these two models appear the same. Once we turn them into smooth models, the difference between them becomes visible, and we can appropriately visualize the edge insertion operation that creates a 10-sided face.  }
\label{fig_Procedure3}
\end{figure*}

\begin{figure*}[htbp!]
    \centering 
        \begin{subfigure}[t]{0.55\textwidth}
        \includegraphics[width=1.0\textwidth]{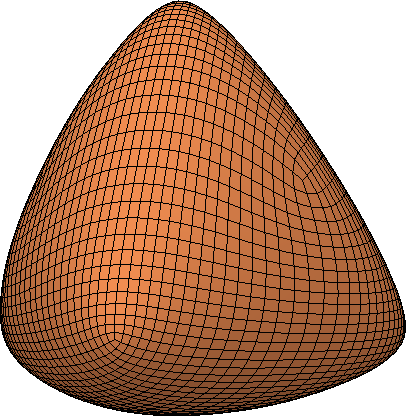}
        \caption{The polygon mesh obtained from surface in Figure~\ref{fig_Procedure/tetrahedron_6}.}
        \label{fig_Procedure/tetrahedron_7}
    \end{subfigure}
    \hfill  
    \begin{subfigure}[t]{0.49\textwidth}
        \includegraphics[width=1.0\textwidth]{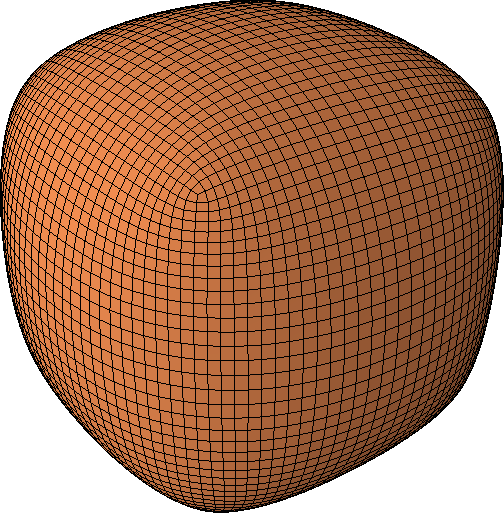}
        \caption{The polygon mesh obtained from surface in Figure~\ref{fig_Procedure2/6}.}
        \label{fig_Procedure2/7}
    \end{subfigure}
    \hfill  
    \begin{subfigure}[t]{0.49\textwidth}
        \includegraphics[width=1.0\textwidth]{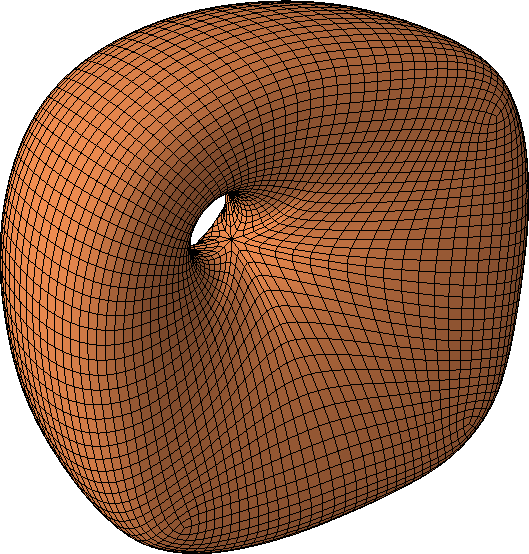}
        \caption{The polygon mesh obtained from surface in Figure~\ref{fig_Procedure3/6}.}
        \label{fig_Procedure3/7}
    \end{subfigure}
    \hfill  
\caption{These examples demonstrate the smoothed polygonal mesh obtained from two-manifold meshes shown in Figures~\ref{fig_Procedure}, \ref{fig_Procedure2}, and~\ref{fig_Procedure2}.  }
\label{fig_Smoother_Polyhedra}
\end{figure*}

\subsection{High-Level Summary of Our Implementation}

Using this theoretical approach, we have developed a simple process to interactively model complicated smooth surfaces. Our process produces hierarchically organized bicubic B\'{e}zier patches that are stitched with $G^1$ continuity in extraordinary vertices and $C^1$ continuity along the edges emanating from extraordinary vertices. A high-level summary of our implementation approach to producing smooth stitched bicubic B\'{e}zier patches from any given two-manifold mesh can be given as a five-step process. Figures~\ref{fig_Procedure}, \ref{fig_Procedure2}, and~\ref{fig_Procedure2} provide examples that demonstrate our process. It also shows the superiority of smooth modeling over polygon modeling: Although there is no visual difference between two polygonal models, since the additional edge is not visible, smooth models clearly show the hole resulting from the insert edge operation.

\begin{enumerate}

\item Start with an orientable two-manifold mesh (see Figure~\ref{fig_Procedure/tetrahedron_0}). 

\item Assign a planar convex polygon with $K=n$ sides to every face with $n$, a planar convex polygon with $K=m$ sides to every $m$ valent vertex, and a polygon with $K=4$ sides to every edge of the initial two-manifold mesh. Polygons with the $4$ sides, i.e. quadrilaterals, do not have to be planar. (see Figure~\ref{fig_Procedure/tetrahedron_1}) 

\item Using the vertex insertion scheme, which is the remeshing scheme of Catmull-Clark subdivision \citep{catmull1978}, obtain a quadrilateral decomposition of the initial two-manifold mesh. Using this quadrilateral decomposition, construct a $4 \times 4$ bicubic B\'{e}zier control points in such a way that every resulting control polyhedron shares the same boundary conditions with its four neighboring control polyhedra (see Figure~\ref{fig_Procedure/tetrahedron_3}).  

\item If the patch is regular, simply render it by evaluating the corresponding B\'{e}zier polynomial. If the patch is not regular, subdivide it into two four patches by applying the modified de Casteljau subdivision that preserves the boundary conditions. Repeat the process until we reach the desired resolution (see Figure~\ref{fig_Procedure/tetrahedron_6}).

\item If desired, convert it back to a two-manifold polygonal mesh (see Figure~\ref{fig_Procedure/tetrahedron_7}).  
\end{enumerate}

\begin{figure}[htb!]
    \begin{subfigure}[t]{0.490\textwidth}
        \includegraphics[width=1.0\textwidth]{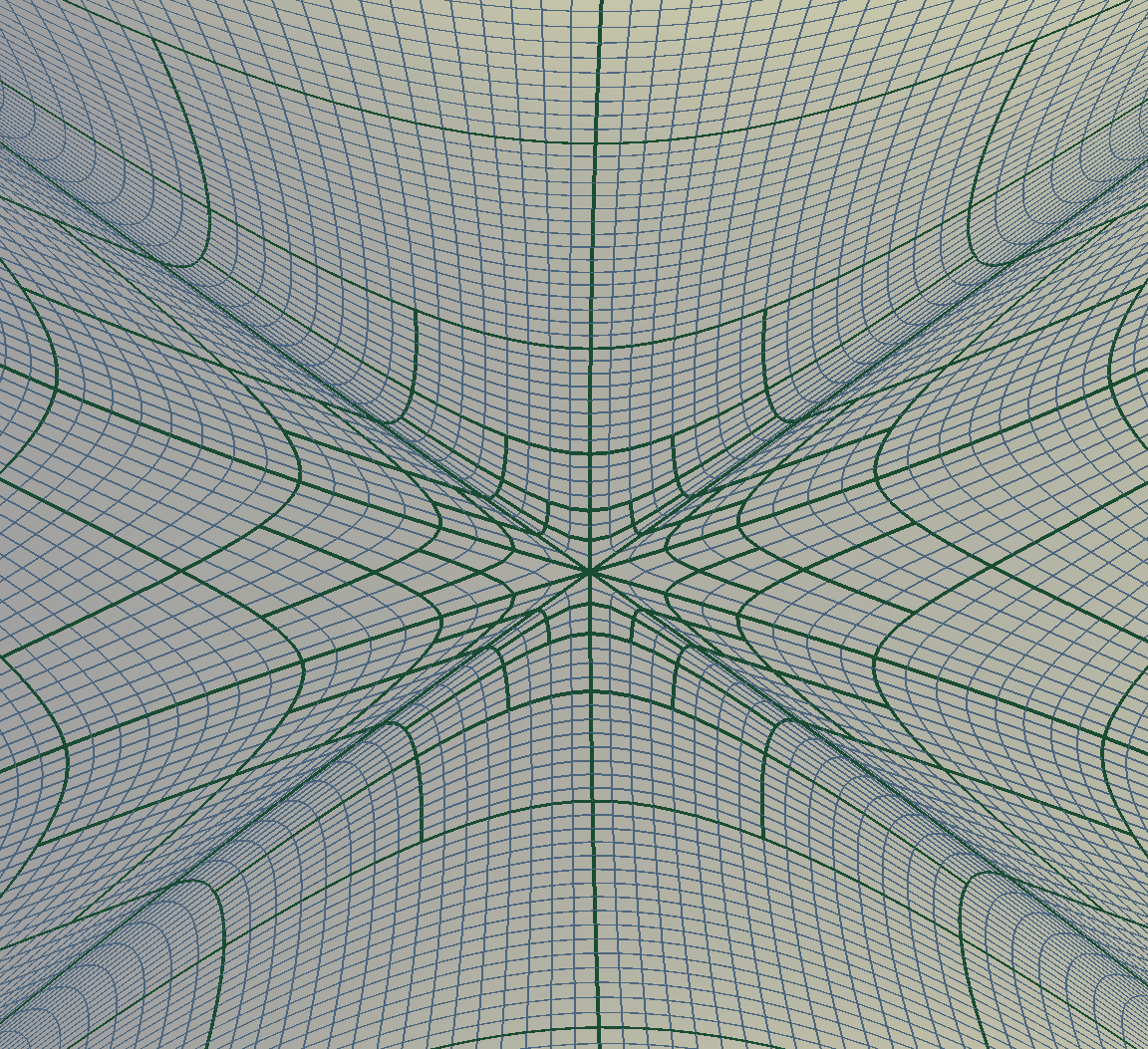}
        \caption{ With a star polygon.}
        \label{fig_cubehole14}
    \end{subfigure}
    \hfill     
    \begin{subfigure}[t]{0.490\textwidth}
        \includegraphics[width=1.0\textwidth]{comparison/cubehole5}
        \caption{ With a convex polygon.}
        \label{fig_cubehole25}
    \end{subfigure}
\caption{An example of the effect of the shape of the $K$-sided polygons over the visual quality of extraordinary neighborhoods: A 10-valent neighborhood obtained by using a $10$-sided (a) star polygon, and (b) convex and more regular polygon.}
\label{fig_10-valent2}
\end{figure}

We included some additional images that can provide more information about the features of our system. Uploaded videos provide more information about the system. These are not essential for this paper.  Therefore, we include these images to provide additional visual information for a clarification of the advantages of smooth modeling over direct polygonal modeling when dealing with an unusual topology.

\section{Discussion, Conclusion and Future Work \label{sec_conclusion}}

The theoretical analysis demonstrated that we can freely choose the sizes, positions, and shapes of $K \neq 4 $-sided polygons as long as they are planar star polygons. Their shapes and sizes play a role similar to Kochaneck-Bartels splines, and the decisions about their sizes and shapes belong to shape designers \cite{kochanek1984}. On the other hand, it is still possible to give some intuitive guidelines similar to Catmull-Rom splines \cite{catmull1974,yuksel2011}. 

We observe that it is better to have an almost-regular convex polyhedron. In other words, we want $\vec{V}_{k} \cdot \vec{V}_{k} \approx |\vec{V}_{k}| |\vec{V}_{k+1}| \cos(2\pi/K)$ and $|\vec{V}_{k}| \approx |\vec{V}_{k+1}|$ for every $k$. Figure~\ref{fig_10-valent2} shows the superiority of such almost regular convex polygons over star polygons. Such regular polygons can usually be obtained with Doo-Sabin refinement rules. Doo-Sabin refinement is known to be powerful in obtaining convex, almost regular, and almost planar polygons with a few iterations \cite{akleman2017}. However, we observed that for some of the unusual configurations, it takes more steps to obtain a reasonably regular polygon. Therefore, we added a second step that takes the dual polygon to be even a number of times. This second step helps us to obtain better-looking initial estimations of the polygons for unusual cases. This stage does not affect continuity $C^1$ or $G^1$, but improves the look around the extraordinary vertices of high valent. For example, in the example given in Figure~\ref{fig_10-valent2}, just a few steps of the Doo-Sabin refinement process produced a star polygon instead of a convex one. Just making it convex improves the quality significantly, as shown in Figure~\ref{fig_10-valent2}.

\begin{figure}[htb!]
    \begin{subfigure}[t]{0.470\textwidth}
        \includegraphics[width=1.0\textwidth]{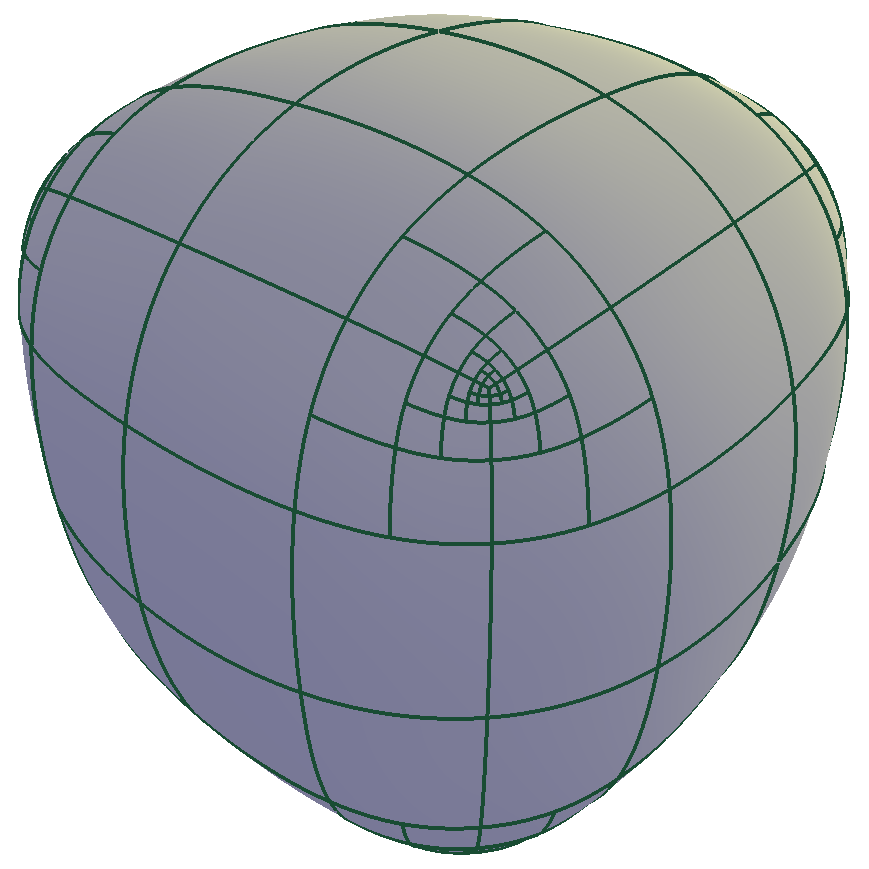}
        \caption{ A bulkier and boxier shape.}
        \label{fig_comparison2cube0}
    \end{subfigure}
    \hfill     
    \begin{subfigure}[t]{0.520\textwidth}
        \includegraphics[width=1.0\textwidth]{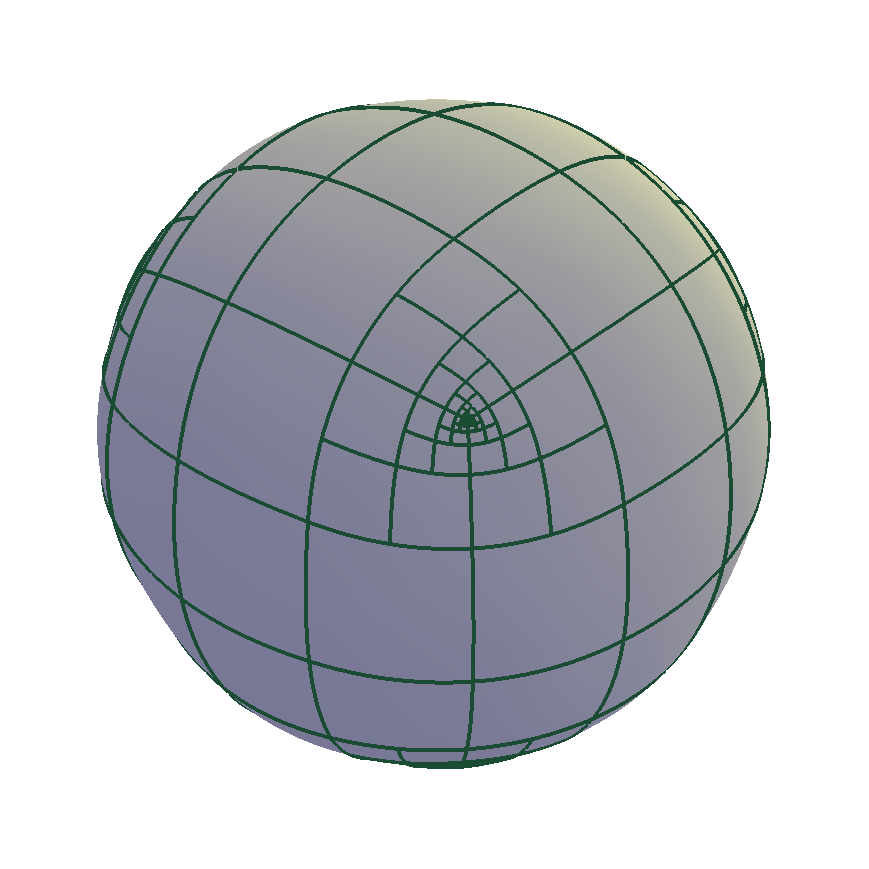}
        \caption{ A smoother shape.}
        \label{fig_comparison2cube1}
    \end{subfigure}
\caption{Effect of changing the positions and sizes of $K$-sided polygons. }
\label{fig_comparison2cubes}
\end{figure}

The positions and sizes of the $K$-sided planar polygons are also important to obtain good results. For estimating polygon centers that are interpolated by curves, it is possible to use the Doo-Sabin or Catmull-Clark refinement to estimate positions and sizes (see Figure~\ref{fig_comparison2cubes}).
As shown in Figure~\ref{fig_comparison2cubes}, it is possible to change the overall shape of the surfaces by changing the sizes and positions of planar polygons on the sides of $K$.  

Another aspect of this method is that complicated shapes can be described with a two-manifold structure that consists of only a few vertices, faces, and edges. To unleash the true power of this method for designers, there is a need for the development of user interfaces that can allow shape designers to control final surfaces more intuitively through $K$-sided polygons.

\begin{figure}[htb!]
   \centering     \includegraphics[width=0.49\textwidth]{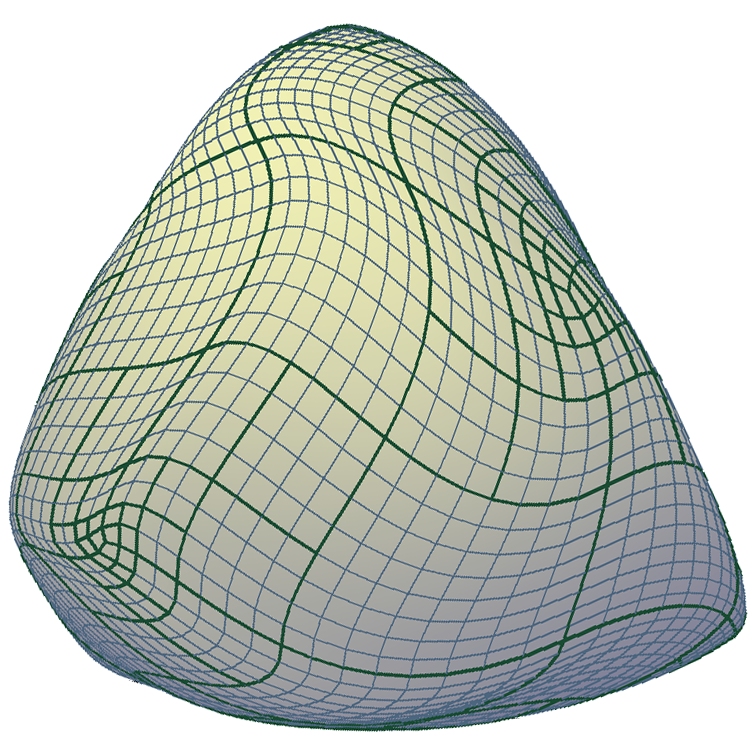}
    \includegraphics[width=0.49\textwidth]{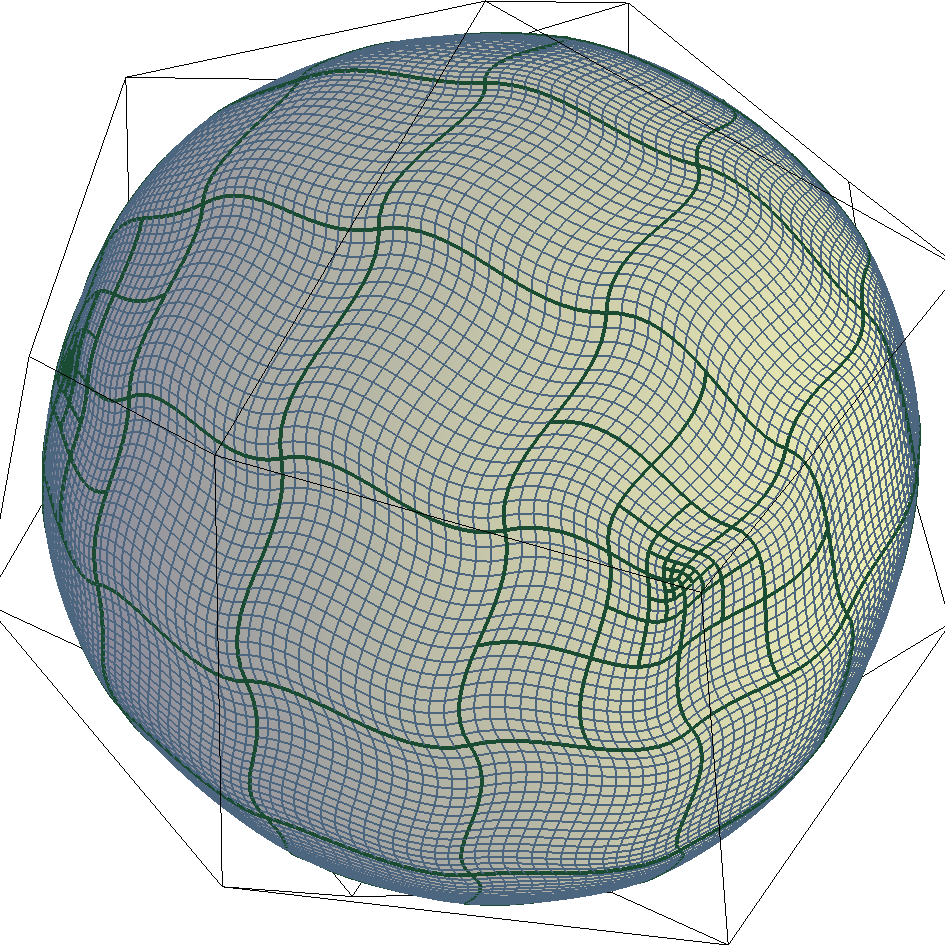}
\caption{Effect of rotating every $K$-sided polygon. }
\label{fig_rotation}
\end{figure}

An additional effect that comes free with this approach is to control the relative orientations of planar polygons with $K$ sides. Figure~\ref{fig_rotation} shows two such examples that are obtained by rotating each polygon on the same side $K$. These shapes cannot be created by any other subdivision method since we explicitly control iso-parameter curves on the surface.

In this particular implementation, to construct bicubic B\'{e}zier control polyhedra we used a vertex insertion scheme. This is not necessary. We can even start with a quad-mesh that is obtained by any quadrangulation method such as \cite{kalberer2007,bommes2009}. It is also possible to define the polygon with $K$ side only for the vertices and faces and construct bicubic B\'{e}zier control polyhedra using the $\sqrt{2}$ subdivision \cite{li2004}, that is, the dual of the simplest \cite{peters1997}. Our initial investigation suggests that the method we have introduced in this paper can also be used to obtain $C^2$-stitched tensor product B\'{e}zier patches. An application of such $C^2$-stitched surfaces can be the approximation of given shapes using Morse-Smale complexes \cite{ni2004}. 

\begin{figure}[htb!]
    \begin{subfigure}[t]{0.490\textwidth}
        \includegraphics[width=1.0\textwidth]{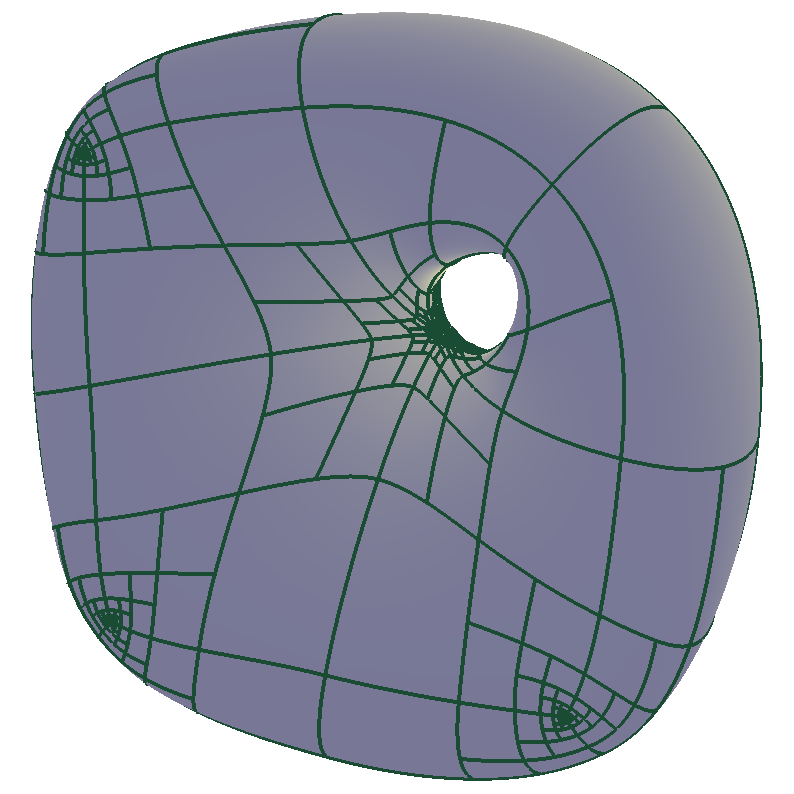}
        \caption{ A bulkier and boxier shape.}
        \label{fig_comparison2cubehole0}
    \end{subfigure}
    \hfill     
    \begin{subfigure}[t]{0.490\textwidth}
        \includegraphics[width=1.0\textwidth]{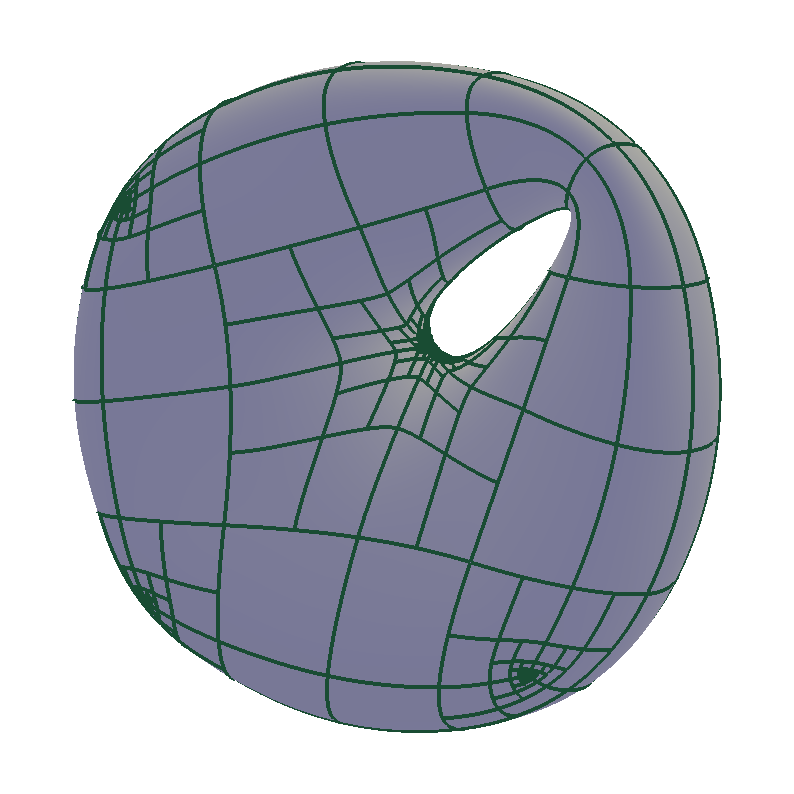}
        \caption{ A smoother shape.}
        \label{fig_comparison2cubehole1}
    \end{subfigure}
\caption{Effect of changing the positions and sizes of $K$-sided polygons. }
\label{fig_comparison2cubeholes}
\end{figure}

\begin{figure}[htb!]
    \begin{subfigure}[t]{0.99\textwidth}
        \includegraphics[width=1.0\textwidth]{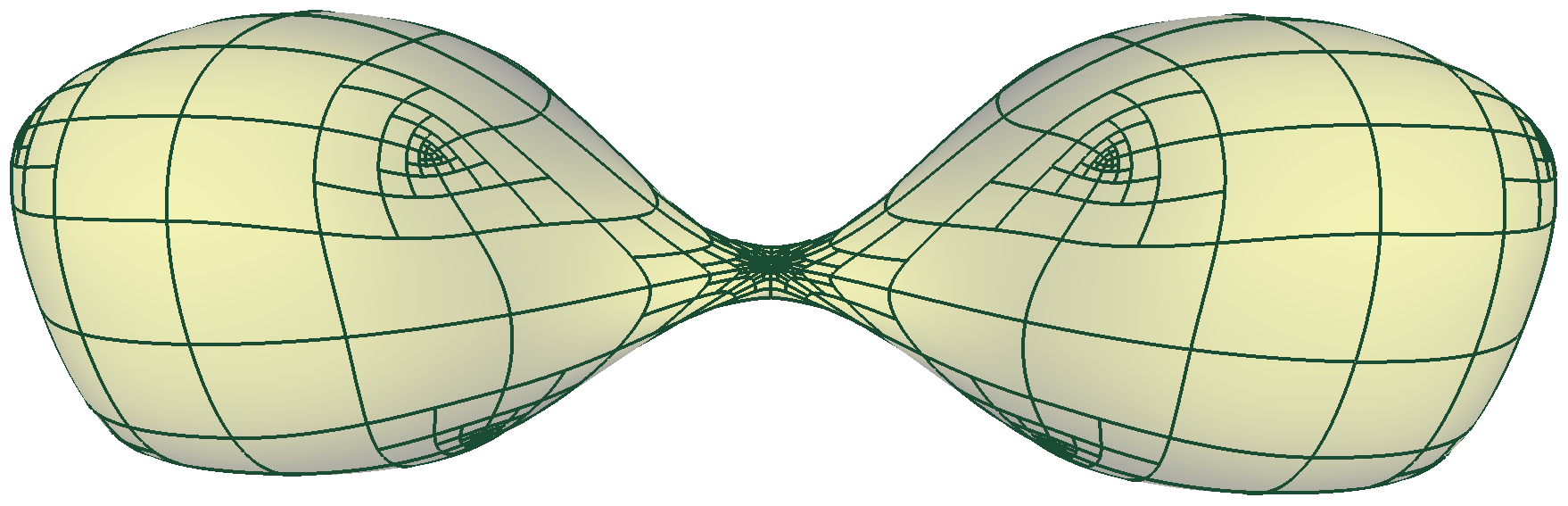}
        \caption{ A bulkier and boxier shape.}
        \label{fig_comparison2twoboxes0}
    \end{subfigure}
    \hfill     
    \begin{subfigure}[t]{0.99\textwidth}
        \includegraphics[width=1.0\textwidth]{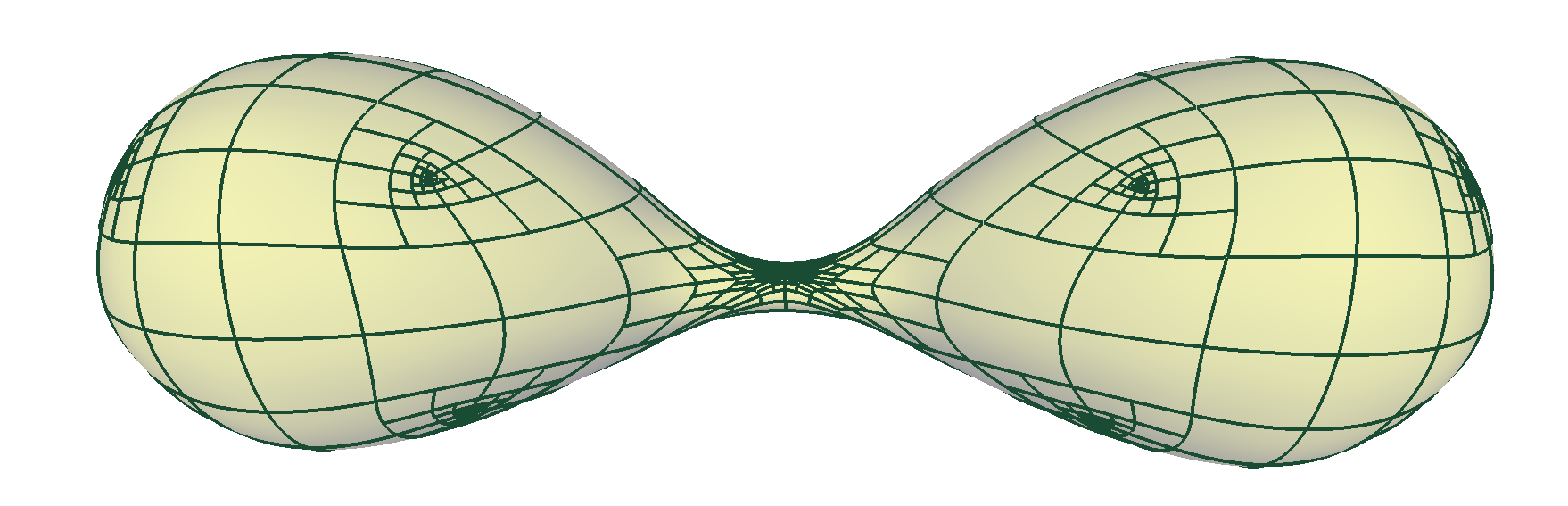}
        \caption{ A smoother shape.}
        \label{fig_comparison2twoboxes1}
    \end{subfigure}
\caption{Effect of changing the positions and sizes of $K$-sided polygons. }
\label{fig_comparison2twoboxes}
\end{figure}

\begin{figure}[htb!]
    \begin{subfigure}[t]{0.49\textwidth}
        \includegraphics[width=1.0\textwidth]{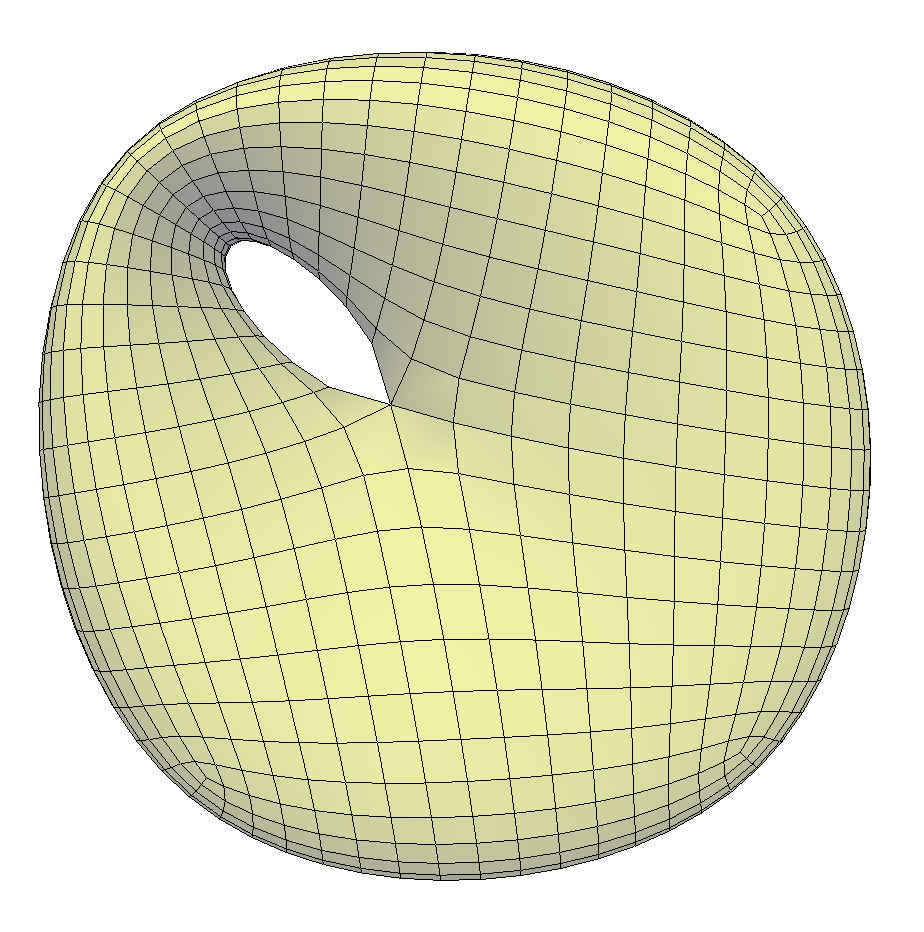}
        \caption{ Catmull-Clark suvbdivision.}
        \label{fig_comparison1/Capture0}
    \end{subfigure}
    \hfill     
    \begin{subfigure}[t]{0.49\textwidth}
        \includegraphics[width=1.0\textwidth]{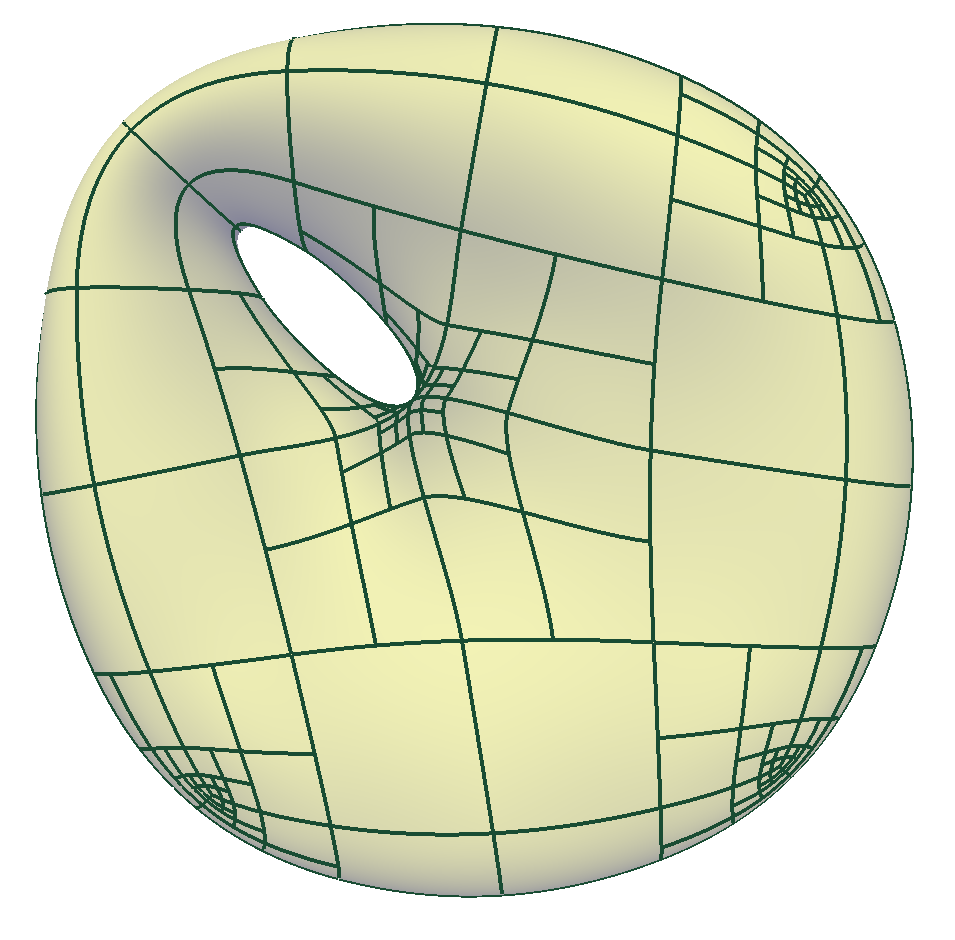}
        \caption{ Our modified de Casteljau Subdivision.}
        \label{fig_comparison1/Capture1}
    \end{subfigure}
\caption{Comparison with Catmull-Clark subdivision. Note that we get almost the same shape with smoother hole. }
\label{fig_comparison1CC1}
\end{figure}

\begin{figure}[htb!]
    \begin{subfigure}[t]{0.99\textwidth}
        \includegraphics[width=1.0\textwidth]{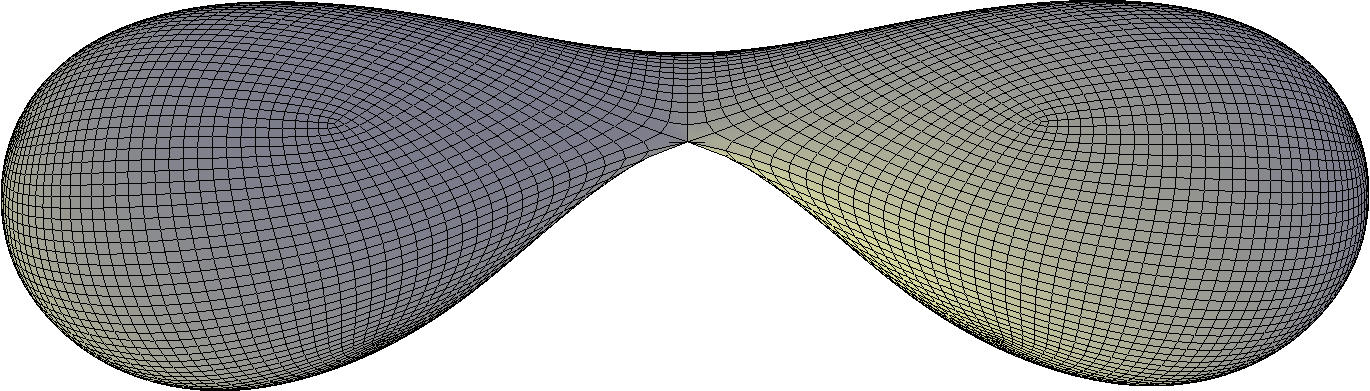}
        \caption{ Catmull-Clark suvbdivision.}
        \label{fig_comparison1/twocubes0}
    \end{subfigure}
    \hfill     
    \begin{subfigure}[t]{0.99\textwidth}
        \includegraphics[width=1.0\textwidth]{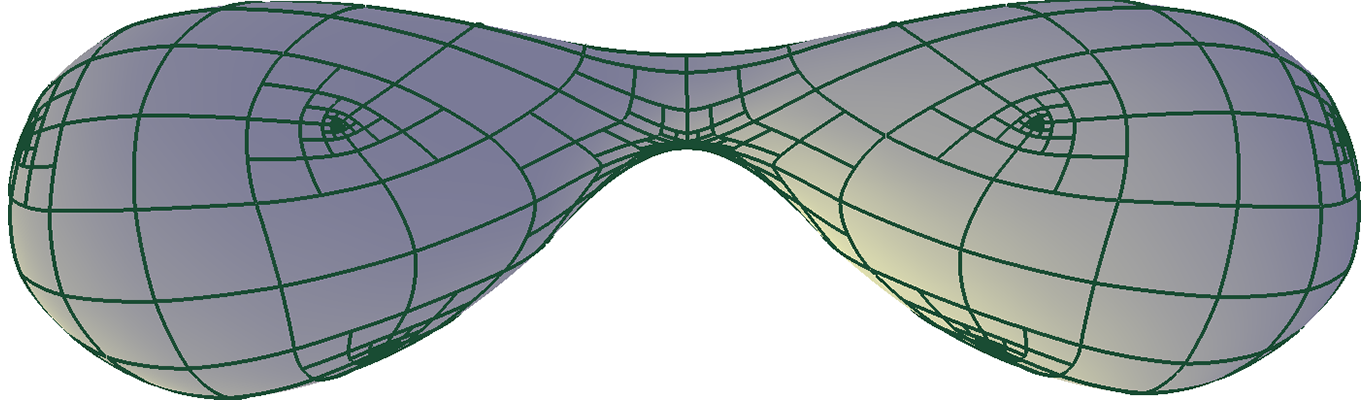}
        \caption{ Our modified de Casteljau Subdivision.}
        \label{fig_comparison1/twocubes1}
    \end{subfigure}
\caption{Another comparison with Catmull-Clark subdivision. Note that we get almost the same shape with smoother handle. }
\label{fig_comparison1CC2}
\end{figure}

\begin{figure}[htb!]
    \begin{subfigure}[t]{0.99\textwidth}
        \includegraphics[width=1.0\textwidth]{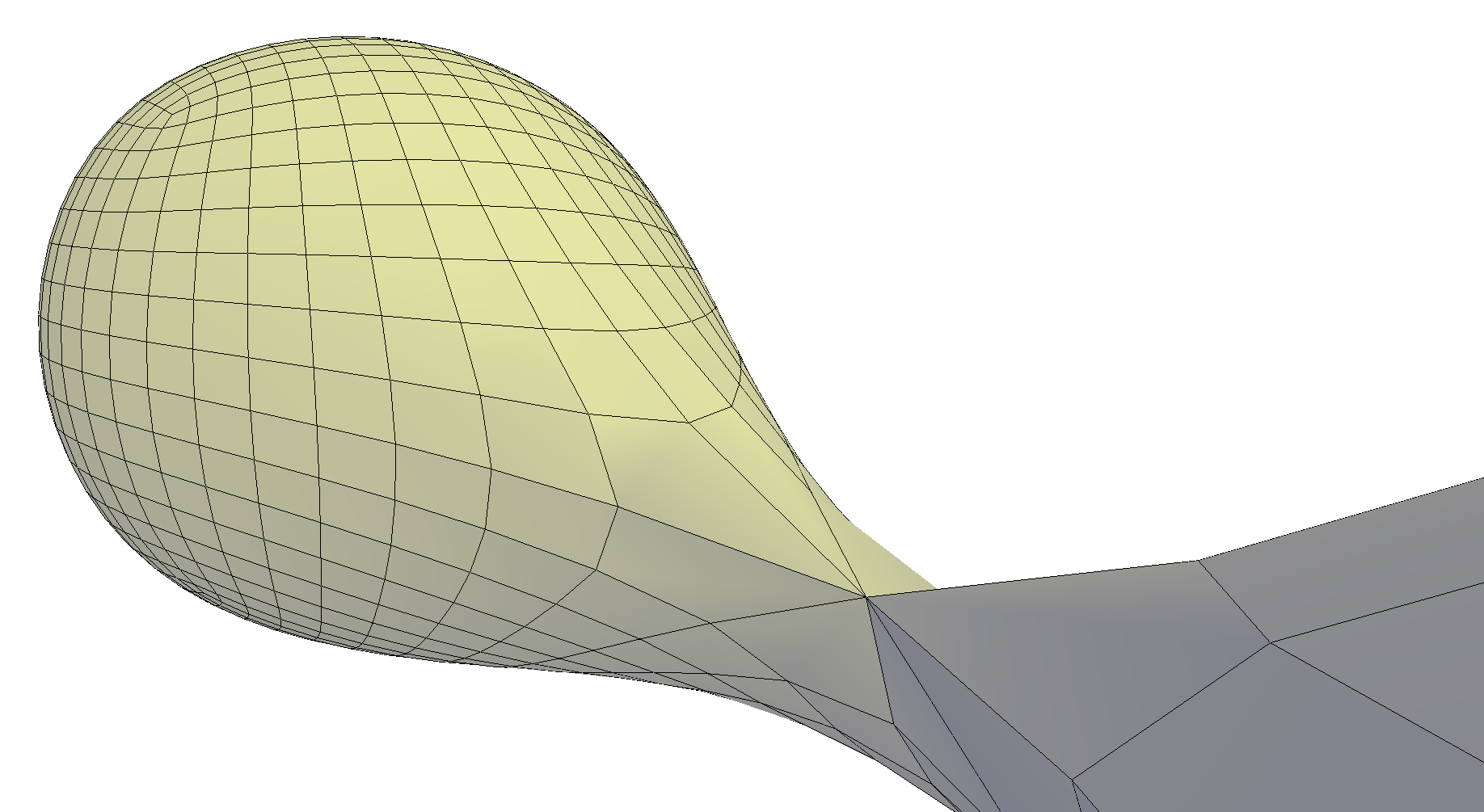}
        \caption{ Catmull-Clark suvbdivision.}
        \label{fig_comparison1/twocubes6}
    \end{subfigure}
    \hfill     
    \begin{subfigure}[t]{0.99\textwidth}
        \includegraphics[width=1.0\textwidth]{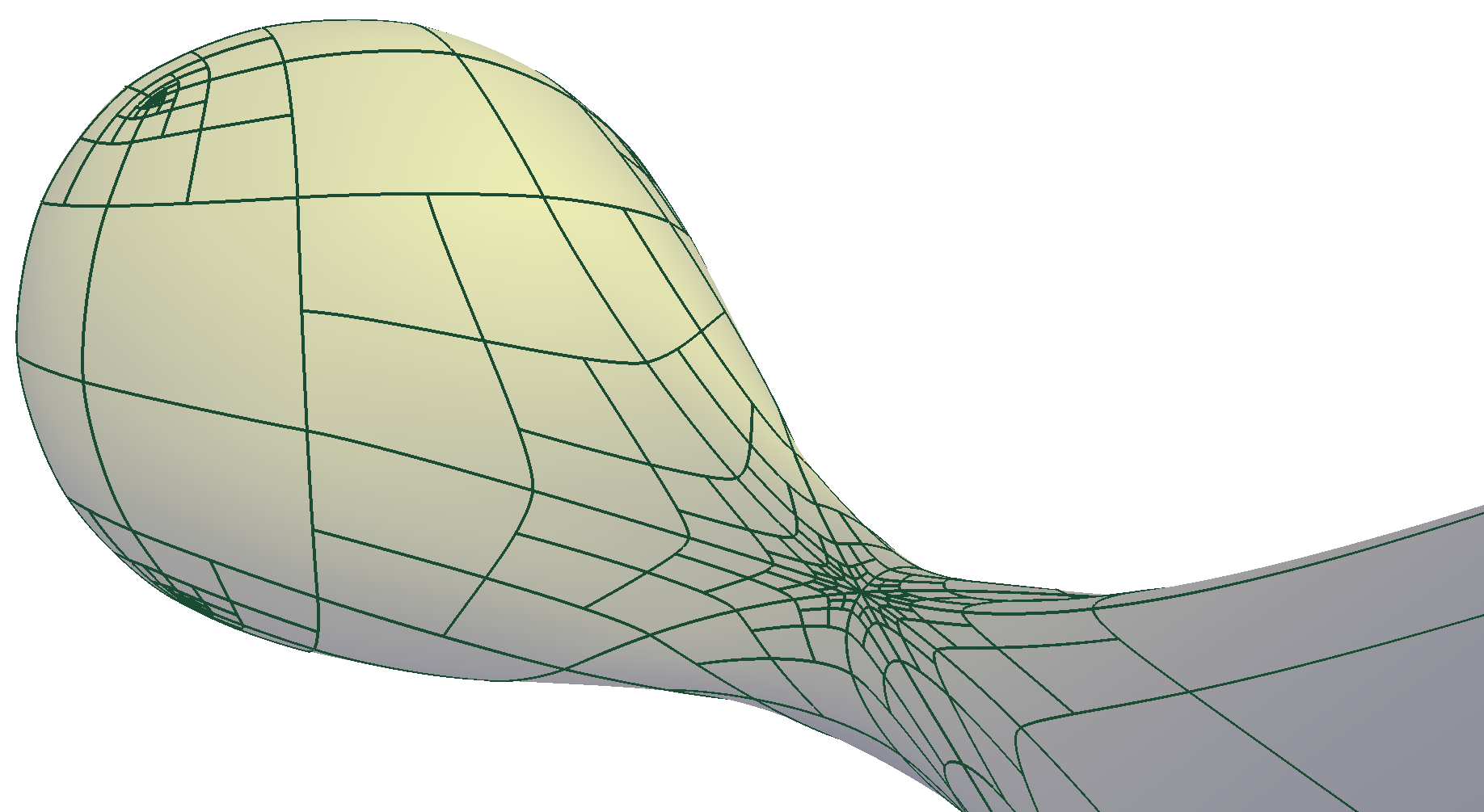}
        \caption{ Our modified de Casteljau Subdivision.}
        \label{fig_comparison1/twocubes8}
    \end{subfigure}
\caption{Another detailed comparison with Catmull-Clark subdivision. Note that it is clear that the handle is smoother. }
\label{fig_comparison1CC2detail}
\end{figure}

Figures~\ref{fig_comparison2cubeholes} and~\ref{fig_comparison2twoboxes} provide two more examples that demonstrate the effect of changing the positions and sizes of polygons with sides $K$. These examples show that by providing more control to designers, it can be possible to control the shapes of holes and handles. This also suggests that we can simply animate the geometry by changing these polygons with $K$ sides. Figures~\ref{fig_comparison1CC1}, \ref{fig_comparison1CC2}, and \ref{fig_comparison1CC2detail} further demonstrate the visual quality resulting from our modified de Casteljau subdivision compared to the Catmull-Clark subdivision. 

\begin{figure}[htb!]
    \begin{subfigure}[t]{0.99\textwidth}
        \includegraphics[width=1.0\textwidth]{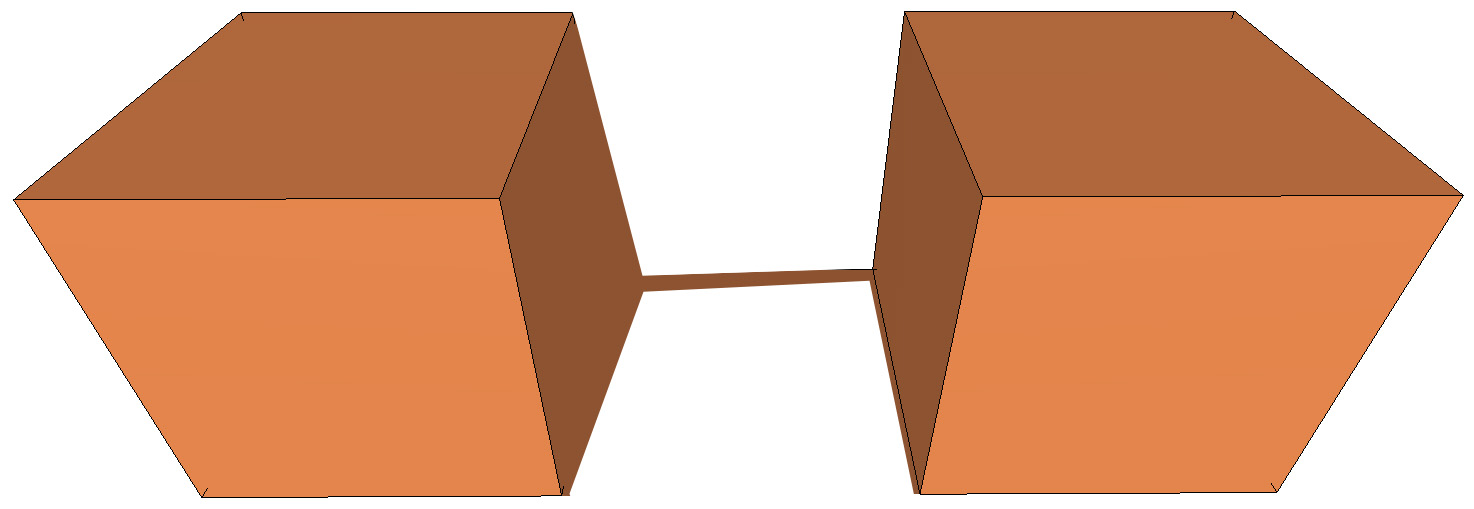}
        \caption{ Initial polygonal mesh that is created by connecting two cubes with one edge, which creates a bridge between two cubes.}
        \label{fig_procedure2TwoCubes_0}
    \end{subfigure}
    \hfill     
    \begin{subfigure}[t]{0.99\textwidth}
        \includegraphics[width=1.0\textwidth]{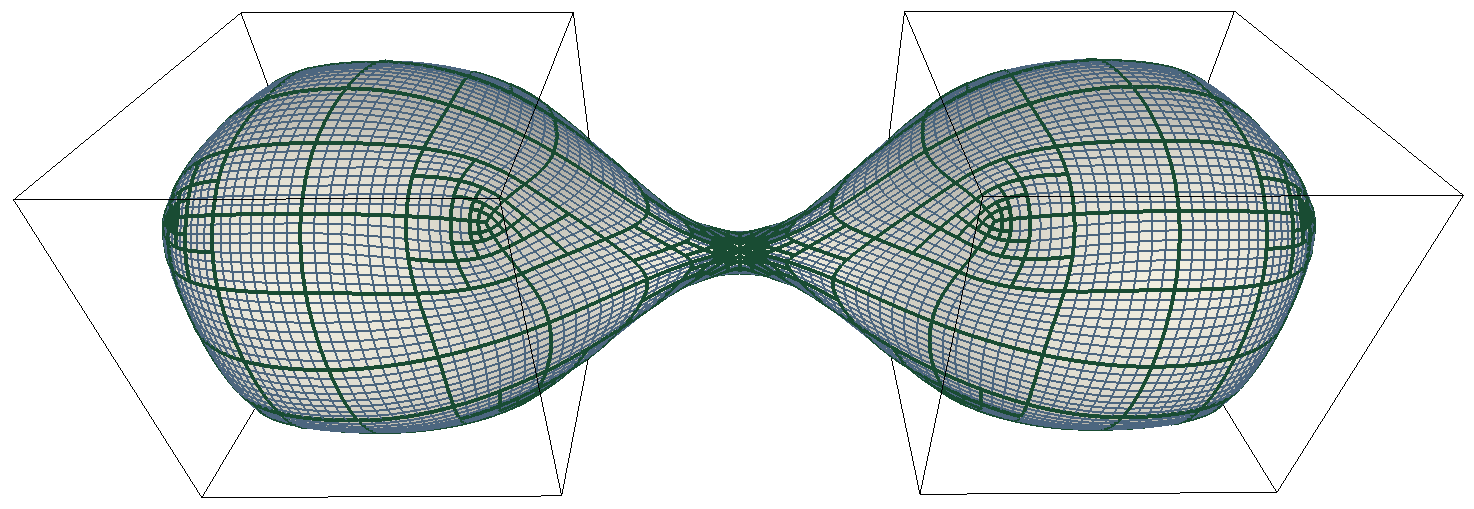}
        \caption{Corresponding smooth surface.}
        \label{fig_procedure2TwoCubes_6}
    \end{subfigure}
\caption{Smoothing two cubical shapes that are connected by one edge, which creates a bridge. }
\label{fig_procedure2TwoCubes}
\end{figure}

\bibliographystyle{unsrtnat}
\bibliography{references}

\end{document}